\documentclass[10pt,preprint]{aastex}

\usepackage{caption}

\def\chandra{{\it Chandra}}

\newcommand{\cgs}{ ${\rm erg~cm}^{-2}~{\rm s}^{-1}$} 
\newcommand{\lum}{\rm erg~s$^{-1}$}
\def\gtrsim{\mathrel{\hbox{\rlap{\hbox{\lower4pt\hbox{$\sim$}}}\hbox{$>$}}}}

\def\lum{{\rm {erg~s$^{-1}$}}}
\usepackage{txfonts}
\usepackage{natbib}

\bibpunct{(}{)}{;}{a}{}{,}
\slugcomment{Version: Accepted on May 27 2014}

\shorttitle{\chandra\ COSMOS elliptical galaxies}
\shortauthors{}

\begin{document}

 \title{Early-type galaxies in the \chandra\ COSMOS Survey}

\author{F. Civano\altaffilmark{1,2,3}, G. Fabbiano\altaffilmark{2}, S. Pellegrini\altaffilmark{4}, D-W. Kim\altaffilmark{2}, A. Paggi\altaffilmark{2}, R. Feder\altaffilmark{5} and M. Elvis\altaffilmark{2}} 

\altaffiltext{1}{Yale Center for Astronomy and Astrophysics, 260 Whitney ave, New Haven, CT 06520, USA}
\altaffiltext{2}{Smithsonian Astrophysical Observatory, 60 Garden st., Cambridge, MA 02138, USA}
\altaffiltext{3}{Department of Physics and Astronomy, Dartmouth College, Wilder Laboratory, Hanover, NH 03855, USA}
\altaffiltext{4}{Universit\'a degli studi di Bologna, Dipartimento di Astronomia, Via Ranzani 1, 40127 Bologna, Italy}
\altaffiltext{5}{Great Neck South High School, 341 Lakeville Road Great Neck, New York 11020, USA}



\begin{abstract}
We study a sample of 69 X-ray detected Early Type Galaxies (ETGs), selected from the \chandra\ COSMOS survey, to explore the relation between the X-ray luminosity of hot gaseous halos (L$_{X, gas}$) and the integrated stellar luminosity (L$_K$) of the galaxies, in a range of redshift extending out to z=1.5. In the local universe a tight steep relationship has been stablished between these two quantities  ($L_{X,gas}\sim L_K^{4.5}$) suggesting the presence of largely virialized halos in X-ray luminous systems. We use well established relations from the study of local universe ETGs, together with the expected evolution of the X-ray emission, to subtract the contribution of low mass X-ray binary populations (LMXBs) from the X-ray luminosity of our sample. Our selection minimizes the presence of active galactic nuclei (AGN), yielding a sample representative of normal passive COSMOS ETGs; therefore the resulting luminosity should be representative of gaseous halos, although we cannot exclude other sources such as obscured AGN, or enhanced X-ray emission connected with embedded star formation in the higher z galaxies.
We find that most of the galaxies with estimated L$_X<$10$^{42}$ \lum\ and z$<$0.55 follow the $L_{X,gas}- L_K$ relation of local universe ETGs. For these galaxies, the gravitational mass can be estimated with a certain degree of confidence from the local virial relation. However, the more luminous (10$^{42}<L_X<$10$^{43.5}$ \lum) and distant galaxies present significantly larger scatter; these galaxies also tend to have younger stellar ages. The divergence from the local $L_{X,gas}- L_K$  relation in these galaxies implies significantly enhanced X-ray emission, up to a factor of 100 larger than predicted from the local relation. We discuss the implications of this result for the presence of hidden AGN, and the evolution of hot halos, in the presence of nuclear and star formation feedback. 

\end{abstract}


\section{Introduction}
The discovery of widespread diffuse X-ray emission from early type galaxies (elliptical and SO, ETG hereafter) with the Einstein X-ray Observatory (Trinchieri \& Fabbiano 1985; Forman et al. 1985) spurred speculations on the nature of this emission and on the amount that could be ascribed to gaseous halos. The latter, if in hydrostatic equilibrium, would provide a unique way for measuring the total gravitational mass of these galaxies. However, with the earlier data, it was virtually impossible to discriminate between gaseous and non-gaseous emission in these galaxies. These observational limitations left the field open to a lively discussion, aimed at understanding the nature of the X-ray emission, and how this emission could constrain physical models of halo evolution, including SNIa and nuclear feedback and interaction with cluster and group hot media (see e.g., Fabbiano 1989; Canizares et al. 1987; Ciotti et al. 1991; David et al. 1991; White \& Sarazin 1991; Matthew \& Brighenti 2003; Ciotti et al. 2010).  

With the high angular resolution and sensitivity of \chandra, coupled with the spectral capabilities of the ACIS detector (Garmire et al. 2003), the study of the X-ray emission of ETGs in the local universe has taken a substantial leap forward. The presence of populations of low-mass X-ray binaries (LMXBs, Trinchieri \& Fabbiano 1985) has been definitely proven, and these LMXBs have been detected and studied in several ETGs, to distances of several ten megaparsecs (see review by Fabbiano 2006).
Using \chandra\ observations, and the information gathered from deep studies of nearby ETGs, which established the LMXB luminosity function (e.g., Kim \& Fabbiano 2004; Gilfanov 2004;  Kim et al. 2009), Boroson, Kim and Fabbiano (2011; BKF), were able to estimate accurately the gaseous component of the X-ray luminosity (L$_{X, Gas}$) of 30 ETGs within a distance of 32 Mpc and establish scaling relations for this emission. 

Kim \& Fabbiano (2013; KF13) took this approach a step further by establishing that L$_{X, Gas} \propto M_{Total}^{\sim3}$, where M$_{Total}$ is the total mass obtained from kinematic measurements of globular clusters and planetary nebulae (located within $\sim$5 effective radii from the center) for a small sample of 14 ETGs. This scaling relation, together with the steep L$_{X, Gas} \propto T_{Gas}^{4.5}$ of BKF suggest that the hot gas is virialized, at least for L$_{X, Gas} > 10^{40}$ \lum. If this result holds in general for ETGs, we may have finally a way to measure the total mass of these galaxies. 

Not many studies of X-ray selected ETGs have been performed beyond the local Universe. 
While Tzanavaris \& Georgantopoulos (2008) mainly focused on the X-ray luminosity function of ETGs in the \chandra\ Deep Field North, finding no evolution out to $z=0.67$, 
Lehmer et al. (2007) and Danielson et al. (2012) studied the X-ray (using both detections and stacking analysis) and multiwavelength properties of optically selected ETGs in the \chandra\ Deep Field South.  
The \chandra\ Deep Fields works consistently find that the mean $L_X$  of $L_B\geq 10^{10}$ L$_{\odot}$ ETGs remains equal or mildly increases
over z=0-0.7 (Lehmer et al. 2007), and that the soft X-ray
luminosity/B-band luminosity evolves mildly as $L(0.5-2$keV)/$L_B\propto [1+z]^{1.1\pm 0.7}$ since $z\approx 1.2$ (Danielson et al. 2012).
This was taken as evidence for some heating mechanism preventing the hot gas from cooling, and this mechanism was found to be 
consistent with mechanical heating from radio AGNs. Indeed, a number of processes, in addition to radiative cooling and radio AGN heating, 
are expected to contribute to the evolution of the hot gas: a continuous, time decreasing, rate of mass input from stellar winds and energy input from
Type Ia supernovae, periodic heating from nuclear radiative outbursts and AGN winds, and galaxy interactions (to quote those expected to be most relevant;
e.g., Kim \& Pellegrini 2012). To quantify the role of each of these processes requires detailed, and self-consistent, hydrodynamical simulations of the gas evolution, as for example those made by Ciotti et al. (2010). Subsequently Pellegrini et al. (2012) derived expected observational properties for the latter models, finding that,
outside nuclear outbursts, the gas keeps at a roughly constant average T, and shows a mild secular decline of Lx, in agreement with current observations.

Jones et al. (2013), starting from a K-band limited and selected sample of 3500 galaxies in the COSMOS field at 0.5$<z<$2, performed X-ray stacking analysis using the \chandra\ data available in the field (see below) dividing the sources according to galaxy types and redshift bins. For older galaxies they find that
nuclear activity or hot gas dominate the X-ray emission, and a slight increase in X-ray luminosity with redshift.

The COSMOS survey (Scoville et al. 2007), resulting in a multi-wavelength characterization of over 1.5 million galaxies at redshift up to 5 in 2 deg$^2$ of the sky, and with \chandra\ X-ray coverage (C-COSMOS, Elvis et al. 2009) over 0.9 deg$^2$, gives us the means to select X-ray ETGs extending local studies to higher redshifts and explore the redshift evolution of hot halos. The small area covered by the deep fields allows to probe higher redshifts and fainter sources, while the larger area of COSMOS is optimal for this study detecting brighter and more rare sources. 

In this paper, we use the sample of ETGs detected in X-rays in the C-COSMOS survey (Section 2), to expand the ETGs ``local sample''\footnote{We will call hereafter ETGs ``local sample'' the sample including the 30 ETGs from BKF plus 8 ETGs added to this sample subsequently by KF13.} to higher redshifts, and explore if the BKF relation is valid at higher redshift and test whether there is evolution of the scaling relation with redshift. We subtract the LMXB contribution from their X-ray luminosity (Section 3.2), obtaining an estimate of the L$_{X, Gas}$ plus an unknown contribution of nuclear emission, which we try to constrain using the observational properties of the sample. The resulting L$_X$-L$_K$ relation is examined in Section 4, as a function of several inferred and observational properties of the ETGs (Section 3). This relation is consistent with the L$_{X, Gas}$ - L$_K$ relation derived in the local universe (BKF; KF13) for sources with X-ray luminosity  L$_X<$ 10$^{42}$ \lum, suggesting that the hot gas in these galaxies may be virialized as well. We define a local strip and analyze the outliers. For the sources in the local strip, we estimate their total masses using the KF13 relation (Section 5).   

We assume a cosmology with H$_0$ = 71~km~s$^{-1}$~Mpc$^{-1}$, 
$\Omega_M$ = 0.27 and $\Omega_{\Lambda}$= 0.73. The AB magnitude system is used in this paper if not otherwise stated.

\section{Sample Selection}
The galaxies in our sample were selected from the C-COSMOS X-ray source identification catalog (Civano et al. 2012) to have a rest frame X-ray luminosity of L$_X <5 \times 10^{43}$ \lum\ in the 0.5-10 keV band (using the spectral assumptions in Elvis et al. 2009) and to be classified as elliptical or S0 from their optical to near-infrared spectral energy distribution (SED). 
The SED identification is from the most recent version of the photometric catalog of Ilbert et al. (2009), which includes the near-infrared photometry from the Ultra Deep Survey with the VISTA telescope (Ultra-VISTA; McCracken et al. 2013). For the SED fit, Ilbert et al. (2009) used 7 elliptical galaxy templates (see their Figure 1) with ages from 2 to 13 Gyr (Polletta et al. 2007). Ilbert et al. (2010) derived galaxy properties (mass, age and star formation rate) for the galaxies in the same COSMOS sample. We also use these properties here. More details on how the properties were derived can be found in their paper. Briefly, the SED templates were generated with the stellar population synthesis package developed by Bruzual \& Charlot (2003), assuming an initial mass function from Chabrier (2003) and an exponentially declining star formation history.

\begin{figure}
\centering
\includegraphics[width=0.45\textwidth, angle=0]{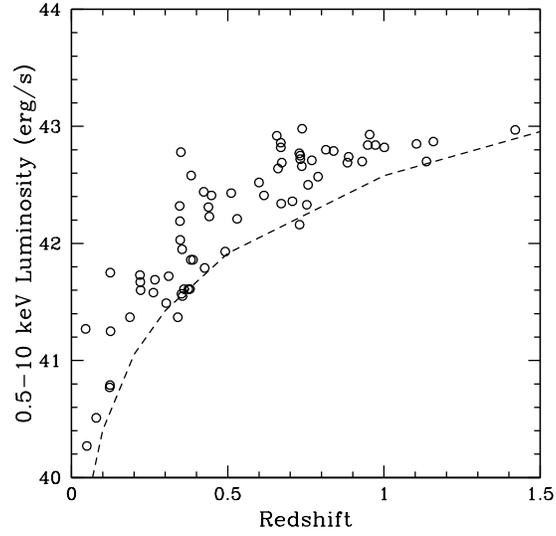}
\caption{\small X-ray luminosity in the 0.5-10 keV band (rest frame) versus redshift for all the X-ray ETGs in the C-COSMOS survey. The luminosity limit of C-COSMOS at 20\% completeness (F$_X$=1.1$\times$10$^{-15}$ \cgs) is reported as dashed line. }
\label{lxz}
\end{figure}

Our sample includes 69 sources; we will refer to it as the X-ray ETG sample hereafter. 
The C-COSMOS survey sensitivity limit 1.1$\times$10$^{-15}$ \cgs\ (at 20\% completeness, dashed line in Figure \ref{lxz}; Puccetti et al. 2009) in the 0.5-10 keV band is such that at z=0.8 the minimum detectable X-ray luminosity is 10$^{42.3}$ \lum. The luminosities in the 0.5-10 keV band (computed assuming a model with Galactic column density N$_{H, Gal}$=2.6$\times$10$^{20}$cm$^{-2}$, Kalberla et al. 2005), and a power law slope  $\Gamma$=1.4 \footnote{This slope value is used here for the sole purpose of comparing the luminosity of the sample with the C-COSMOS survey limit.} as used in the C-COSMOS catalog) for the 69 X-ray ETGs are plotted in Figure \ref{lxz}. The flux limit applied to the sample is consistent with the signal-to-noise ratio thresholds chosen by Puccetti et al. (2009), on the basis of extensive simulations, to avoid the Eddington bias in the computation of the number counts of the entire C-COSMOS sample. Thus, Eddington bias is not affecting this sample and analysis.

As a comparison sample, we selected from the COSMOS photometric catalog, covering the full 2 deg$^2$ of the COSMOS field, all the elliptical and S0 galaxies with reliable Ultra-VISTA photometry in the J and K band and photometric redshifts. To be consistent with the X-ray selection, we consider only the sources in the C-COSMOS area. The comparison sample comprises $\sim$6600 galaxies and we will refer to it as COSMOS ETG sample.

\section{Sample properties}
\subsection{Optical properties}

Of the 69 galaxies in the X-ray ETG sample, 53 have spectroscopic redshifts. For the remaining 16 galaxies, we used the available photometric redshifts reported in the C-COSMOS identification catalog (Civano et al. 2012, Salvato et al. 2011). Given the high quality of the photometry available for the sources in the COSMOS field (see Salvato et al. 2011), we are confident that we can use photometric redshifts when spectroscopic ones are not available. the Although not uniform, an optical spectroscopic classification is  available for the C-COSMOS sources (see Civano et al. 2012 for the details on observing programs). 
According to the spectroscopic classification, our sample includes: 24 absorption line galaxies (ALG) and 29 narrow emission line (NL) objects. Moreover, using standard diagnostic diagrams (Bongiorno et al. 2010), we find that of the 29 NL sources, 5 are classified as NL active galactic nuclei (i.e. Type 2 AGN), 7 as star forming galaxies, while 17 spectra remain unclassified.

The K band luminosity was computed from the Ultra-VISTA K-band aperture magnitude of the COSMOS photometric catalog. We applied aperture corrections to derive the total magnitude using the appropriate extension parameter available in the COSMOS catalog (called {\it ext} below), which was computed from the FWHM measured for all the COSMOS extended sources in the {\it Hubble} ACS data (Leauthaud et al. 2007). To evaluate rest frame K-band luminosities, we assumed a spectral shape of the type $f_\nu \propto \nu^{-\alpha} $ with $\alpha$=$- \frac{J-K}{lg(\nu_J/\nu_K)}$, where J and K are taken from the COSMOS catalog. The luminosity expressed in solar luminosity is then
\begin{equation} L_K/L_{\odot}=10^{-(K+ext-5.19)/2.5} \times (1+z)^{\alpha-1} \times (D_L/10)^2 \end{equation} 
where K is the AB magnitude from the COSMOS photometric catalog, {\it ext} is the aperture correction parameter, $z$ is the redshift, $D_L$ is the luminosity distance in parsec.

\begin{figure}
\centering
\includegraphics[width=0.45\textwidth, angle=0]{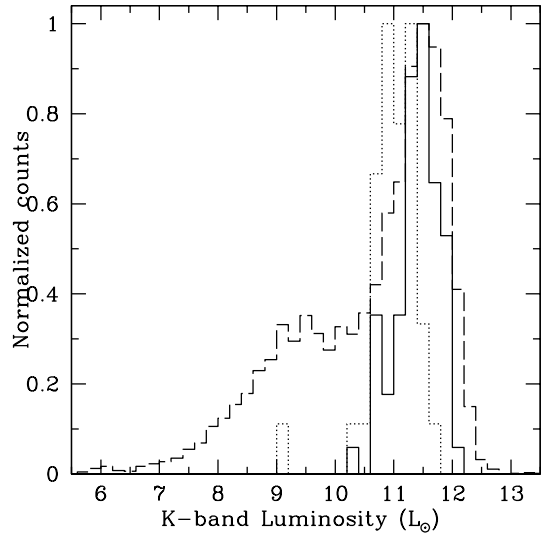}
\caption{\small Normalized rest frame K band luminosity histogram for the X-ray ETGs (solid line), for all the COSMOS ETGs (dashed line) and for the ETGs in the local sample (dotted line).  X-ray ETGs represent the most luminous tail of the whole COSMOS ETG luminosity distribution.}
\label{lk_isto}
\end{figure}

In Figure \ref{lk_isto}, the K-band luminosity histograms for the X-ray ETGs (solid line), the comparison sample of COSMOS ETGs (dashed line) and the BKF local sample are plotted. Given the X-ray flux limit of the C-COSMOS survey, the X-ray ETGs have luminosities consistent with the high luminosity tail of the entire ETG population. Note that the luminosity threshold is a function of redshift because of the uniform magnitude limit of the Ultra-VISTA COSMOS survey (K$_s$ $\sim$ 24, McCracken et al. 2013). The main peak of the K-band luminosity histogram for X-ray detected and undetected ETGs is consistent with the distribution of the local sample, although slightly shifted to higher luminosities.

\begin{figure}
\centering
\includegraphics[width=0.45\textwidth]{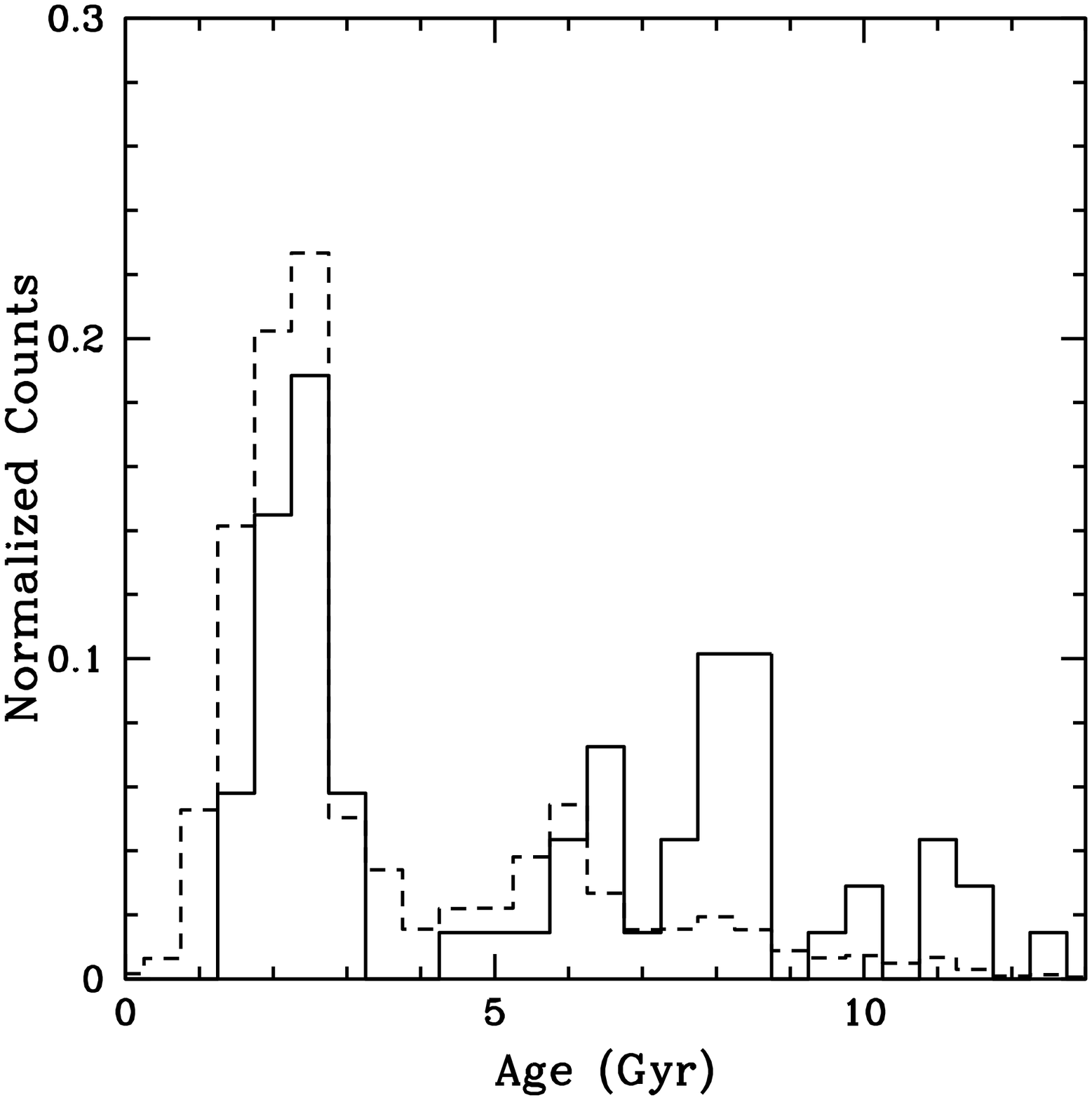}
\includegraphics[width=0.45\textwidth]{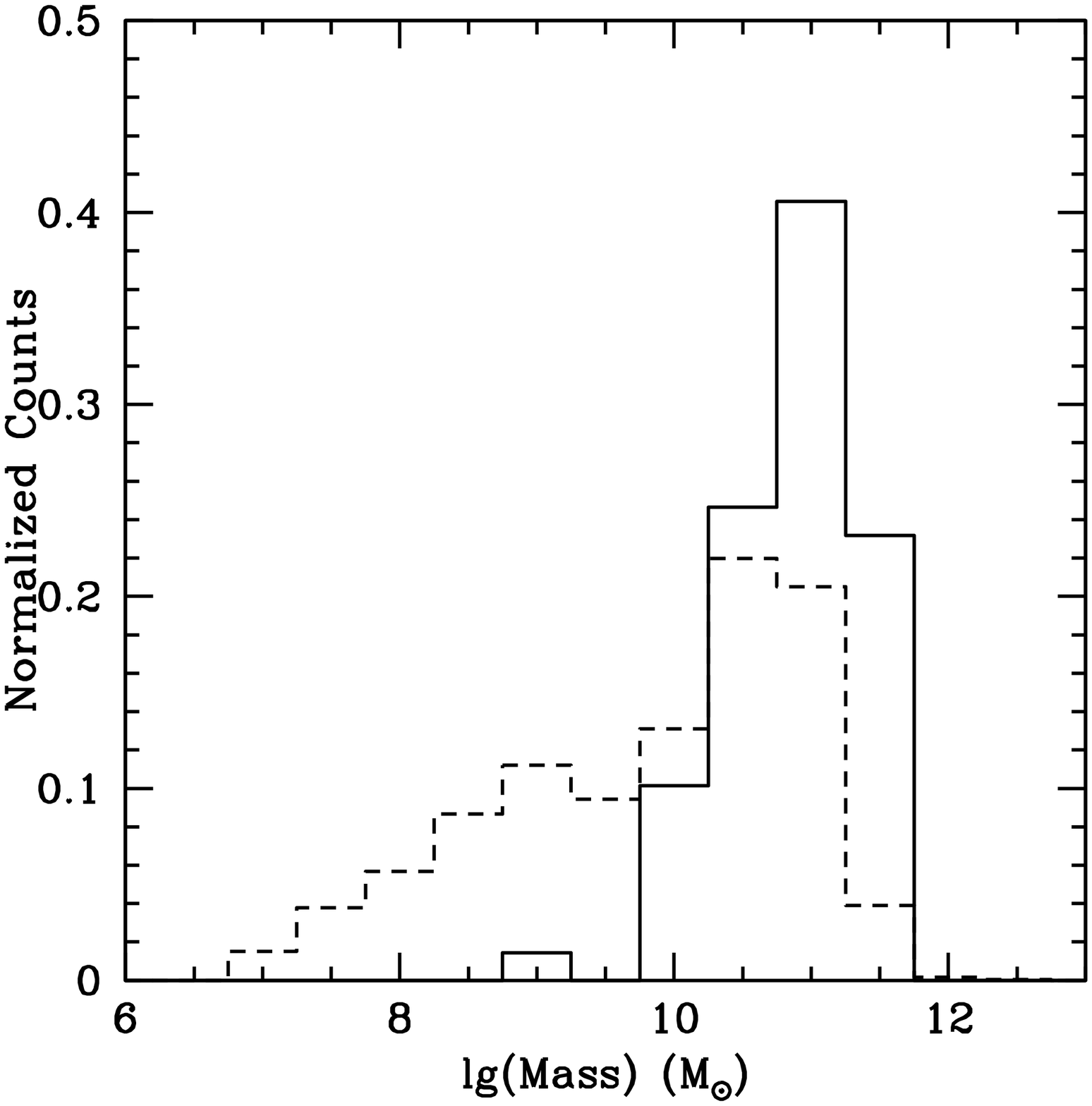}
\caption{\small Normalized histograms of age (in Gyr) and stellar mass (in solar masses unit) for the X-ray ETGs (solid line) and for the COSMOS ETG comparison sample (dashed line).  X-ray detected ETGs are on average older and more massive than non X-ray detected ETGs.}
\label{age_mass}
\end{figure}

In Figure \ref{age_mass}, the age and stellar mass (Ilbert et al. 2010) of the X-ray ETGs are compared to the values of the entire COSMOS ETG population. The age was derived from Bruzual \& Charlot (2003) template fitting. Given the possible errors on age estimation, we will later divide the sources in 3 wide age bins ($<$5 Gyrs, 5--9 Gyrs and $>$9 Gyrs).
The age spread of the X-ray ETGs is representative of the entire ETG sample, although the X-ray detected sources present a larger fraction of old galaxies with ages $>$ 6 Gyrs, relative to the entire population. As can be seen in Figures  \ref{lk_isto} and \ref{age_mass}, X-ray ETGs are among the most massive COSMOS ETGs. 
A two sample Kolmogorov-Smirnov (K-S) test demonstrates that the X-ray ETGs and the COSMOS ETGs are not drawn from the same parent distributions of luminosity, age, and mass distributions with p-values of $1\times10^{-9}$, $8\times10^{-8}$, and $2\times10^{-14}$, respectively.

\begin{figure}
\centering
\includegraphics[width=0.45\textwidth]{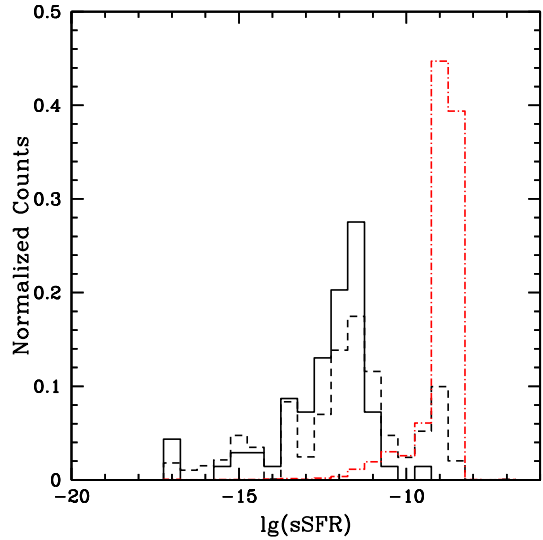}
\caption{\small Normalized histograms of the specific star formation rate for the X-ray ETGs (solid line), for the COSMOS ETGs (dashed line) and for a selected sample of spiral galaxies in COSMOS (dot-dashed red line).  X-ray ETGs and COSMOS ETGs have similar sSFR distribution, significantly lower than spiral galaxies.} 
\label{ssfr}
\end{figure}

Figure \ref{ssfr} compares the specific star formation rates (sSFRs) of X-ray ETGs with those of COSMOS ETGs and with a sample of $\sim$130,000 spiral galaxies selected from the same Ilbert et al. (2010) catalog using SED classification. The X-ray ETGs have consistent sSFR values with the
COSMOS ETGs. About 95\% of the spiral galaxies have $\log$(sSFR)$>$-10.5 while only $\sim$20\% of the COSMOS ETGs have similar sSFR and only one X-ray ETG has such high value. The K-S test result shows that X-ray ETGs and COSMOS ETGs could be drawn from different parent sample though with a lower confidence than the above tests on mass, age and luminosity (K-S p-value = 6$\times$10$^{-5}$). On the contrary, we can state with high confidence that both X-ray ETG and COSMOS ETG sSFR are not sampled from the distribution of spiral galaxy sSFR  (K-S p-value $<$1$\times$10$^{-54}$). The sSFR of the X-ray ETGs is in the regime where Ilbert et al. (2010) defines galaxies as quiescent.
Even if there is SF in ETGs (as most likely happens in young ETGs), their sSRF is about 1000 times lower than typical spiral galaxies (Figure \ref{ssfr}).

The properties (luminosity, mass, star formation rate and age) of the COSMOS ETGs are in agreement with those reported by Moresco et al. (2013) for a sample of ETGs selected using multiple criteria including color-color diagrams, spectra, star formation rate and SED classification. 
Overall, masses and luminosities are in agreement with those of the K-selected elliptical 
galaxies presented by Jones et al. (2013) selected from a small area of the COSMOS field, for which individual X-ray detection is not available but only a stacked signal.

\begin{figure}
\centering
\includegraphics[width=0.45\textwidth]{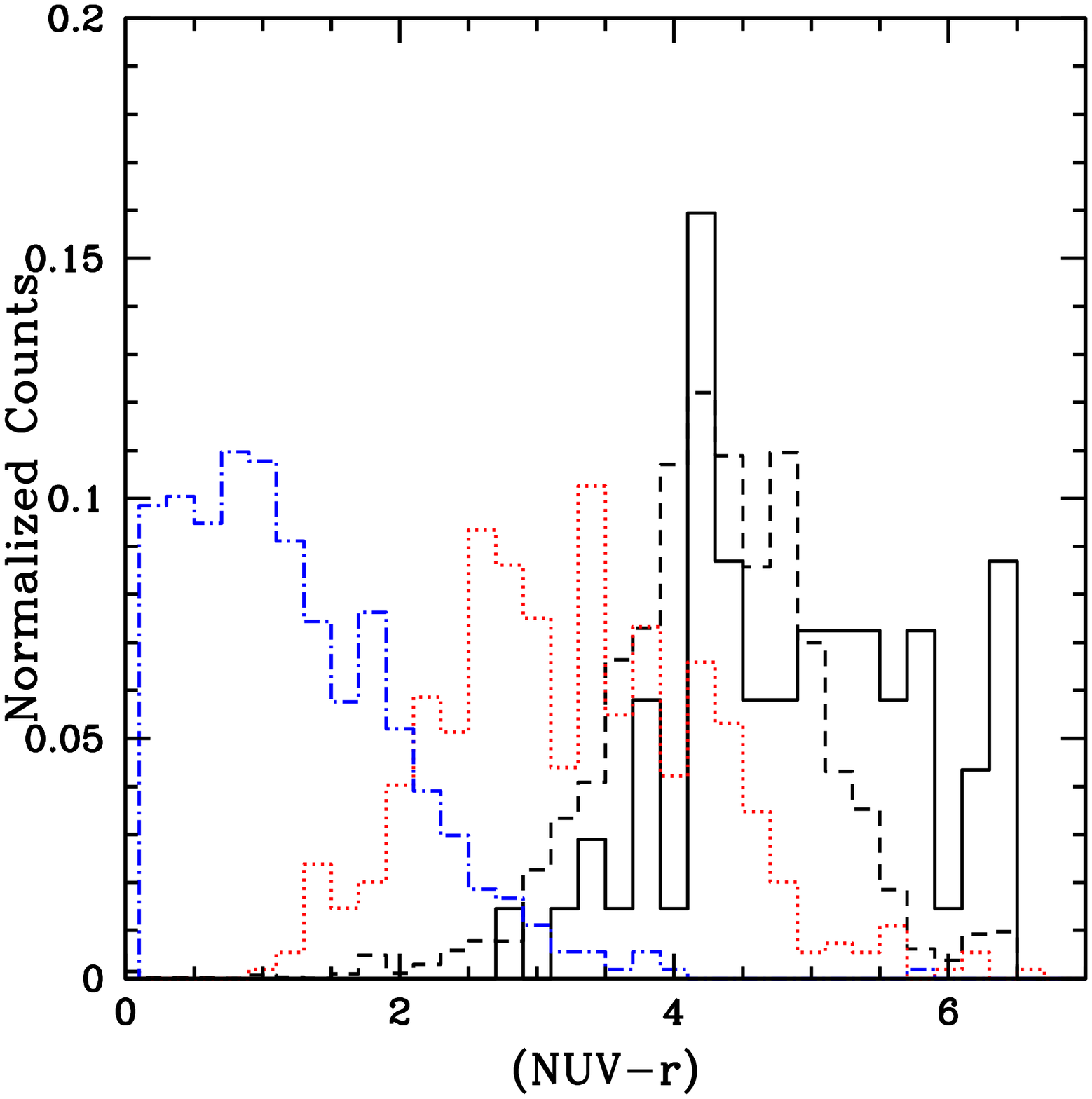}
\caption{\small Normalized histograms of (NUV--r) color for the X-ray ETGs (solid black line), the COSMOS ETGs (black dashed), X-ray detected Type 1 (dot-dashed blue line) and Type 2 (dotted red line) AGN.}
\label{nuvr}
\end{figure}

Figure \ref{nuvr} shows the (NUV--r) rest-frame color distribution as used by Ilbert et al. (2010) to classify galaxies according to their current and past star formation activity: quiescent galaxies with (NUV--r)$>$3.5, intermediate with (NUV--r)= 1.2 to 3.5, and high activity galaxies with (NUV--r) $<$1.2. The rest frame color is reported in the photometric catalog of lbert et al. (2010). 
Here, we compare the distributions of the X-ray ETGs, the COSMOS ETGs and also spectroscopically and photometrically identified X-ray detected COSMOS Type 1 and Type 2 AGN (Lusso et al. 2010, 2011, 2012). 
X-ray ETGs have the reddest, highest (NUV--r) color and show an excess of extremely red galaxies with (NUV--r)$>$5 with respect to the X-ray undetected population. Type 1 AGN (538 sources) have the colors of high activity galaxies, Type 2 AGN (546 sources) lie in an intermediate region with a tail of the distribution overlapping with the X-ray ETG distribution. When looking at the spectroscopically identified Type 2 AGN only, the distribution becomes narrower showing less overlap with the ETG sample. This effect is due to the degeneracy of some obscured AGN templates with galaxy templates. This color diagnostic suggests  very little or no contamination from Type 2 and Type 1 AGN, respectively, is expected to affect the X-ray ETG sample.

All the sources included in the X-ray ETG sample have been classified as extended in the optical band using the {\it Hubble} ACS data (filter FW814) (Leauthaud et al. 2007). A further visual inspection of the sources confirms this finding, even for the sources at the higher redshifts (see Figure \ref{fc1}).

\subsection{X-ray properties}
\label{xraypro}

The goal of this analysis is to compare the X-ray luminosity due to the hot gas emission with the K-band luminosity following the analysis of BKF and KF13. 
In order to do so, all the possible contribution to the X-ray luminosity must be isolated and removed.
BKF gives a relation that can be used to estimate the integrated LMXB contribution to the X-ray emission of an 
ETG. This relation was empirically derived from their study performing emission decomposition of a sample of nearby ETGs observed with \chandra. However, this relation is strictly valid only in the nearby universe, where it was derived. Fragos et al. (2013) showed how the luminosity of field LMXBs evolves with redshift and is a function of the galaxy stellar mass and of the age of the parent stellar population. We therefore used the formulation of Fragos et al. (2013) to estimate the LMXB contribution to the luminosity in our galaxies. We note that their formulation is only valid for native field LMXBs. For LMXBs formed dynamically in globular clusters, this age correction will not apply, so the Fragos et al. (2013) correction may over estimate the X-ray fading of a given galaxy. We keep this possibility in mind when examining our correlation (see Section \ref{scaling}). We note that, the age correction is small enough to not offset the distribution of points significantly. 

To subtract the LMXB contribution from the total X-ray luminosity (L$_X$), we used the count rates in the 0.5-7 keV band from the C-COSMOS X-ray catalog and converted them into luminosities in the 0.3-8 keV  rest frame band using a power law model with a slope of $\Gamma$=1.8, consistent with the typical spectrum of LMXBs, and Galactic N$_{H}$. For these parameters, the conversion factor used from \chandra\ count rates to fluxes is 1.24$\times$10$^{-11}$ counts$^{-1}$ \cgs. 
From these X-ray luminosities, we then subtracted the contribution of the LMXBs using the following relation in the 0.3-8 keV band 
\begin{equation}
 \log(L_X)=\log(M_{\star})+40.259-1.505 \times \log(age_{Gyr})-0.421 \times (\log age_{Gyr})^2+0.425\times (\log age_{Gyr})^3+0.135\times(\log age_{Gyr})^4-10
\end{equation}
reported by Fragos et al. (2013). 
The contribution of the LMXB ranges from $<$1\% for bright galaxies to 70\% for faint galaxies. 

The X-ray luminosity can also be contaminated by high mass X-ray binaries (HMXBs), in particular in young X-ray ETGs. To determine HMXBs contribution, we use the relation between the HMXBs luminosity and the star formation rate reported by Mineo, Gilfanov \& Sunyaev (2012), which has been derived for star-forming galaxies and will therefore return an upper limit on this population luminosity. The typical SFR of the X-ray ETGs is $\sim$0.1 M$_{\odot}$/yr, which implies a full band X-ray luminosity of $\sim$10$^{38}$\lum for HMXBs. This value is two orders of magnitude (or more) smaller than the typical luminosity of the X-ray ETGs reported here, so we conclude that HMXBs do not affect the measured X-ray luminosity.

Once the LMXB contribution is removed and the contribution of HMXBs assessed to be negligible, we converted the remaining count rates into a luminosity using a thermal model (APEC in Sherpa, Freeman et al. 2001), adopting different temperatures depending on the total X-ray luminosity of the source, plus Galactic N$_{H}$. We used kT=0.7, 1 and 2 keV for X-ray luminosities of lg(L$_X)<$41, 41--42 and $>42$ \lum, respectively, according to BKF and Dai et al. (2007).
The X-ray luminosity was K-corrected using the {\it calc\_kcorr} tool in Sherpa adopting the above spectral models. The final rest frame X-ray luminosity computed and reported in Table 1 and in the following Figures is in the 0.3-8 keV band. 
This luminosity, besides the emission from hot gaseous halos, may include a contribution from AGN 
emission.

\begin{figure}
\centering
\includegraphics[width=0.45\textwidth]{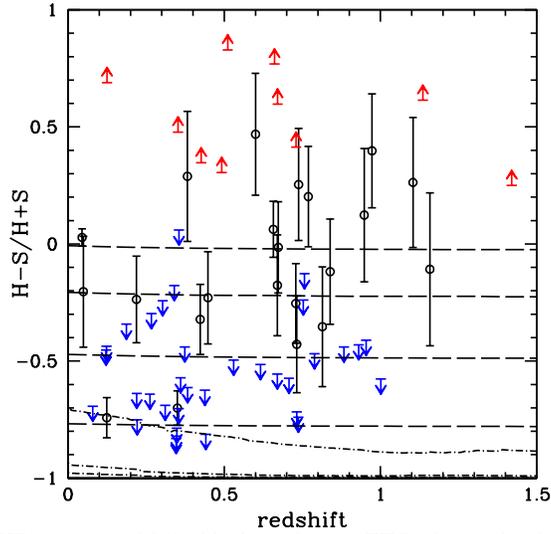}
\vspace{-0.5cm}\caption{\small X-ray hardness ratio (HR) versus redshift: black circles are ETGs detected in both soft and hard band; in blue ETGs detected in soft and full bands; in red ETGs detected in full and hard bands. Dashed lines represent power law models with $\Gamma$=1, 1.4, 2, 3 (from top to bottom). Dot-dashed lines represent a thermal model with increasing temperatures from bottom to top (kT=0.7, 1, 2 keV). For all models Galactic N$_{H}$ was assumed.}
\label{hr}
\end{figure}

X-ray spectral analysis could help in separating the hot gas from the AGN contribution. The low number of counts of these sources (around 30 counts in the full band) prevents us to perform single source spectral analysis. Therefore, we computed the hardness ratio, defined as HR=$\frac{H-S}{H+S}$, where H and S are the number of net counts in the hard and soft band respectively, to study the spectral shape of the sources. 
The 69 X-ray ETGs are all detected in the C-COSMOS 0.5-7 keV band: 22 sources are detect in both soft, hard and full bands; 34 sources in soft and full band; 10 sources in hard and full band; 3 sources are detected only in the full band. 
The HR versus redshift for the X-ray ETGs is plotted in Figure \ref{hr}. Upper limits, for those sources which are only detected in the soft and full band, are reported as downwards arrows, lower limit for sources detected in the hard and full band only are reported as upwards errors.
Curves representing a model of power law with $\Gamma$=1, 1.4, 2 and 3 (from top to bottom) absorbed by Galactic N$_{H}$ have been used. 
In the same plot, curves obtained from an APEC model with the same temperatures used to compute the X-ray luminosity above (kT=0.7, 1, 2 keV from bottom to top) are represented as dash-dotted line. The last set of curves are all clustered at HR$\sim$-0.7 and lower. The power law better represents the nuclear emission while the APEC model the hot gas emission. 

All the sources with no detection in the hard band (the HR upper limit) are consistent with thermal or very steep power-law models, resembling the thermal model. Their 2-10 keV luminosity upper limits are consistent with the LMXB emission computed using the Fragos et al. (2013) relation as above in the hard band. The sources with no detection in the soft band have an harder spectrum, representable by a flat power-law resembling a steep power-law spectrum plus additional obscuration (N$_{H} >$N$_{H, Gal}$). Sources detected in both bands have HRs in the range -0.5 to 0.5 and theirs spectrum could be represented by a combination of both models. 

In order to determine whether the X-ray emission in the 69 X-ray ETGs is extended or it is dominated by a nuclear point like source, we also analyzed the spatial distribution of the X-ray counts in each source. This analysis can be done only for sources with enough counts in a single observation and with redshift z$<$0.5 in order to avoid resolution problems (at higher redshifts, if the source is not perfectly on axis it is hard to resolve an extended source). Detailed simulations, using the CIAO software Chart\footnote{http://cxc.harvard.edu/chart/runchart.html} and MARX\footnote{http://cxc.harvard.edu/chart/threads/marx/}, allow us to determine the minimum number counts needed at different off-axis angle to recognize when a source is point like or extended. Given the limit of 6 counts needed in a single observation and that many sources in our sample have a low count statistics, this analysis could be done only for 27 sources out of 69. Using the CIAO tool {\it dmellipse}, we compared the spatial distribution of the counts with the shape of the \chandra\ point spread function (PSF). We produced a PSF image at the same position on the detector of each analyzed source using Chart and MARX. Using this information we derive that 14 of 27 analyzed sources have a size comparable to the one of the PSF at their position (size$_{source}$ $<$ 2$\times$size$_{PSF}$). The remaining 13 sources resulted to be extended (size$_{source}$ $>$ 2$\times$size$_{PSF}$). 

Overall, the hot gas and AGN contributions to the X-ray emission of the X-ray ETGs are hard to disentangle and all the above analysis will be subject of the discussion in Section 4.

We have also matched the X-ray ETGs with the catalog of X-ray groups in the COSMOS field (Giodini et al. 2010) and we find that 14 sources lie within 6$^{\prime\prime}$ from the center of the group they belong to, meaning they most probably are the central galaxies of these groups. Given the large number of structures (groups and clusters) in the COSMOS field (see Scoville et al. 2013), we kept the matching radius small to find only central galaxies.

Table 1 summarizes the properties of the X-ray ETGs used in this work: C-COSMOS identifier, spectroscopic redshift when available or photometric, optical spectral type, 0.5-7 keV count rates and errors from the C-COSMOS catalog, the rest frame L$_{X,LMXB}$, the remaining rest frame X-ray luminosity after LMXB contribution subtraction, HR, extension of X-ray emission, L$_K$ and the error derived from the fading, M$_{\star}$, age, group relevance and radio emission flag.

\begin{center}
\begin{table*}
\caption{COSMOS X-ray ETGs properties.}
\tiny
\begin{tabular}{ccccccccccccccc}
\hline
\hline
        CID      &        redshift &      spec. type      &          Count rates$^a$ (cts/s)  &          Count rates err$^a$  (cts/s)      &        $log$  L$_{X,lmxb}^b$     &      $log$ 	L$_{X}^c$      &       HR        &       extension      &       $lg$ L$_{K}$   &	   $lg$ L$_{K}$ err   &	   $lg$ M$_{\star}$	 &	 Age	  &	  Group Flag	  &	 Radio Flag\\
    &                &        &     0.5-7 keV    &    0.5-7 keV    &    0.3-8 keV    &     0.3-8 keV    &        &        &     in L$_{\odot}$    &    &  in M$_{\odot}$    &     Gyr    &        &    \\
\hline
          73      &       0.045        &	  ALG	   &	   2.99e-03   &  1.1e-4    &	38.72	    &	    40.91     &       0.03	 &	    1	   &	   10.87     	&      0.02  &        9.85    &       1e+10	 &	  \dotfill  &	       1 \\ 
          91      &         0.657      &	  NLAGN      &     4.52e-04   &  5.5e-5      &  40.3	      &        43	&	0.06	   &	      0      &       11.12   	  &    0.09    &      10.52	 &	 2e+09      &	     \dotfill  &	\dotfill \\ 
         123      &     0.6$^p$       &       \dotfill     &	   2.19e-04  &   6.3e-5   &	39.99	  &	    42.57   &	    0.47	&	   0	  &	  10.70       &        0.09   &        10.06  &       2e+09	 &	  \dotfill  &	     \dotfill \\ 
         196      &     0.738	       &	 NL	 &	   3.98e-04   &  1.00e-4   &	40.84	  &	     43.1   &	    0.25      & 	0      &       11.81	      &        0.09  &  	    11.16      &       2e+09  &        \dotfill  &	     1 \\ 
         384      &     0.448	       &	 NL	 &	   3.18e-04   &  6.3e-5  &	39.91	  &	    42.34   &	    -0.23      &	  1	 &	 10.61        &        0.10  &  	    10.06      &       2e+09	&	     \dotfill  &     \dotfill \\ 
         633      &       0.350        &	  ALG	   &	   1.31e-03   &  1.2e-4    &	40.33	    &	    42.66     &       -0.70	 &	    1	   &	   11.70     	&      0.10  &        11.30    &       7e+09	  &	   \dotfill  &  	1 \\ 
         634      &       0.124        &	  ALG	   &	   1.10e-03   &  1.1e-4    &	40.17	    &	     41.4     &       -0.74	 &	    6	   &	   12.01     	&      0.04  &        11.30    &       1e+10	  &	     1  	&	   1 \\ 
         651      &      0.423         &	 ALG	  &	   3.93e-04   &  5.5e-5   &	40.04	   &	    42.36    &       -0.32	&	   2	  &	  11.29        &       0.10  &  	11.05	&	8e+09	   &	      \dotfill  &		1 \\ 
         685      &     0.839	       &	 NL	 &	   1.97e-04   &  4.2e-5  &	40.34	  &	    42.97   &	    -0.12      &	  0	 &	 11.46        &        0.10  &  	    10.84      &       3e+09	&	     \dotfill  &     \dotfill \\ 
         750      &     0.769$^p$     &        \dotfill  &	   1.94e-04  &   4.2e-5  &	40.45	  &	    42.85   &	    0.20      & 	  0	 &	 11.39        &        0.09  &  	    10.70      &       2e+09	&	     \dotfill  &     \dotfill \\ 
         765      &      0.814         &	 ALG	  &	   2.11e-04   &  4.7e-5   &	39.74	   &	    42.96    &       -0.35	 &	    0	   &	   11.04       &       0.10  &        10.68    &       6e+09	  &	   \dotfill  &        \dotfill \\ 
         996      &      0.670         &	 ALG	  &	   3.44e-04   &  7.1e-5   &	40.11	   &	    42.91    &       -0.18	 &	    0	   &	   11.50       &       0.09  &        11.13    &       8e+09	  &	   \dotfill  &        \dotfill \\ 
        1059      &       0.049        &	  ALG	   &	   2.39e-04   &  5.6e-5    &	 38.9	    &	    39.92     &       -0.20	  &	     1      &	    10.29    	&      0.02   &        10.05	&	1e+10	   &	    \dotfill  &        \dotfill \\ 
        1087      &     0.673	       &	 NL	 &	   2.48e-04   &  4.5e-5  &	40.32	  &	    42.77   &	    -0.01      &	   0	  &	  11.25       &        0.09  &        10.64   &       2e+09	 &	  \dotfill  &	     \dotfill \\ 
        1117      &     0.948$^p$     &        \dotfill  &	   1.67e-04  &   4.7e-5  &	40.46	  &	    43.07   &	    0.12      & 	 0	&	11.46	      &        0.09  &  	   10.71      &       2e+09    &	\dotfill  &	   \dotfill \\ 
        1176      &     0.973$^p$     &        \dotfill  &	   1.60e-04  &   4.1e-5  &	40.59	  &	    43.09   &	    0.40      & 	 0	&	11.60	      &        0.09  &  	    10.91      &       2e+09   &	\dotfill  &	   \dotfill \\ 
        1188      &     1.158	       &	 NL	 &	   1.15e-04   &  3.7e-5  &	39.84	  &	    43.22   &	    -0.11	 &	    0	   &	   11.74      &        0.09  &        10.22    &       2e+09	  &	   \dotfill  &        \dotfill \\ 
        1189      &     0.732	       &	 NL	 &	   2.25e-04   &  4.5e-5  &	40.49	  &	    42.84   &	    -0.43	 &	    0	   &	   11.82      &        0.09  &        11.46    &       7e+09	  &	   \dotfill  &  	1 \\ 
        1217      &     1.104$^p$     &        \dotfill  &	   1.23e-04  &   3.7e-5  &	40.45	  &	    43.17   &	    0.26      & 	 0	&	11.56	      &        0.10  &  	    10.77      &       2e+09   &	\dotfill  &	   \dotfill \\ 
        1289      &       0.219        &	  ALG	   &	   3.18e-04   &  5.3e-5    &	40.33	    &	    41.42     &       -0.24	  &	     3      &	    11.72    	&      0.08   &        11.40	&	9e+09	   &	    \dotfill  & 	 1 \\ 
        1310      &      0.729         &	 ALG	  &	   2.56e-04   &  4.3e-5   &	40.63	   &	     42.9    &       -0.25	  &	     0      &	    11.85      &       0.09   &        11.51	&	6e+09	   &	    \dotfill  & 	 1 \\ 
        1495      &     0.382$^p$     &        \dotfill  &	   1.29e-04  &   3.8e-5   &	40.24	  &	    41.82   &	    0.29      & 	 1	      &	11.58	      &        0.09   & 	 11.30      &	    8e+09      &	\dotfill  &	   \dotfill \\ 
          13      &         0.187      &	  NLAGN      &     1.94e-04   &  5.5e-5      & 39.92	      &     41.04	  &	  $<$-0.34	&	   2	  & 11.26    	  &    0.09   &        11.05  &       1e+10	 &	  \dotfill  &	       1 \\ 
         133      &       0.220        &       SF gal	   &	   2.77e-04   &  7.0e-5    &	40.19	    &	    41.36	&	$<$-0.64      & 	 1	&	11.68	&      0.09   &        11.25   &       8e+09	  &	     1      &	     \dotfill \\ 
         288      &       0.346        &	  ALG	   &	   4.55e-04   &  6.8e-5    &	40.51	    &	    42.17	&	$<$-0.83      & 	 3	&	11.97	&      0.10   &        11.55   &       8e+09	  &	     1      &	       1 \\ 
         364      &       0.438        &	  ALG	   &	   2.71e-04   &  4.0e-5    &	40.46	    &	    42.23	&	$<$-0.62      & 	 1	&	11.76	&      0.10   &        11.50   &       8e+09	  &	     1      &	       1 \\ 
         680      &     0.954	       &	 NL	 &	   2.01e-04   &  6.2e-5  &	41.27	  &	    43.16     &       $<$-0.41      &	       0      &       11.96   &        0.09   &        11.17	  &	  1e+09      &  	1      &	  1 \\ 
         690      &     0.670$^p$     &        \dotfill  &	   3.75e-04  &   8.7e-5  &	40.56	  &	    42.94     &       $<$-0.56      &	       0      &       11.89   &        0.09   &        11.57	  &	  7e+09      &  	1      &	  1 \\ 
         783      &       0.122        &	  ALG	   &	   1.20e-04   &  3.6e-5    &	39.65	    &	    40.42	&	$<$-0.45      & 	 3	&	11.31	&      0.05    &	10.75	   &	   1e+10      &        \dotfill  &	  \dotfill \\ 
         827      &       0.354        &	  ALG	   &	   1.85e-04   &  4.1e-5     &	40.49	    &	    41.81	&	$<$-0.70      & 	 2	&	11.77	&      0.09    &	11.55	   &	   8e+09      &        \dotfill  &	    1 \\ 
         898      &      0.347         &	 ALG	  &	   3.39e-04   &  5.6e-5   &	40.31	   &	    42.06      &       $<$-0.82      &  	2      &       11.53   &       0.09    &       11.10	  &	  5e+09      &  	1      &	  1 \\ 
         930      &     0.930	       &	 NL	 &	   1.27e-04   &  3.5e-5   &	40.76	  &	    42.93     &       $<$-0.43      &	       0      &       11.73   &        0.09    &      11.01	 &	 2e+09      &	     \dotfill  &	\dotfill \\ 
         983      &       0.221        &	  ALG	   &	   2.34e-04   &  6.1e-5    &	40.29	    &	    41.29	&	$<$-0.75      & 	 0	&	11.81	&      0.09    &      11.40	 &	 1e+10      &	       1      &        \dotfill \\ 
         993      &       0.348        &       SF gal	   &	   2.31e-04   &  5.3e-5    &	40.43	    &	    41.89	&	$<$-0.79      & 	 0	&	11.71	&      0.09    &      11.47	 &	 8e+09      &	     \dotfill  &	\dotfill \\ 
        1241      &     0.737$^p$     &        \dotfill  &	   1.93e-04  &   3.9e-5  &	40.18	  &	    42.79     &       $<$-0.74      &	       0      &       11.50   &        0.10     &      11.15	  &	  7e+09      &  	1      &	\dotfill \\ 
        1243      &     0.732	       &	 NL	 &	   2.37e-04   &  4.5e-5  &	40.29	  &	    42.87     &       $<$-0.72      &	       0      &       11.24   &        0.09     &      10.55	  &	  2e+09      &        \dotfill  &	 \dotfill \\ 
        1292      &       0.530        &	  ALG	   &	   1.42e-04   &  3.8e-5    &	40.33	    &	    42.23	&	$<$-0.50      & 	 0	&	11.63	&      0.09	&	 11.35      &	    8e+09      &	  1	 &	    1 \\ 
        1301      &     0.757	       &	 NL	 &	   1.25e-04   &  3.8e-5  &	40.25	  &	    42.63     &       $<$-0.13      &	       0      &       11.14   &        0.09     &      10.57	  &	  2e+09      &        \dotfill  &	 \dotfill \\ 
        1364      &       0.311        &       SF gal	   &	   1.47e-04   &  4.6e-5    &	40.27	    &	    41.51	&	$<$-0.69      & 	 4	&	11.61	&      0.10	 &	 11.24      &	    7e+09      &	  1	 &	  \dotfill \\ 
        1401      &      0.707         &	 ALG	  &	   1.05e-04   &  3.5e-5   &	39.61	   &	    42.46      &       $<$-0.57      &  	0      &       11.31   &       0.09      &	10.58	   &	   7e+09      &        \dotfill  &	  \dotfill \\ 
        1478      &     0.616$^p$     &        \dotfill  &	   1.60e-04  &   3.9e-5  &	39.73	  &	    42.47     &       $<$-0.52      &	       0      &       10.61   &        0.09     &      9.95	 &	 2e+09      &	     \dotfill  &	\dotfill \\ 
        1500      &       0.079        &	  ALG	   &	   1.62e-04   &  4.1e-5    &	39.77	    &	    40.09     &       $<$-0.69      &	       4      &       11.23  	&      0.02     &      10.84	  &	  9e+09      &  	1      &	\dotfill \\ 
        1521      &     0.360	       &	 NL	 &	   8.30e-05   &  4.05e-5  &	 39.8	  &	    41.49     &       $<$-0.57      &	       2      &       11.14   &        0.09     &      10.84	  &	  8e+09      &        \dotfill  &	 \dotfill \\ 
        1541      &       0.789        &       SF gal	   &	   1.33e-04   &  4.6e-5    &	40.18	    &	    42.71	&	$<$-0.47      & 	 0	&	11.52	&      0.09	  &	 11.12      &	    6e+09      &	\dotfill  &	   \dotfill \\ 
        1583      &     0.383$^p$     &        \dotfill  &	   6.68e-04  &   1.52e-4  &	39.68	  &	    42.47     &       $<$-0.61      &	       0      &       11.07   &        0.09      &     10.72	  &	  8e+09      &        \dotfill  &	 \dotfill \\ 
        1811      &       0.123        &	  ALG	   &	   1.23e-04   &  4.5e-5    &	39.83	    &	    40.42	&	$<$-0.44      & 	10	&	11.46	&      0.05	  &	 10.97      &	    1e+10      &	\dotfill  &	     1 \\ 
        1871      &       0.267        &       SF gal	   &	   1.88e-04   &  6.0e-5    &	40.11	    &	    41.41	&	$<$-0.30      & 	 2	&	11.55	&      0.09	  &	 11.21      &	    1e+10      &	\dotfill  &	     1 \\ 
        2122      &         0.340      &	  NLAGN      &     5.40e-05    &  4.29e-5      & 40.06	      &     41.19	  &	  $<$-0.18	&	   0	  & 11.46    	  &    0.09	 &	  11.12       &       8e+09	 &	  \dotfill  &	     \dotfill \\ 
        2633      &     0.441	       &	 NL	 &	   2.21e-04   &  3.9e-5  &	40.39	  &	    42.15     &       $<$-0.81      &	       1      &       11.27   &        0.09       &    10.89	  &	  3e+09      &  	1      &	\dotfill \\ 
        2797      &      0.262         &	 ALG	  &	   1.52e-04   &  3.9e-5   &	39.61	   &	    41.32      &       $<$-0.64      &  	4      &       10.70   &       0.09      &	10.43	   &	   5e+09      &        \dotfill  &	  \dotfill \\ 
        2822      &      1.001         &	 ALG	  &	   1.42e-04   &  4.4e-5   &	 40.9	   &	     43.1      &       $<$-0.58      &  	0      &       11.81   &       0.09      &	11.28	   &	   2e+09      &        \dotfill  &	  \dotfill \\ 
        2876      &     0.355	       &	 NL	 &	   7.40e-05   &  5.72e-5  &	40.14	  &	    41.42     &       $<$0.06	   &	      0      &       11.47    &        0.09      &    11.18	 &	 8e+09      &	     \dotfill  &	\dotfill \\ 
        3060      &         0.753      &	  NLAGN      &     8.60e-05   &  3.94e-5      & 40.32	      &     42.47	  &	  $<$-0.24	&	   0	  & 11.22    	  &    0.09	 &	  10.57       &       2e+09	 &	  \dotfill  &	     \dotfill \\ 
        3247      &     0.883	       &	 NL	 &	   1.39e-04   &  4.3e-5  &	40.55	  &	     42.9     &       $<$-0.44      &	       0      &       11.49   &        0.10      &     10.87	  &	  2e+09      &        \dotfill  &	 \dotfill \\ 
        3564      &     0.303$^p$     &        \dotfill  &	   9.00e-05  &   4.75e-5  &	39.49	  &	    41.27     &       $<$-0.24      &	       2      &       10.80   &        0.10      &     10.55	  &	  8e+09      &        \dotfill  &	 \dotfill \\ 
        3665      &      0.374         &	 ALG	  &	   7.50e-05   &  4.91e-5   &	40.48	   &	    41.48      &       $<$-0.44      &  	1      &       11.93   &       0.10      &	11.53	   &	   8e+09      & 	 1	&	 \dotfill \\ 
         368      &         0.511      &	  NLAGN      &     2.55e-04   &  4.6e-5      & 40.27	      &     42.44	&	$>$0.83      &  	0      &       10.95 	  &    0.09      &	10.34	   &	   2e+09      &        \dotfill  &	    1 \\ 
         514      &       0.125        &	  ALG	   &	   3.41e-04   &  7.6e-5    &	39.82	    &	    40.88      &       $>$0.69      &	       2      &        11.49 	&      0.05      &	10.94	   &	     1e+10	&	 \dotfill	&	 \dotfill \\ 
         929      &       0.671        &       SF gal	   &	   1.13e-04   &  3.6e-5    &	39.98	    &	    42.42      &       $>$0.60      &	       0      &        10.97 	&      0.09      &	10.36	   &	     2e+09	&	 \dotfill	&	 \dotfill \\ 
        1016      &     0.661	       &	 NL	 &	   2.33e-04   &  5.7e-5  &	40.05	  &	    42.73   &	     $>$0.77	  &	     0      &	    11.36     &        0.09	 &     10.99	&	6e+09	   &	    \dotfill  &        \dotfill \\ 
        1803      &     0.352$^p$     &        \dotfill  &	   7.80e-05  &   4.07e-5  &	40.37	  &	    41.41     &       $>$0.48	  &	     0      &	    11.25     &        0.09	 &     10.87	&	3e+09	   &	    \dotfill  &        \dotfill \\ 
        1807      &     0.426$^p$     &        \dotfill  &	   8.7e-05  &   4.3e-5   &	38.93	  &	    41.88     &       $>$0.35	  &	     0      &	    10.72     &        0.10	 &     9.03    &       2e+09	  &	   \dotfill  &        \dotfill \\ 
        2113      &      0.730         &	 ALG	  &	   6.20e-05   &  3.49e-5   &	40.38	   &	    42.28     &       $>$0.41	  &	     0      &	    11.43      &       0.09	 &     10.75	&	2e+09	   &	    \dotfill  &        \dotfill \\ 
        2471      &     0.492	       &	 NL	 &	   8.60e-05   &  4.06e-5  &	40.49	  &	    42.13    &       $>$0.31	  &	     1      &	    11.45     &        0.10	 &     11.00	&	3e+09	   &	    \dotfill  &        \dotfill \\ 
        2692      &     1.136$^p$     &        \dotfill  &	   8.30e-05  &   3.04e-5  &	40.25	  &	    43.05    &       $>$0.61	  &	     0      &	    11.22     &        0.09	 &     10.57	&	2e+09	   &	    \dotfill  &        \dotfill \\ 
        3270      &     1.420$^p$     &        \dotfill  &	   9.50e-05  &   4.08e-5  &	40.91	  &	    43.48    &       $>$0.25	  &	     0      &	    11.71     &        0.06	 &     11.22	&	2e+09	   &	    \dotfill  &        \dotfill \\ 
         803      &     0.379$^p$     &        \dotfill  &	   7.30e-05  &   4.21e-5  &	39.98	  &	    41.52    &       0.00      &	  0	 &	 11.00        &        0.10	 &	  10.24      &       2e+09	&	     \dotfill  &     \dotfill \\ 
        1630      &     0.887	       &	 NL	 &	   1.54e-04   &  7.2e-5  &	40.49	  &	    42.94   &	     0.00      &	  0	 &	 11.29        &        0.10	 &	  10.75      &       2e+09	&	     \dotfill  &     \dotfill \\ 
        3029      &       0.389        &       SF gal	   &	   1.25e-04   &  5.0e-5    &	40.46	    &	    41.82     &        0.00	 &	    0	   &	   11.38     	&      0.09	 &     10.84   &       2e+09	  &	   \dotfill  &        \dotfill \\ 
\hline
\hline
\end{tabular}
$^a$: The count rates and error in the 0.5-7 keV band. $^b$: L$_{X,lmxb}$ is the X-ray luminosity computed using Fragos et al. (2013) relation. 
$^c$: L$_{X}$ is the remaining X-ray luminosity after subtracting the LMXB contribution to the total and it is computed using a thermal model as in Section 3.1.
$^p$: Photometric redshift.
\label{tab1}
\end{table*}
\end{center}

\begin{figure}
\centering
\includegraphics[width=0.52\textwidth, angle=180]{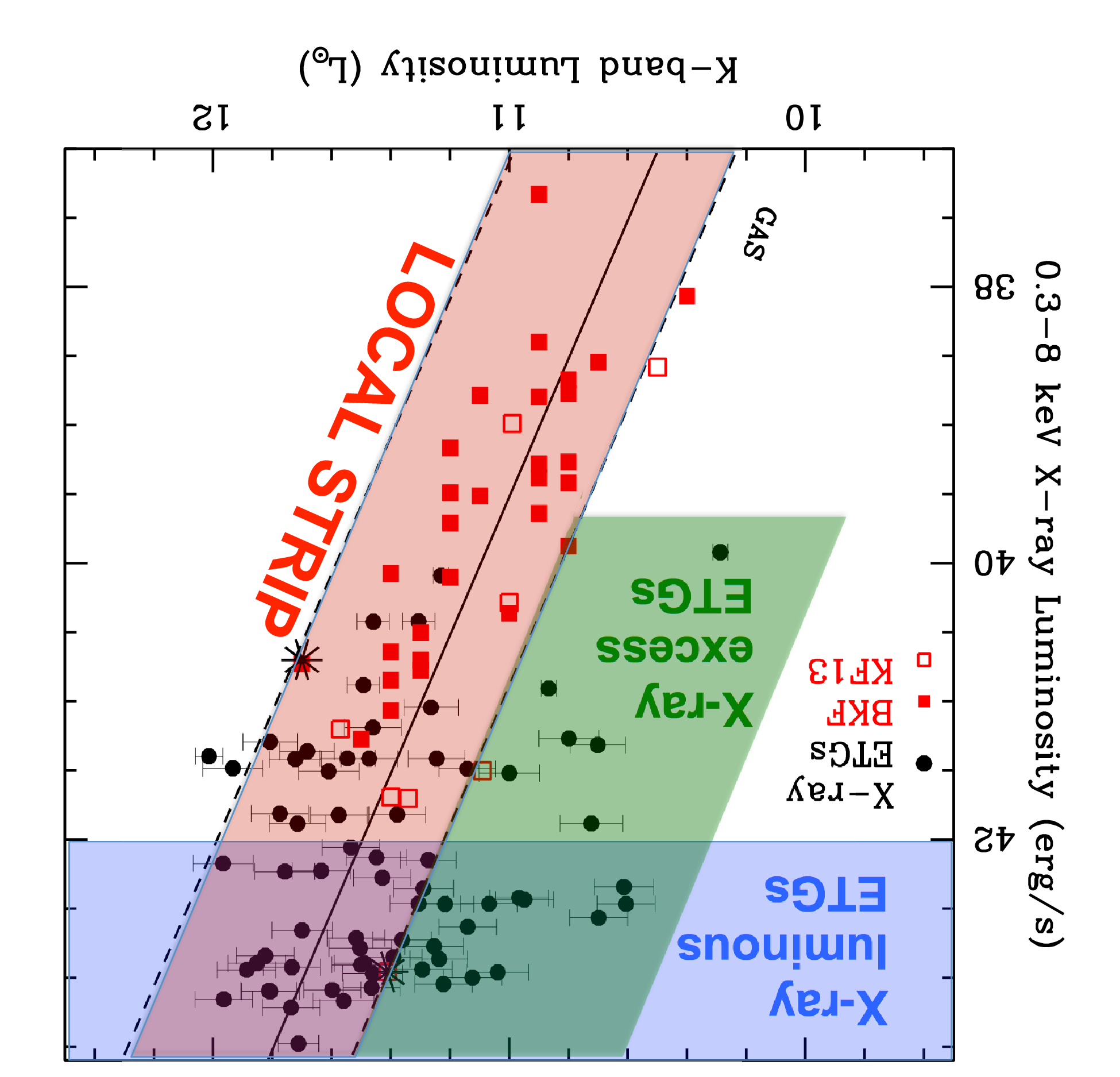}
\caption{\small Rest frame X-ray luminosity (after LMXB contribution subtraction) versus rest frame K-band luminosity for the X-ray ETGs in the C-COSMOS sample and for the local sample ETGs in red (BKF and KF13). The uncertainty on the X-ray luminosity is $\sim$25\% corresponding to $\sim$0.1dex. The solid line represent the KF13 relation and the dashed lines the limits of the local strip defined by M87 and NGC1316 (starred symbols). The local strip is marked as a red shaded area, the regions including X-ray luminous ETGs and X-ray excess ETGs are the shaded blue and green regions, respectively.}
\label{lxlk_bis}
\end{figure}

\begin{figure}
\centering
\includegraphics[width=0.48\textwidth]{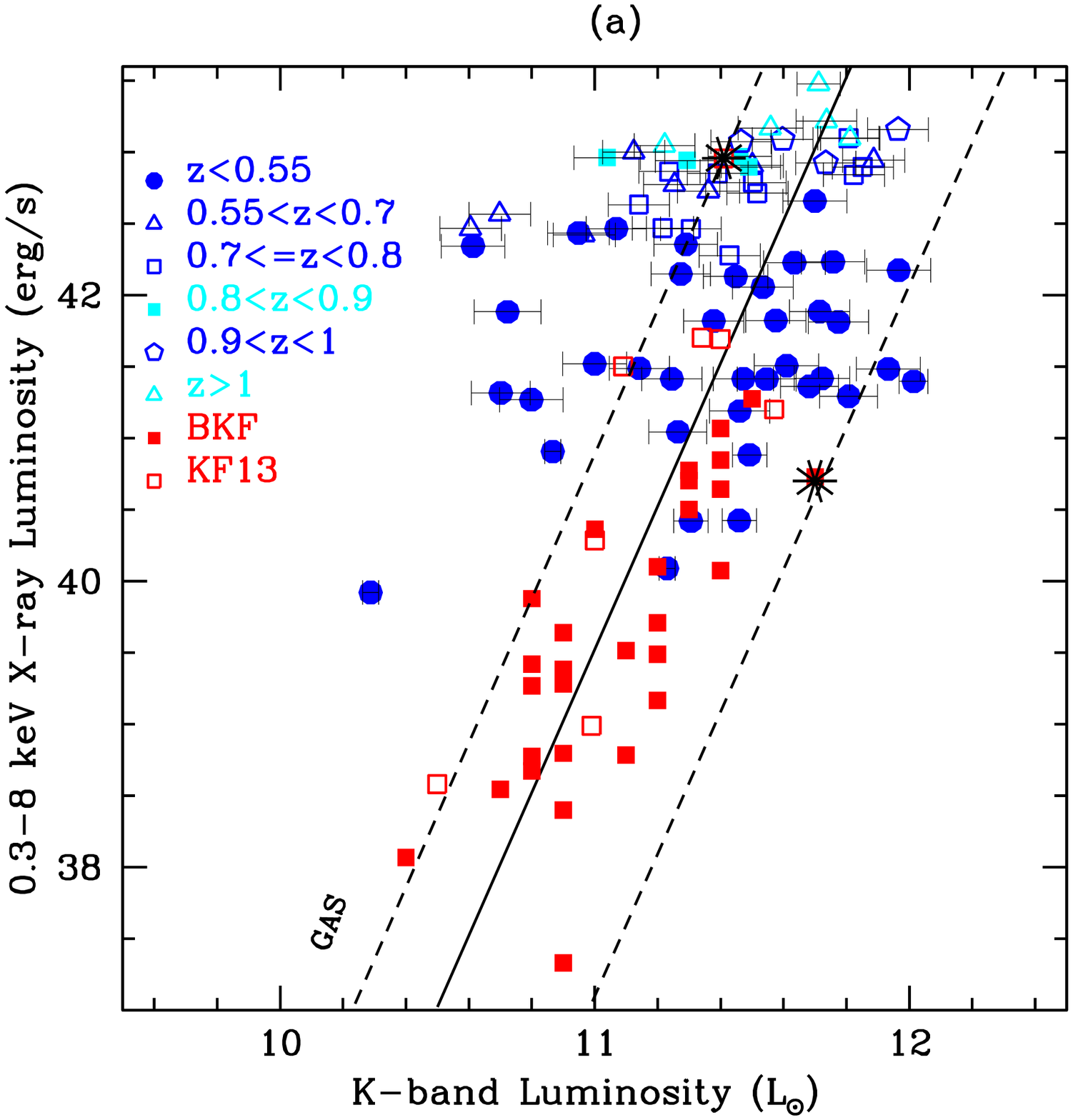}
\includegraphics[width=0.48\textwidth]{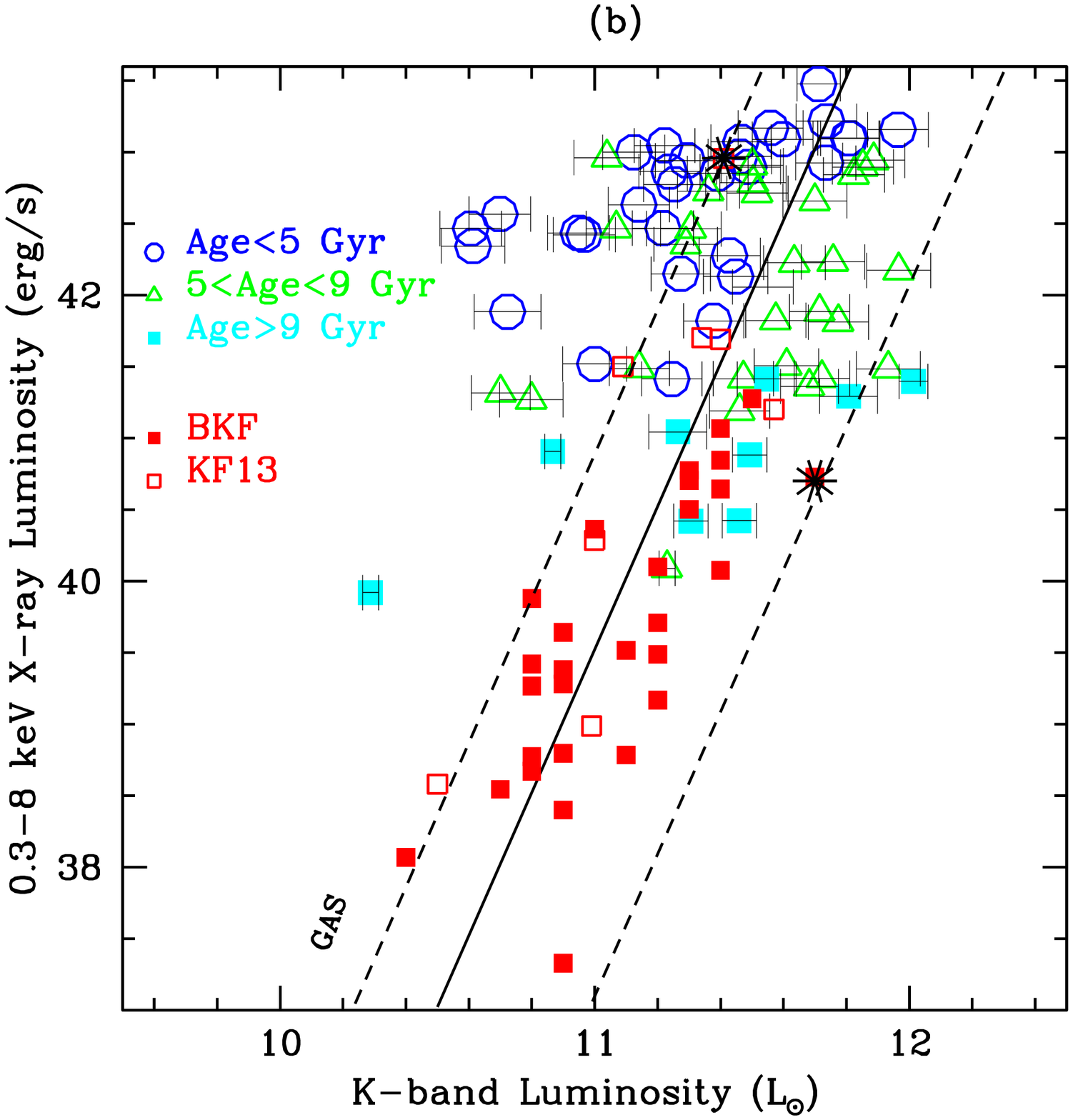}
\includegraphics[width=0.48\textwidth]{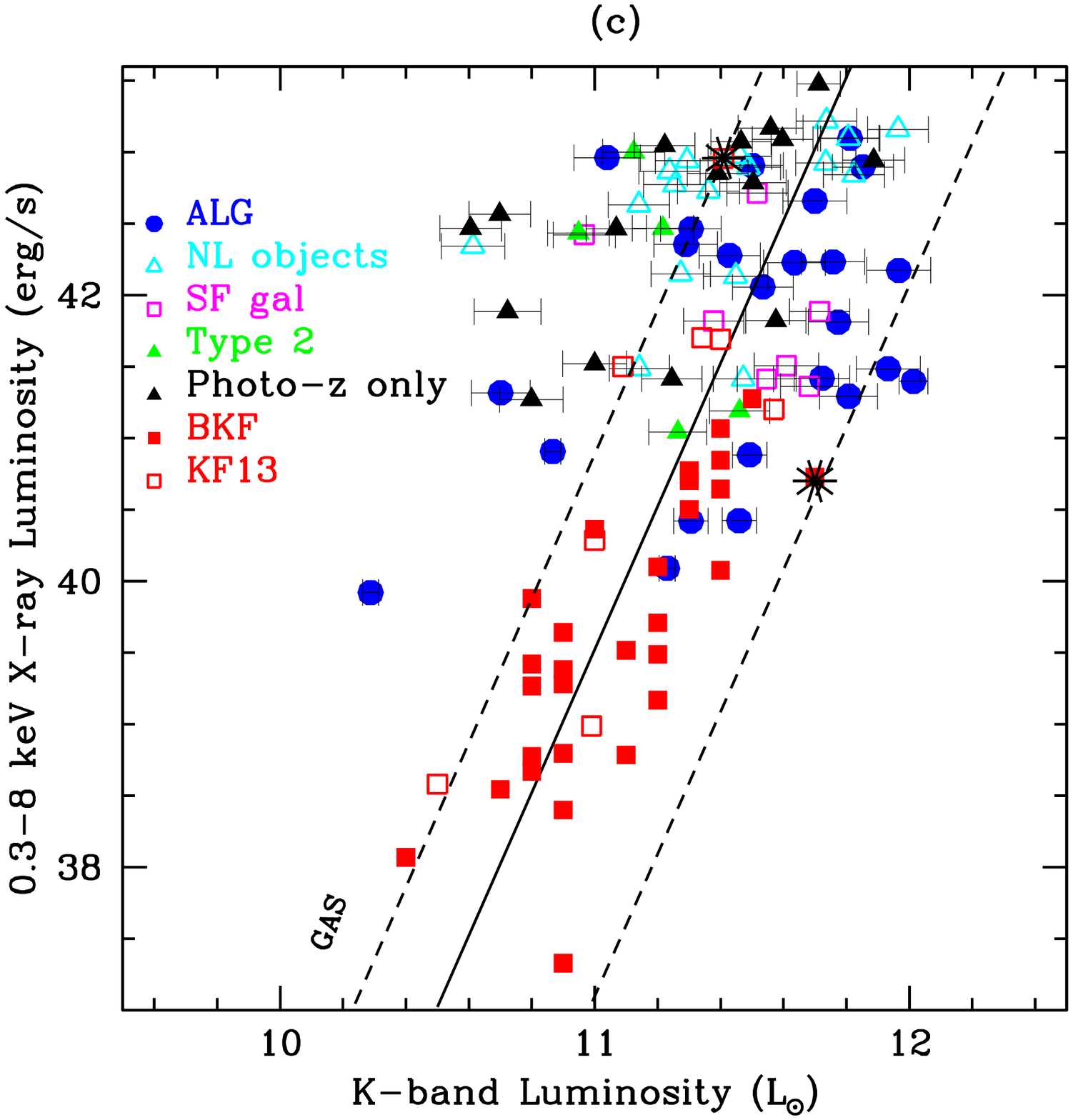}
\caption{\small Rest frame X-ray luminosity (after LMXB contribution subtraction) versus rest frame K-band luminosity for the X-ray ETGs in the C-COSMOS sample and for the local sample ETGs in red (BKF and KF13). The X-ray ETGs are labelled according to their redshift (a), age (b), spectroscopic type (c). The $L_K$ uncertainties reflect the age uncertainties which affect the fading of the stellar population with age. The uncertainty on the X-ray luminosity is $\sim$25\% corresponding to $\sim$0.1dex. The solid line represent the KF13 relation and the dashed lines the limits of the local strip defined by M87 and NGC1316 (starred symbols).}
\label{lxlk}
\end{figure}

\begin{figure}
\centering
\ContinuedFloat
\includegraphics[width=0.48\textwidth]{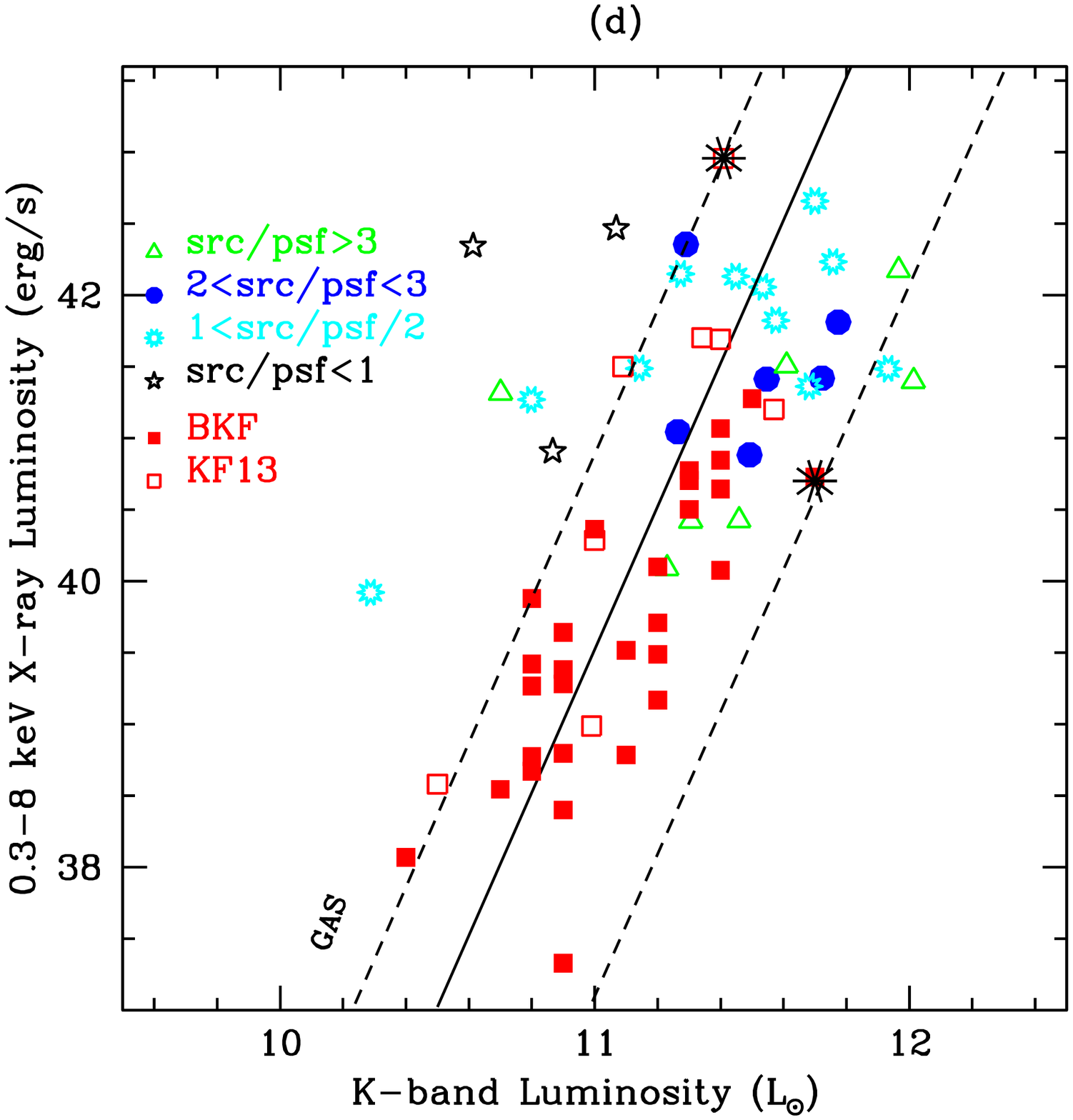}
\includegraphics[width=0.48\textwidth]{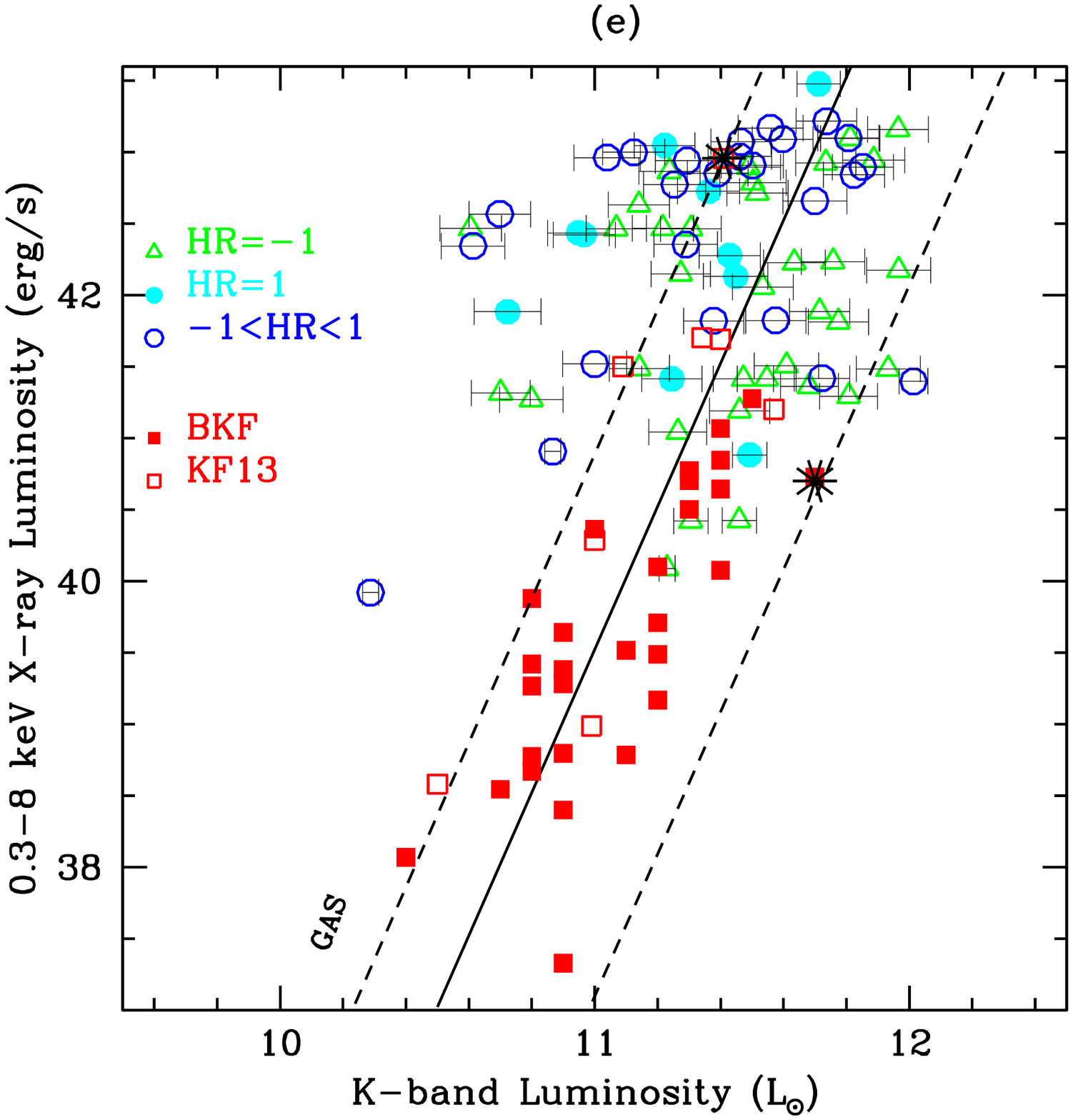}
\includegraphics[width=0.48\textwidth]{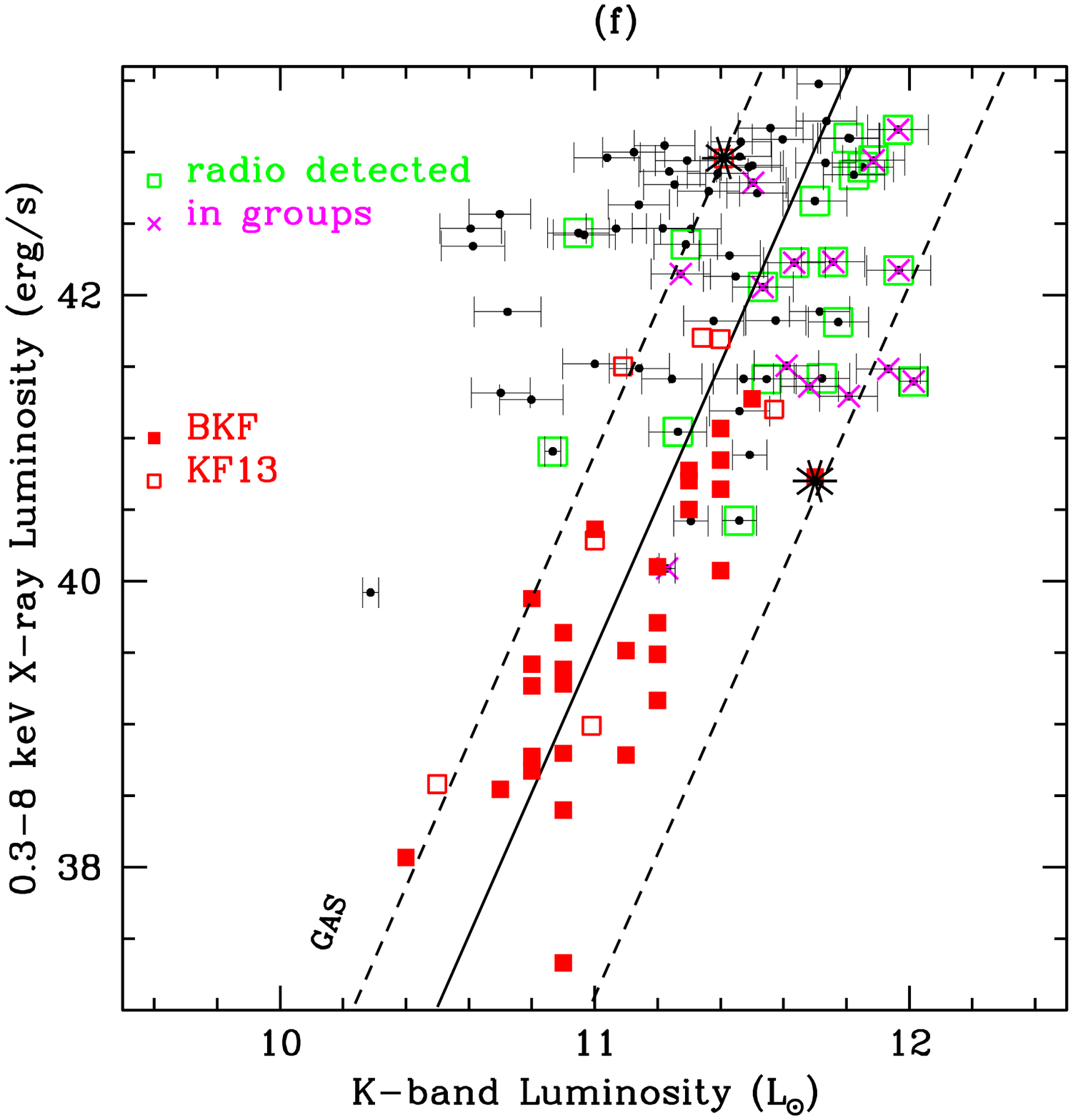}
\caption{\small Rest frame X-ray luminosity (after LMXB contribution subtraction) versus rest frame K-band luminosity for the X-ray ETGs in the C-COSMOS sample and for the local sample ETGs in red (BKF and KF13). The X-ray ETGs are labelled according to the extension of their X-ray emission (d), hardness ratio (e), radio detection and group occupancy (f). The $L_K$ uncertainties reflect the age uncertainties which affect the fading of the stellar population with age. The uncertainty on the X-ray luminosity is $\sim$25\% corresponding to $\sim$0.1dex. The solid line represent the KF13 relation and the dashed lines the limits of the local strip defined by M87 and NGC1316 (starred symbols). (Continued)}
\label{lxlk}
\end{figure}

\section{The local $L_X-L_K$ relation and its high redshift version}
\label{scaling}

In Figure \ref{lxlk_bis}, the X-ray ETGs are plotted together with the local sample. 
In order to compare the K-band luminosity of the X-ray ETGs with those of the local sample, we have taken into account the evolution of the stellar population (i.e., the fading effect). We computed the correction factor for ages in the range 2 to 10 Gyr at different redshifts starting from an elliptical galaxy template produced using a Chabrier (2003) initial mass function with an instantaneous burst of star formation. At a given z$_{obs}$, younger galaxies have larger fading correction than older ones. 
For all the sources, given the uncertainties on the age, we report the possible range of K-band luminosities corrected for fading as an error bar on the x-axis. The K-band luminosity plotted in Figure \ref{lxlk_bis} and reported in Table 1 is the value at the center of this range.
As discussed in Section \ref{xraypro}, the LMXB contribution was subtracted from the total X-ray luminosity, and the remaining luminosity should now represent the gas component plus some contribution from the nuclear point source, if any. 
The ``typical'' error bar on the X-ray luminosity corresponds to 25\% of the X-ray luminosity ($\sim$0.1dex). 

BKF found a $L_{X,gas}-L_K$ relation with best fit slope of $2.6\pm 0.4$, considering ETGs in the local sample with 0.3--8 keV luminosities of the hot gas $L_{X,gas}\leq 2\times 10^{41}$ erg s$^{-1}$, and excluding central dominant (cD) galaxies. 
Including in the local sample a few more ETGs reaching
$L_{X,gas}\sim 10^{42}$ erg s$^{-1}$, and one ETG (M87) with $L_{X,gas}=9\times 10^{42}$ erg s$^{-1}$, KF13 found a steeper $L_{X,gas}-L_K$
relation, with best fit slope of $4.5\pm 0.8$ (with M87), or $4.0\pm0.7$ (without M87). Both samples of BKF and KF13 are shown in Figure \ref{lxlk_bis},
with different symbols (solid and open red squares). 
In Figure \ref{lxlk_bis}, for reference, we have plotted two lines (dashed black lines), parallel to the local relation of KF13, indicating the most extreme ranges (to the left and right) of the relation at bright X-ray (passing through M87) and bright K-band (passing through NGC1316) luminosities. We will refer to this in the text as the ``local strip'' (red shaded area).

The first remarkable evidence from Figure \ref{lxlk_bis} is that most of COSMOS X-ray ETGs lie along
the local $L_{X,gas}-L_K$ relation, and seem to confirm and extend its validity to $L_X$ values larger than observed locally. This finding suggests that the main factors regulating $L_{X,gas}$ of local ETGs (e.g., gravitational
attraction, heating from supernovae and AGN feedback) were already the main factors at $z>0$, and created a state for the gas that is
long-lasting, since it corresponds to $L_{X,gas}-L_K$ relations that are consistent at $z=0$ and at $z=0-1$. For example, the
solid line in Fig. \ref{lxlk_bis} can be considered as a fiducial ``virial relation'' for the hot gas coronae of the local universe (KF13): 
this local $L_{X,gas}-L_K$ relation is consistent with 
a tighter $L_{X,gas} \propto M_*^{3}$ and $L_{X,gas} \propto T_{gas}^{4.5}$ relation (see KF13 for more details).
Figure \ref{lxlk_bis} shows that, at least for the most massive X-ray ETGs, the gas is bound by the dark matter potential, at a temperature close to $T_{vir}$, and in an overall equilibrium state
where similar average gas properties are kept on a secular timescale, and evidently heating (from supernovae and massive black hole accretion feedback) 
and radiative cooling balance. 

The panels in Figure \ref{lxlk} highlight seven properties of the X-ray ETGs: redshift (a), age (b), spectroscopic classification (c), X-ray extension (d), HR (e), radio detection (see Section \ref{xraybrightetgs}) and group occupancy (f).
The redshift bins in Fig. \ref{lxlk} (a) have been selected in order to have equal comoving volume in each redshift bin. 

The majority of COSMOS X-ray ETGs with $L_X <10^{42}$ erg s$^{-1}$ are less distant (all except one have $z<0.5$; Fig. \ref{lxlk} (a)), with $HR=-1$ (Fig. \ref{lxlk} (e)), many are extended in the X-rays (Fig. \ref{lxlk} (d)) and optically classified as ALG (Fig. \ref{lxlk} (c)).
Based also on their moderate $L_X$ values, the X-ray emission can be entirely due to hot ISM.  With respect to the more distant and X-ray
more luminous X-ray ETGs, they populate the region around the $L_{X,gas} - L_K$ relation more uniformly, with roughly the same number of objects to
the left and to the right of it. Overall, these X-ray ETGs follow nicely the local $L_{X,gas}-L_K$, just adding an enhanced scatter to it.
We also note that all those in groups are within the local strip (Fig. \ref{lxlk} (f))

Three further points emerge from comparing the distribution of the ETGs from the local sample and COSMOS X-ray ETGs: (1) There is a large number (43 sources) of X-ray ETGs with $L_X\geq 10^{42}$ erg s$^{-1}$, a range of values that is
not so frequent in ETGs of the local universe, if cases of AGN contamination
are excluded (e.g., O'Sullivan et al. 2001). (2) There is a substantial group of X-ray ETGs
lying to the left of the strip, due to an excess of $L_X$ for their $L_K$ with respect to what found locally. Most (but not all) of these X-ray ETGs have also 
$L_X\geq 10^{42}$ erg s$^{-1}$. (3) There are no ETGs to the top right of the local strip at $L_X\geq 10^{42}$ erg s$^{-1}$
because their L$_K$ would be higher than the typical K-band luminosity of the ETGs in the COSMOS sample. 
As shown in Figure \ref{lxlk_bis}, we define the X-ray ETGs in the first group as X-ray luminous ETGs and in the second group as X-ray excess ETGs. Below we discuss possible origins for both groups.

\subsection{The X-ray luminous ETGs}
\label{xraybrightetgs}

The X-ray emission of the X-ray luminous ETGs with $L_X\geq
10^{42}$ \cgs\ (in the blue shaded area of Figure \ref{lxlk_bis}) can be explained by three
possibilities: (a) these galaxies are central dominant ETGs in groups
or clusters; (b) the emission includes a major contribution from an AGN; (c)
these galaxies are experiencing a specially bright phase for the hot gas, due
to an evolutionary effect (indeed, most of these sources are at $z>0.55$; Fig. \ref{lxlk} (a)).

In the local universe, hot gas coronae in galaxies that are isolated
or in hot gas-poor environments have luminosity values $L_X<10^{42}$
erg s$^{-1}$, as shown by observational evidences (e.g., O'Sullivan et
al. 2001, BKF) and by numerical simulations (e.g., Pellegrini \&
Ciotti 1998, Mathews \& Brighenti 2003).  Luminosities $L_X\geq
10^{42}$ erg s$^{-1}$ for hot gas coronae, instead, are typical of
cD ETGs in groups or clusters (Helsdon et al. 2001,
Matsushita 2001, Nagino \& Matsushita 2009), where the conditions for
gas retention, or even accretion from outside the galaxies, are
favorable; the X-ray emission could be also contaminated by the
group/cluster emission. In Fig. \ref{lxlk}, the representative case of a
cD galaxy is used (M87, black starred symbol on top left) in the local sample. Central
galaxies thus show exceptional $L_X/L_K$ ratios, since they couple an
optical luminosity typical of a bright\footnote{For example M87 is not
  the K-brightest ETG in Virgo: its log$L_K$=11.4, while two other
  giant ETGs in Virgo have log$L_K$=11.50 (NGC4472) and log$L_K$=11.48
  (NGC4649).} ETG to X-ray luminosities that can reach $\sim 2\times
10^{43}$ \lum\ (Helsdon et al. 2001). Such high $L_X$ values
cannot be reproduced by models for isolated ETGs (e.g., Brighenti \&
Mathews 1998, Pellegrini \& Ciotti 1998).

A few facts lend support to the idea of the presence of some central ETGs among the X-ray luminous ETGs. 
First, $\sim$40\% X-ray luminous ETGs have $HR=-1$, indicative of dominant soft X-ray emission from a
thermal plasma (Fig. \ref{hr}, Sect. 3.1, Fig. \ref{lxlk} (e)). Second, in Fig. \ref{lxlk} (a)
number of X-ray luminous ETGs are found around the position of M87, slightly to the
left of the local $L_{X,gas}-L_K$ relation, and could be M87-like $z>0$
counterparts.  Finally, for a number of cases (8 sources), we have found that our X-ray luminous ETGs 
lie at the center of groups (Fig. \ref{lxlk} (f)). The number of C-COSMOS 
X-ray sources (from the identification catalog of Civano et al. 2012) lying within 6$^{\prime\prime}$ from the center of a 
group is instead only $\sim$25 of 841 X-ray sources 
up to z=1.4 (the limit in redshift of the Giodini et al. 2010 catalog), 
which makes the number of ETGs at the center of groups very significant.
Further constraints on this idea could come from the
extension of the X-ray sources: the extension could be studied just
for few (8) cases of the X-ray luminous ETGs, that are also the most distant
galaxies (Fig. \ref{lxlk} (a)); in two cases size$_{source}$/size$_{PSF}>$2, two sources are
point-like, and the remaining 4 have 1$<$size$_{source}$/size$_{PSF}<$2 (Fig. \ref{lxlk} (d)).  If
there are M87-like hot haloes of central ETGs among the X-ray luminous ETGs, since
many X-ray luminous ETGs seem to stay close to the local virial relation out to the
largest $z$ in our sample, then already at a redshift $z\sim 1$ there
were haloes that have reached an equilibrium similar to that seen
locally.

Support for an AGN origin of the high $L_X$ is given by the ``hard'' HR value for seven of the X-ray luminous ETGs: of
the 10 X-ray ETGs with HR=1, 7 are X-ray luminous (Fig. \ref{lxlk} (e)). Even in many of the
other softer X-ray luminous ETGs (those with $-1<HR<1$) there could be some
AGN contamination, from an AGN of moderate luminosity ($L_X\la
10^{43}$ erg s$^{-1}$, Ho 2008). Some AGN contamination can be present even in
the hypothesized central ETGs among the X-ray luminous ETGs discussed above.
A hint for nuclear activity is also found using radio emission. Of the 19 X-ray ETGs (27\% of the total sample) 
with a counterpart in the Very Large Array 1.4 GHz COSMOS radio catalog (Schinnerer et al. 2010), 
12 are X-ray luminous (Fig. \ref{lxlk} (f)). This percentage is larger than what found when matching 
the COSMOS ETGs with comparable K band luminosities to our sample with the VLA-COSMOS catalog which 
returns only a 3\% of matches.

The third hypothesis for
the origin of the X-ray luminous ETGs is that of an evolutionary effect.  Gas-dynamical numerical
models for the study of the hot gas behavior during the ETGs'
lifetime, that take into account the evolution of the stellar
population and the effects of AGN feedback, show that the gas coronal
luminosity on average should have been just mildly larger in the past, due to the
combined effects of a stellar mass loss rate that was larger,
and larger duty-cycle of activity
(Ciotti et al. 2010).
Models without feedback predict that the gas was much more X-ray luminous
in the past (up to $L_X\sim 10^{42}$ erg s$^{-1}$ for isolated ETGs,
Ciotti et al. 1991), and for a prolonged time (a few Gyr), with the
accumulation of large amounts of cooled gas mass; 
the introduction of feedback from accretion on the massive black hole (radiative
plus mechanical, triggered by a high mass accretion rate $\dot
 M\ga 0.01 \dot M_{Edd}$) produces recurrent cycles (Ciotti et al. 2010).
During each cycle the hot ISM accumulates and its luminosity slowly
increases, reaching suddenly values even larger than $10^{42}$ erg s$^{-1}$
for a brief time (a few $\times 10^7$ yr), when the AGN turns
on.  After having removed the gas from its surroundings
with its feedback action, the AGN fades, the hot gas
emission decreases, and a new cycle starts. Outside nuclear outbursts, the gas
shows a mild secular decline in $L_X$, of a factor of a few; also the gas 
temperature remains roughly constant (Pellegrini et al. 2012). 

Some X-ray luminous ETGs could then be experiencing the consequences of an accretion episode at high
mass accretion rate, with a brightening of the nucleus, and the
increase of the hot ISM luminosity due to various feedback effects (as
described in Ciotti et al. 2010). The optical classification of the
X-ray luminous ETGs is a mixed bag, not revealing the clear presence of an active source. 
ETGs dominated by the brief and very luminous AGN phase are not included in
the X-ray ETG sample, by construction, but the flare of the hot gas emission
lasts longer than that of the nucleus, and could be at the origin of a
few X-ray luminous ETGs (with their $L_X$ due to mostly to hot gas). A number of the
X-ray luminous ETGs could have such an origin: the duty-cycle of
the nuclear (bolometric) activity is in the range 0.006-0.048 
since $z\approx 0.8$, thus we expect a duty-cycle just larger
for the hot gas ``activity'' (above a level of $L_X=10^{42}$ erg
s$^{-1}$; Pellegrini et al. 2012).
Therefore, just a few such
cases in a sample of $\sim 100$ ETGs at $z=0-0.8$ are expected. Indeed, considering all the X-ray luminous ETGs 
and the COSMOS ETGs at bright K-band luminosities ($>10^{11}$ L$_{\odot}$) and z$<$0.8, 
we find a duty cycle of 0.02 consistent with what expected above.
The expected range of the duty-cycle given above refers to an isolated ETG of $L_K\approx 2\times
10^{11}$  L$_{\odot}$ formed at $z\ga 3$ and with a stellar mass of 2.9$\times10^{9} M_{\odot}$; these numbers become larger for later formation epochs,
and more massive ETGs, or in dense environments (where the gas
retention is larger). 

Recently, for galaxies in the X-ray AEGIS survey at 0.3$\leq z \leq
1.3$, the extended emission of the hot ISM in 96 active and a large
sample of non-active galaxies was characterized with a stacking
analysis, to study possible effects of feedback from AGN on the
diffuse interstellar gas (Chatterjee et al. 2013). By comparing the
average stacked X-ray surface brightness profiles of the two classes
of active and non-active objects, disturbances were found in the
profile of AGN host galaxies, qualitatively similar to the predictions
of the feedback models described above (Pellegrini et al. 2012).
Thus feedback could cause an evolution 
for the hot gas as predicted in these models, and it remains a possibility that
it is at the origin of a small fraction of the X-ray luminous ETGs of Fig. \ref{lxlk_bis}.

In conclusion, the high $L_X$ of X-ray luminous ETGs could be the mixed result of
(in order of decreasing importance) the presence of
many M87-like ETGs, which are rare within 32 Mpc, but have been found
in a large number in the large volume surveyed by COSMOS (the comoving volume within z$<1$ is $\sim$150 Gpc$^3$);
contamination from AGNs, dominating or not the total $L_X$; 
evolution in the X-ray properties of ETGs.

We note that the number of X-ray luminous ETGs ($L_X >10^{42}$ erg
s$^{-1}$) remains constant when looking at similar volumes showing no evolution in the source number: 
there are 11 sources at z$<$0.55, 8 in the redshift range 0.55-0.7 and 11 in the range 0.7-0.8. This lack of evolution is different from that found for the growth 
of X-ray selected AGN (Lehmer et al. 2007). However, we should keep in mind that this sample is small, and for many of its objects the X-ray luminosity 
could be contaminated by low luminosity AGNs, or effects of age/interactions. 
Moving to higher redshift, the survey start to lose sensitivity so we cannot perform a substantial comparison (see Fig. \ref{lxz}). 
The same happens when considering lower X-ray luminosities.

\subsection{X-ray ETGs to the left of the local strip: an age effect?}

We now turn to consider the other major evidence provided by Fig. \ref{lxlk_bis} 
mentioned above: most X-ray ETGs overlap with the local $L_{X,gas}-$strip,
except for a substantial group of X-ray ETGs located to the left of the
strip, that have an excess of $L_X$ for their $L_K$; we call the X-ray ETGs in
this group X-ray excess ETGs.  These include 6 X-ray ETGs with $L_X< 10^{42}$ \cgs, 
plus those 15 X-ray luminous ETGs occupying the leftmost positions
away from the strip.

A few X-ray excess ETGs are found close to M87, and could be galaxies similar to
it and for these the same solutions suggested in the previous section can apply. 
However, the $L_K$ of the X-ray excess ETGs are lower than
that of M87, thus it is not likely that most of them are cD
galaxies. As shown in Figure \ref{lxlk} (f), none of them reside in the central part of a group. 
Central galaxies in the local universe have
log$L_B\ga 10.8$ (Helsdon et al. 2001); taking the $L_K/L_B$ ratio of
M87 as representative, this means log$L_K\ga 11.4$ (in fact M87 
is already one of least luminous central galaxies in the optical).

The optical classification of X-ray excess ETGs is mixed (Fig. \ref{lxlk} (c)), but with just
three ALG. Their HR values are also mixed (Fig. \ref{lxlk} (e)): there are 4
HR=1 cases,  suggesting these could be
AGN-dominated, from their hardness ratio; there are 5 HR=-1 cases, and so likely 
dominated by soft gaseous emission; while the remaining 7 have intermediate HRs.

Figure \ref{lxlk} (b) shows clearly that most X-ray excess ETGs are distinguished by having
ages in the lowest ``age-bin'' considered in this work ($<5$ Gyr).
However, there are many X-ray ETGs with similarly low age and lying closer to the ``virial'' relation; 
as described in the previous Section, the hot gas luminosity (and temperature) is expected to evolve slowly on average going to the past. AÊ few of the
youngest X-ray excess ETGs, though, could be living in an epoch when the duty-cycle of activity was larger, and have been X-ray
detected thanks to the huge increase of the hot gas (and nuclear) luminosity taking place during a nuclear outburst. Note that 
the X-ray excess ETGs lie on the lower-$L_K$-side of the strip, and, due to downsizing in galaxy formation (Cowie et
al. 1996, Thomas et al. 2005), could be experiencing the same hot gas
evolution of the higher-$L_K$-side with some delay.

Another possibility, emerging from the number of X-ray excess ETGs with a
younger age, is that they could be young remnants of major mergers;
during mergers the X-ray emission can reach values of $L_X\sim
10^{43}$ erg s$^{-1}$, depending on the progenitors' mass, lasting for
$\sim 1-2$ Gyr (Cox et al. 2006). 
As for example, the X-ray luminosity of the hot halo in the
merging galaxy NGC 6240 is 10 times higher than the X-ray luminosity expected 
from its stellar mass (Nardini et al. 2013), probably due to a superwind originated by a recent, 
nuclear starburst or more likely to star formation enhanced by the merger. The large number of narrow line objects among the X-ray excess ETGs (7 sources, Fig. \ref{lxlk} (e)) could be indeed a signature of nuclear star formation.
Moreover, if the merger is recent, an increased number of Ultra Luminous X-ray sources (ULXs) could 
be present in the galaxy, as seen for example in the Antennae and Cartwheel galaxies (Fabbiano, Zezas \& Murray 2001; Wolter \& Trinchieri 2004). 
The expected number of ULXs in ETGs, from local studies, is of the order of 1 ULX  per 10$^{11}$ M$_{\odot}$ galaxy mass, thus we expect to see $<10$ in our X-ray ETGs, contributing up to 10$^{39}$ erg/s in X-ray luminosity (Gilfanov 2004, Swartz et al. 2004). In case of a recent merging event, it is possible to have a larger number, up to 1 ULX  per 10$^{10}$ M$_{\odot}$, thus $<$100 ULXs contributing to the overall X-ray luminosity with more than $L_X\sim
10^{40}$ erg s$^{-1}$.
The X-ray excess ETGs would be similar to, but 
more evolved than, NGC6240 in the local universe. 
A qualitative analysis of the {\it Hubble} ACS (filter FW814) images of the 68 X-ray ETGs shows that 35\% 
have a companion which could be interacting or merging, while 60\% are isolated (see Fig. \ref{fc1}). 
Of the X-ray excess ETGs to the left of the local strip, 25\% could be interacting 
or merging. The high X-ray luminosity could be partially explained by this merging/interaction effects which could boost it even by a factor of ten.

More evolved merger remnants show instead a
deficit of $L_X$ with respect to what is typical of ETGs with similar
$L_K$. A \chandra\ survey of interacting galaxies showed that
the X-ray luminosity peaks $\sim 300$ Myr before nuclear coalescence,
and then drops, so that $\sim 1$ Gyr after coalescence, the merger
remnants are X-ray fainter than typical massive, evolved ETGs
(Brassington et al. 2007; as first noted by Fabbiano \& Schweizer 1995).  Evolved merger remnants could account for
the X-ray under-luminosity of the 2 X-ray ETGs to the right of the local strip, none of which has a young age; for example, NGC1316 (the local
ETG with the highest $L_K$ in Fig. 7, black starred symbol) is a nearby merger remnant, and
is under-X-ray-luminous with respect to the local $L_{X,gas}-L_K$ relation.

In conclusion, the excess-$L_X$ of X-ray excess ETGs is likely linked to 
AGN-contamination, or evolution, since most X-ray excess ETGs are particularly 
young. Thus, they could be still living in an earlier phase
of hot gas flow evolution, and represent the rare X-ray brightest peaks 
reached during the cycles of activity, or
could be interacting or merging objects.


\section{The local and high redshift $L_X-M_{tot}$ relations}

The KF13 investigation revealed an intriguing smaller scatter in the
$L_X-M_{tot}$ relation with respect to the $L_X-L_K$ one, such that the $L_X-M_{tot}$
could be used to infer total mass values knowing the X-ray luminosity produced by the hot gas. 
Provided that the X-ray ETGs are the progenitors of local sample ETGs, and given that
many seem to lie on the same local $L_X-L_K$ relation, one could try to
derive the total mass for them, assuming they also follow the local $L_X-M_{tot}$ 
relation. 

\begin{figure}
\centering
\includegraphics[width=0.44\textwidth]{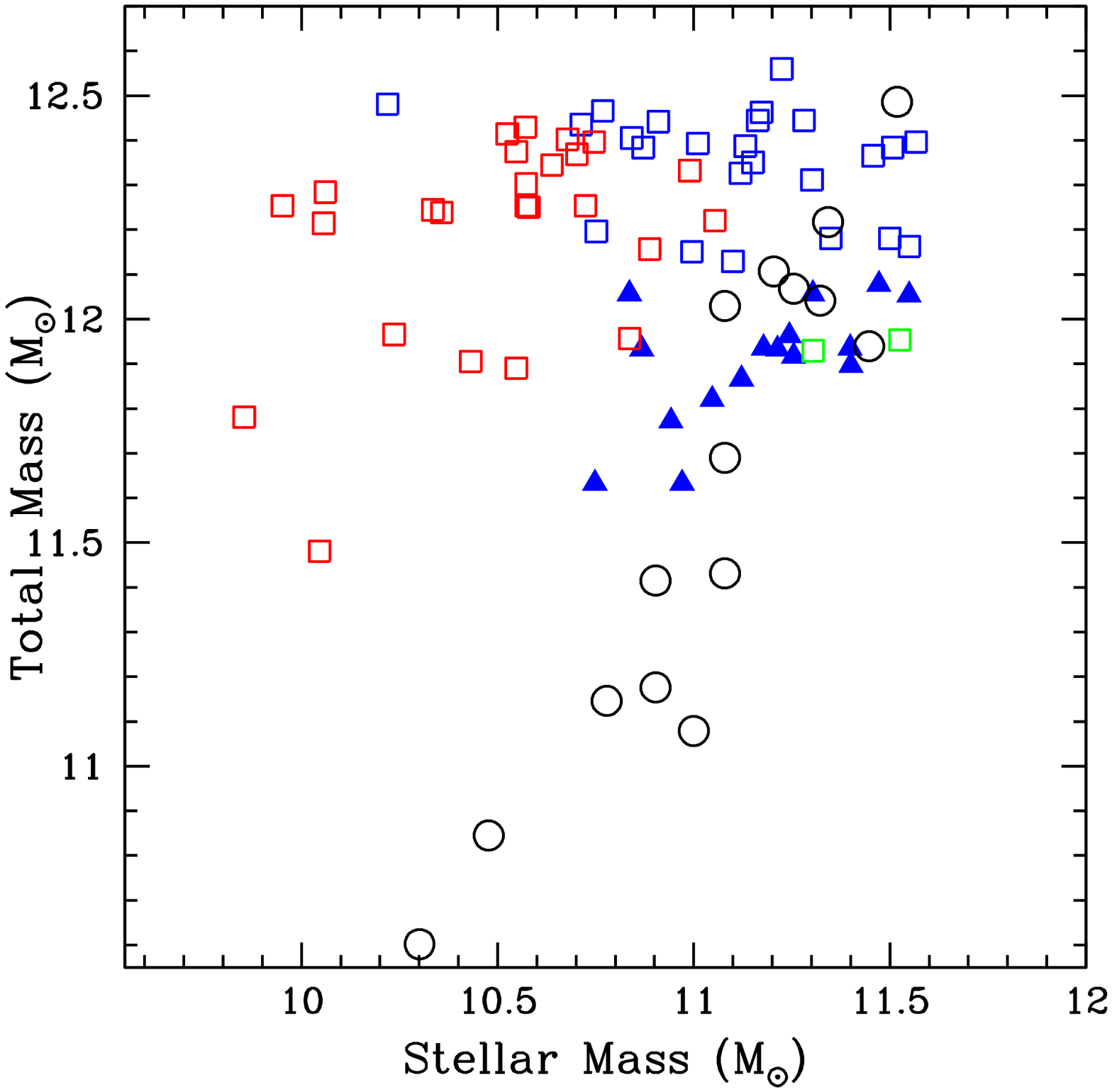}
\caption{\small Total mass versus stellar mass. Black circles are the sources from KF13. In blue X-ray ETGs in the local strip (triangles sources with L$_X<$42, squares with L$_X<$42),  in red X-ray ETGs to the left of the local strip, in green X-ray ETGs to the right of the local strip. }
\label{dark}
\end{figure}

The $M_{tot}$ values derived using KF13 relation are shown in Fig. \ref{dark}, for local and X-ray ETGs
galaxies and are compared with their stellar masses. When dividing X-ray ETGs according to their location with respect to 
the local strip, the sources outside the strip (red squares) on average 
deviate from the trend shown by local ETGs,
while the X-ray ETGs inside the strip (blue triangles and squares) seem consistent. This behavior
reflects the features of the $L_X-L_K$ plot, where a deviation with
respect to the local relation is shown by a group of X-ray excess ETGs (i.e., on
the low-$L_K$ side). The excess-$L_X$ translates in an
``overestimate'' of $M_{tot}$, and indeed in Fig. \ref{dark} the low-$L_K$ X-ray ETGs have
an excess of  $M_{tot}$ for their $M_*$.  
When dividing the sources in the local strip between high X-ray luminosity ($L_X>10^{42}$ \lum, blue squares) 
and low X-ray luminosity ($L_X<10^{42}$ \lum, blue triangles), it is clear that more ETGs in the first group deviate 
from the local relation than those in the second group. 
For an ideal use of the $L_X-M_{tot}$ relation, one should definitely 
disentangle whether $L_X$ is from AGN or large hot haloes (the $L_X-M_{tot}$ relation works for hot haloes).

Another possibility is that $M_*$ has been underestimated for the low-$L_K$ X-ray ETGs although 
their mass is consistent with the overall distribution of ETGs in the COSMOS field.
Finally, one could also note that the local $L_X-M_{tot}$ relation is tight
for 7 ETGs only, with L$_K \geq 10^{11}$ L$_{\star}$.

\section{Summary}

We have built a sample of 69 ETGs at $0.05<z<1.4$, all X-ray detected in the 0.5--10 keV band in the C-COSMOS survey. This sample, by selection, is representative of the population of red, passive ETGs detected in the COSMOS survey (Moresco et al. 2013), with which they share consistent distributions of stellar age, color and sSFR (Fig. 3, 4 and 5). The X-ray ETGs have stellar masses greater than $10^{10}M_{\odot}$ (Fig. 3), thus include the most massive end of COSMOS ETGs at all K-band luminosities. Their optical to infrared SEDs and spectra show no sign of strong contamination from nuclear AGNs. By imposing an X-ray luminosity cut of  $L_X< 5\times 10^{43}$ erg s$^{-1}$, we excluded the cases of X-ray emission dominated by bright unobscured and obscured AGN. 

In all cases, to obtain as close as possible an estimate of the X-ray luminosity of hot gaseous halos, we have subtracted the LMXBs contribution to the X-ray luminosity, following the scaling relations derived in the local universe (BKF), augmented by a prescription that keeps into account the evolution of the LMXB population (Fragos et al. 2013).  Faint nuclear emission is harder to isolate, but we can use the spectral shape of the X-ray emission as a guide, by looking at the \chandra\ hardness ratios (HR). All the X-ray ETGs not detected in the hard band and those with HR$<$0 could have emission dominated by the gas component. The X-ray luminosity of sources with positive HR (Fig. \ref{hr}) could still be contaminated by a nuclear component, possibly a low luminosity AGN.

The sample of X-ray detected ETGs at z $>$ 0 is large enough for a direct comparison with the local sample
of 38 ETGs from BKF and KF13. We note that previous works (Lehmer et al. 2007, 2012; Danielson et al. 2012), which mapped the evolution of the ratio L(0.5-2 keV)/$L_B$ over z=0.1--1.2, used mostly X-ray stacking analysis of optically selected ETGs in the \chandra\ Deep Fields. 

We find that, after having taken into account the different z range covered, the optical properties (as stellar mass, $L_K$, colors, ages) of the X-ray ETGs are similar to those of the local sample, except for the presence of 17 sources brighter than the $L_K$-brightest ETG of the local sample (with $L_K= 10^{11.7}L_{K,\odot}$), which can be explained by the larger volume surveyed in COSMOS.

With the exception of a few X-ray over-luminous objects, which may harbor hidden AGN, the $L_{X,gas}- L_K$ scatter plot of the COSMOS X-ray ETGs with $L_{X,gas} <10^{42}$ erg s$^{-1}$ and z$<$0.55 is largely consistent with that of the local sample (Fig. \ref{lxlk_bis} and \ref{lxlk}). These X-ray ETGs typically have the oldest stellar ages in the sample and absorption line optical spectra (ALG); their X-ray emission tends to be soft and spatially extended, as it would be expected from hot gaseous halos. 

Using stellar age as a discriminant, we find that all the X-ray ETGs with age $>$ 5 Gyr follow reasonably well the locus in the scatter diagram defined by the local sample (local strip), showing that the hot halos are similar to those observed in the local universe. This result is consistent with the predictions of evolutionary gas-dynamical models including stellar mass losses, supernova heating, and AGN feedback (Pellegrini et al. 2012). In these models, outside very short nuclear outbursts, the hot gas luminosity secularly decreases mildly, and the average emission weighted temperature remains roughly constant. Thus, one expects little variation in the $L_{X,gas}- L_K$ relation, as observed. For these galaxies, we conclude that total masses could be derived using the KF13 local sample virial relation (Fig. \ref{dark}).

Younger stellar age galaxies typically are found at higher redshift (z$ >$ 0.9) and they tend to have $L_X >10^{42}$ erg s$^{-1}$ and be over-luminous in X-rays for their $L_K$, when compared to the local sample and older stellar age galaxies. As suggested by their radio detection or hard HR, several of these X-ray luminous ETGs may harbor hidden AGNs. Given the young stellar age of these galaxies, the high X-ray luminosity could be reconnected with merging phenomena, which would enhance the X-ray luminosity of the halo (Cox et al. 2006) and also produce a population of ULXs (as e.g., in the Antennae; Fabbiano, Zezas \& Murray 2001, Wolter \& Trinchieri 2004), not modeled by LMXB stellar populations. Nuclear accretion could also be responsible for awakening an AGN during merging (Cox et al. 2006). The HST images of several of these galaxies show indeed close companions (Fig. \ref{fc1}).

We also notice a group of intermediate age z $>$ 0.55 galaxies (5-7 Gyr) that follows the local strip. Some of these galaxies are at the center of galaxy groups or clusters (as M87), where the condition of gas retention or accretion from outside galaxies are favorable, enhancing their X-ray luminosity, which could also be contaminated by the presence of hot gas in the group.  An evolutionary effect could explain the large X-ray luminosity 
of these ETGs: a SMBH accretion episode could have turned on the AGN and illuminated the gas by feedback effects
but, while the AGN phase is short, the flare of the hot gas emission can last longer (Ciotti et al. 2010).

To better constrain evolutionary models, it would be useful to derive a firm estimate of the duty-cycle from
the observations, with larger samples and more secure estimates of the hot gas properties and of the environmental
conditions. Although this sample of X-ray selected ETGs is already 25\% larger with respect to previous samples at
high redshifts, the \chandra\ COSMOS Legacy Survey, a 2.8 Ms X-ray Visionary Project approved to survey the whole
COSMOS field with \chandra\ at the same C-COSMOS depth, will provide an even larger sample of X-ray ETGs
at both faint and bright X-ray luminosity to improve the current statistic. Moreover, X-ray stacking analysis of the
currently undetected COSMOS ETGs could probe the lower X-ray luminosity end of the population.

\begin{figure}
\centering
\includegraphics[width=0.25\textwidth]{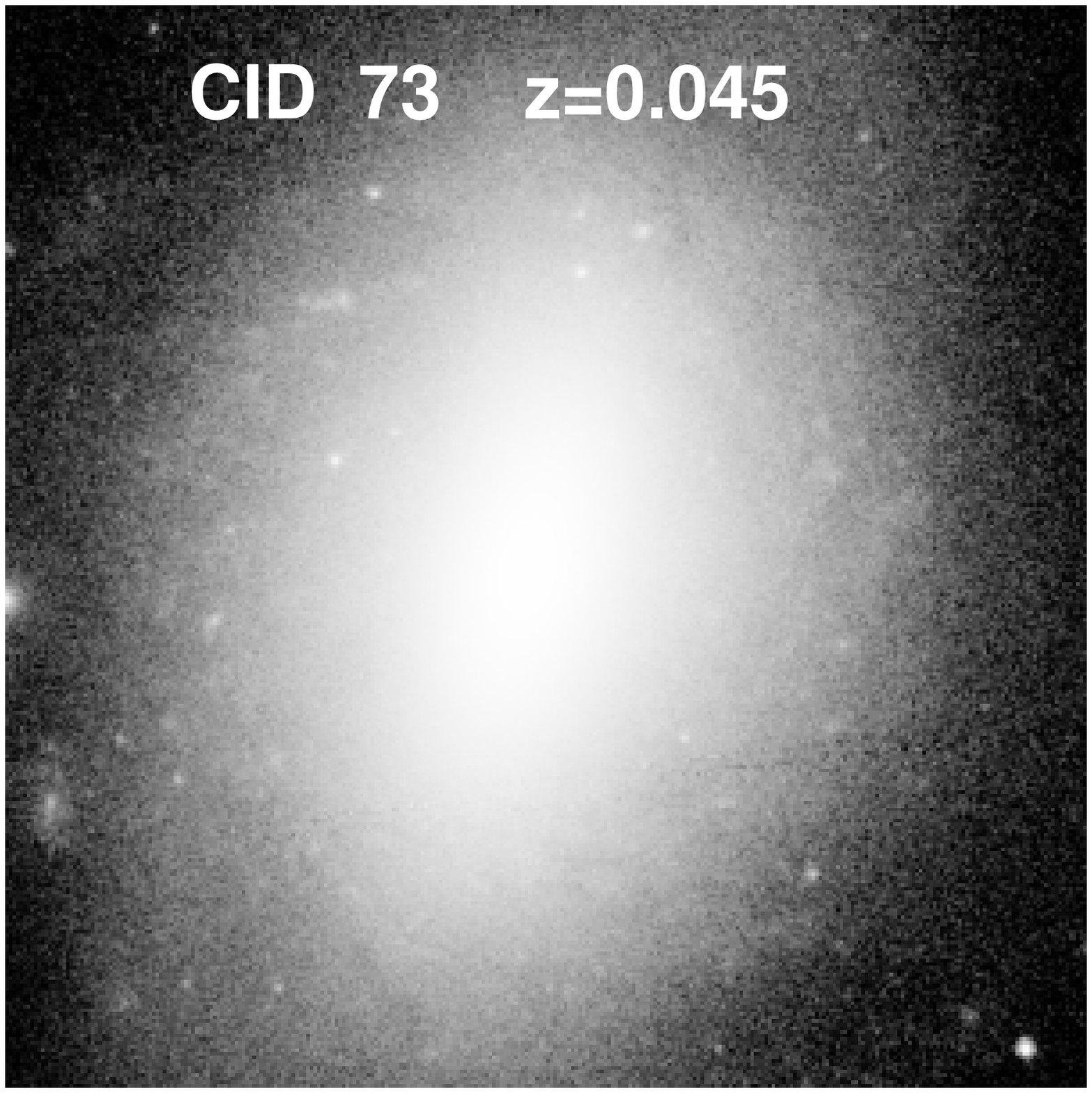}
\includegraphics[width=0.25\textwidth]{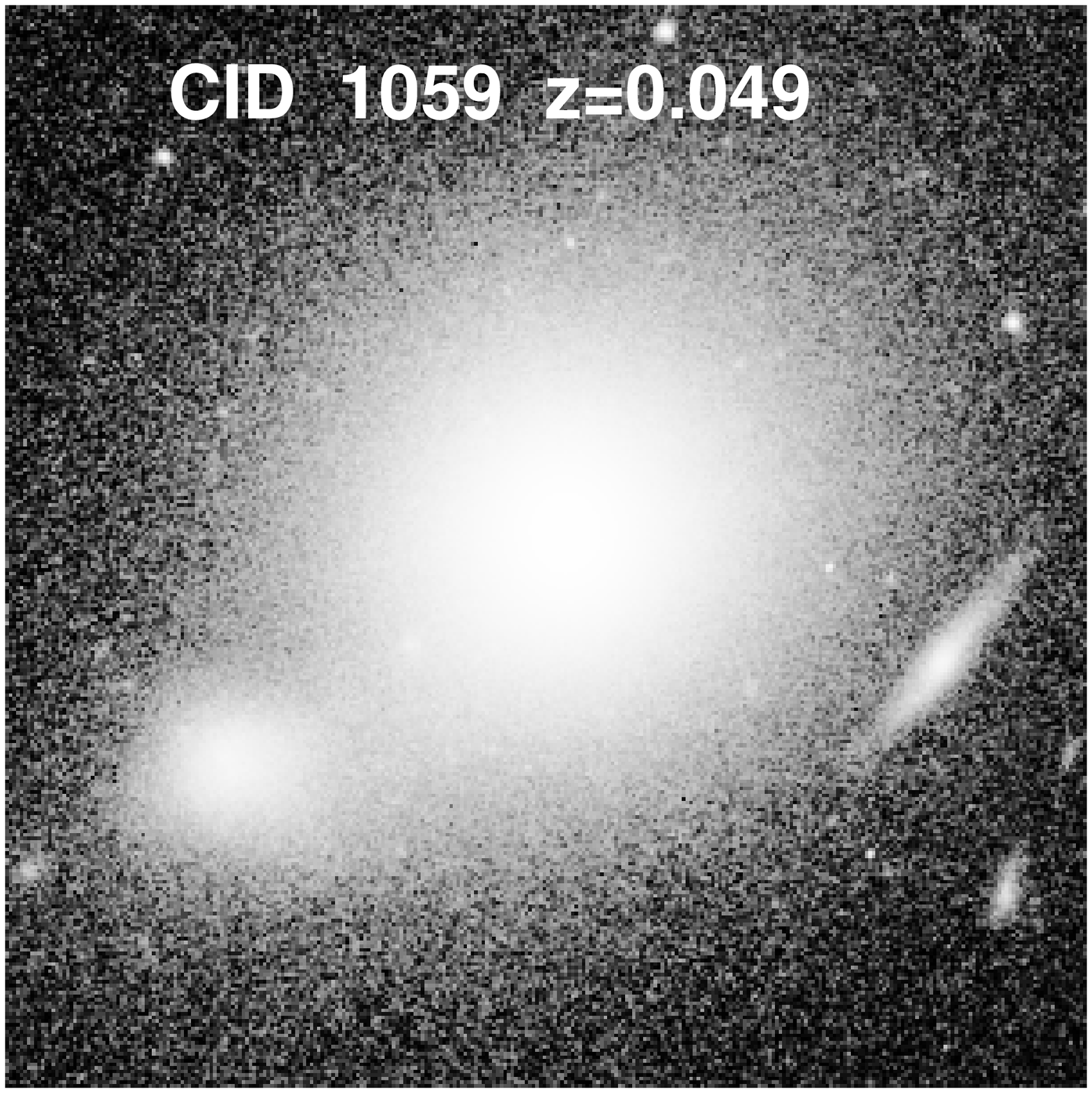}
\includegraphics[width=0.25\textwidth]{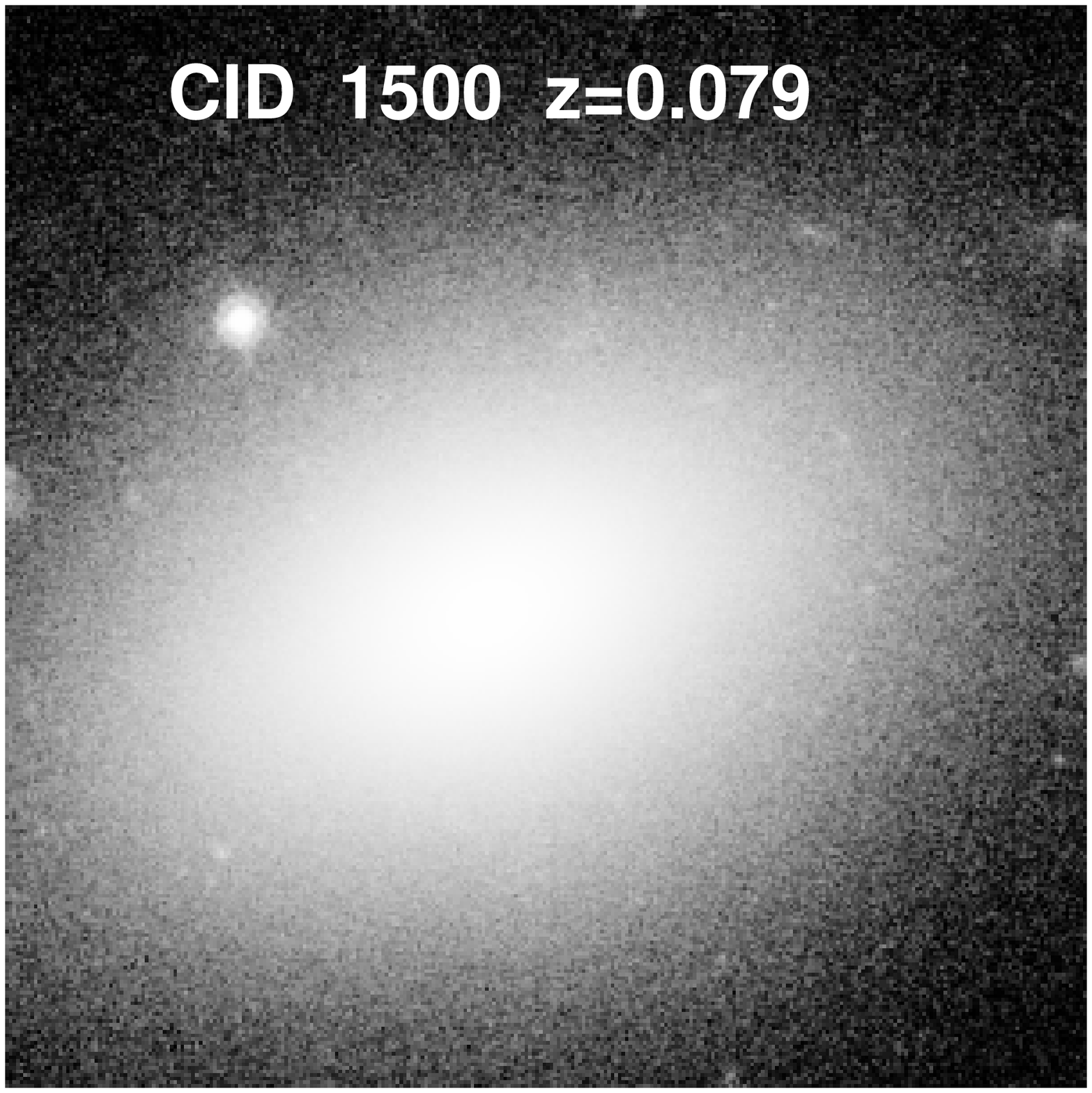}
\includegraphics[width=0.25\textwidth]{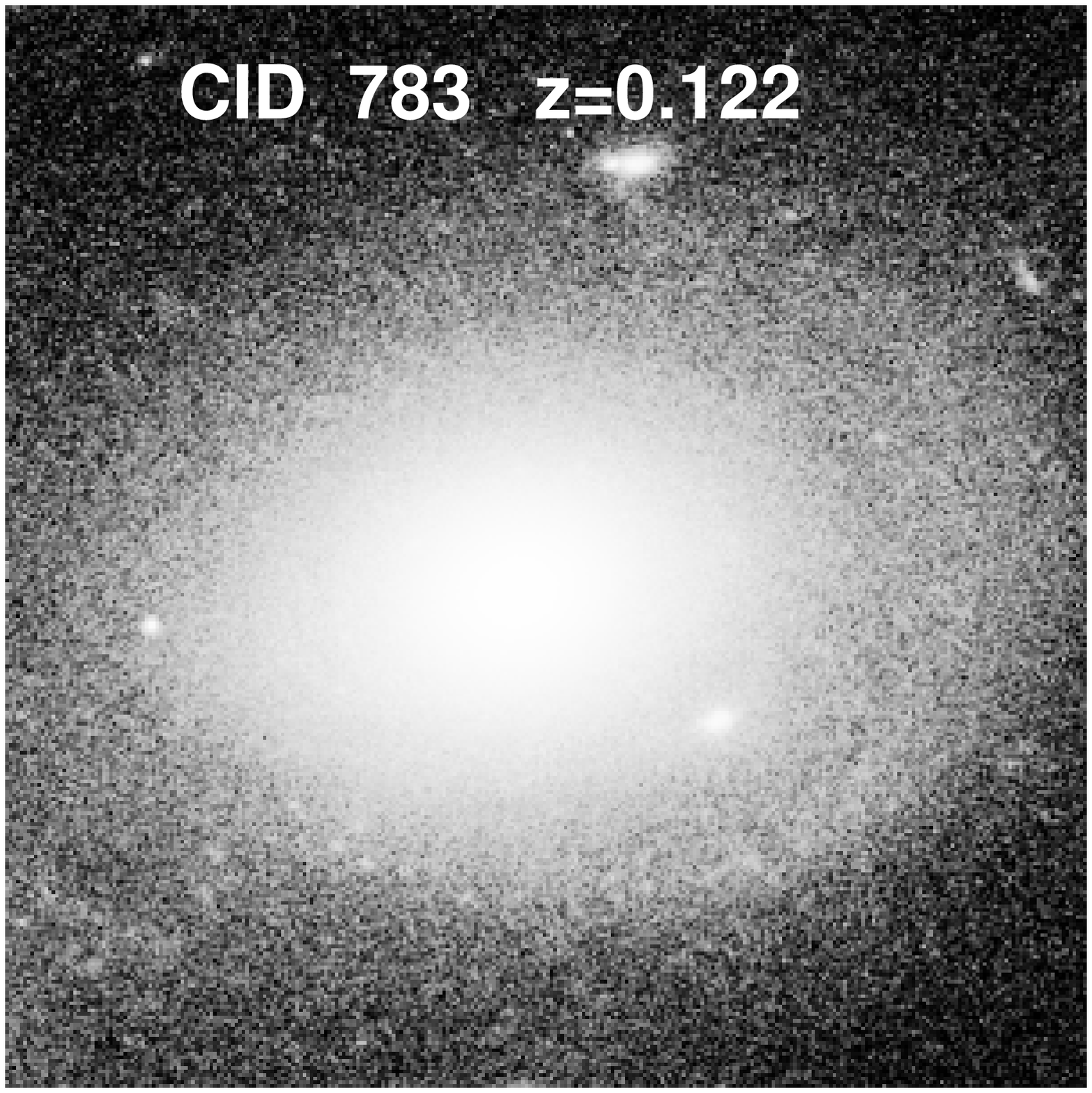}
\includegraphics[width=0.25\textwidth]{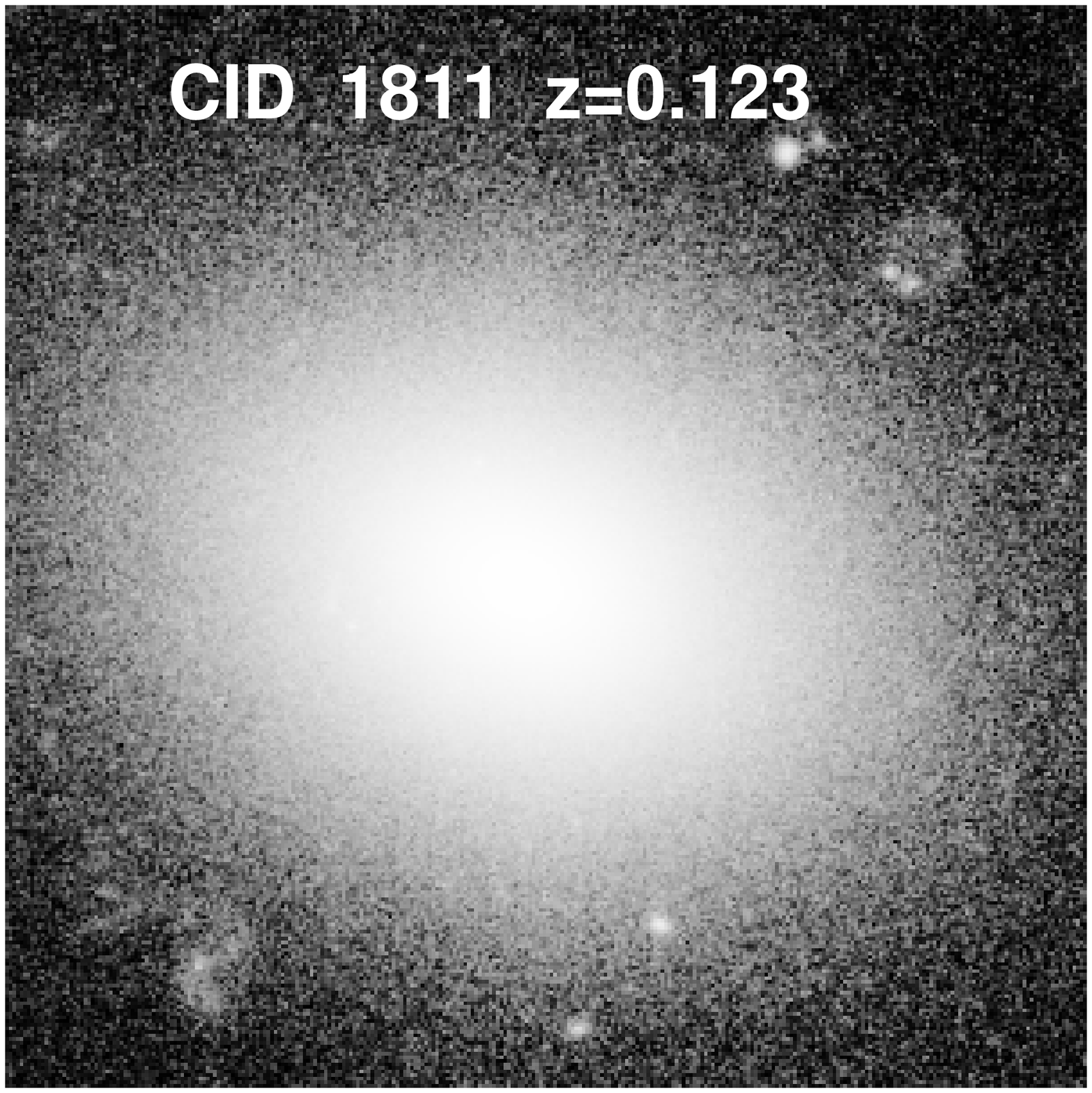}
\includegraphics[width=0.25\textwidth]{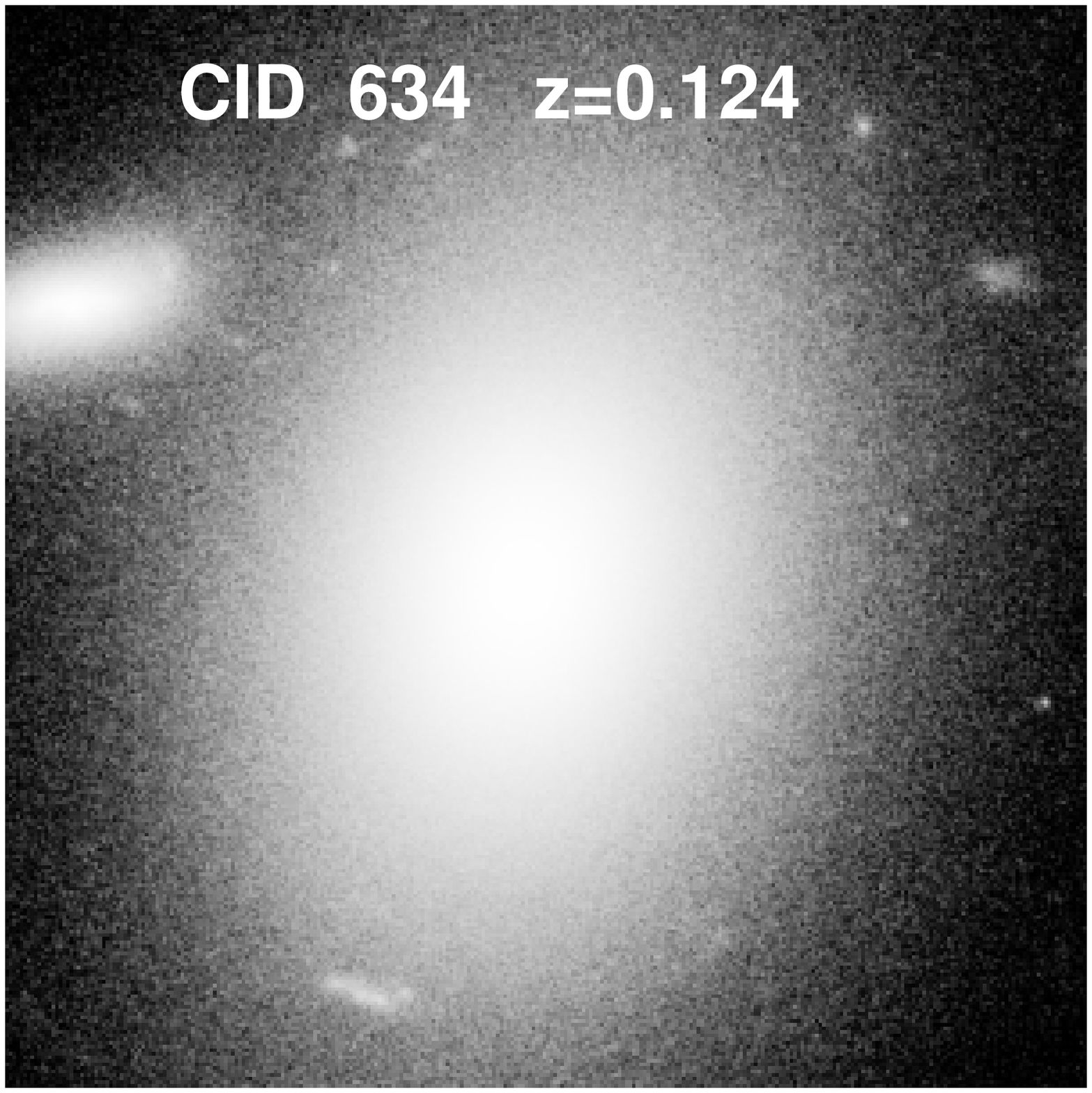}
\includegraphics[width=0.25\textwidth]{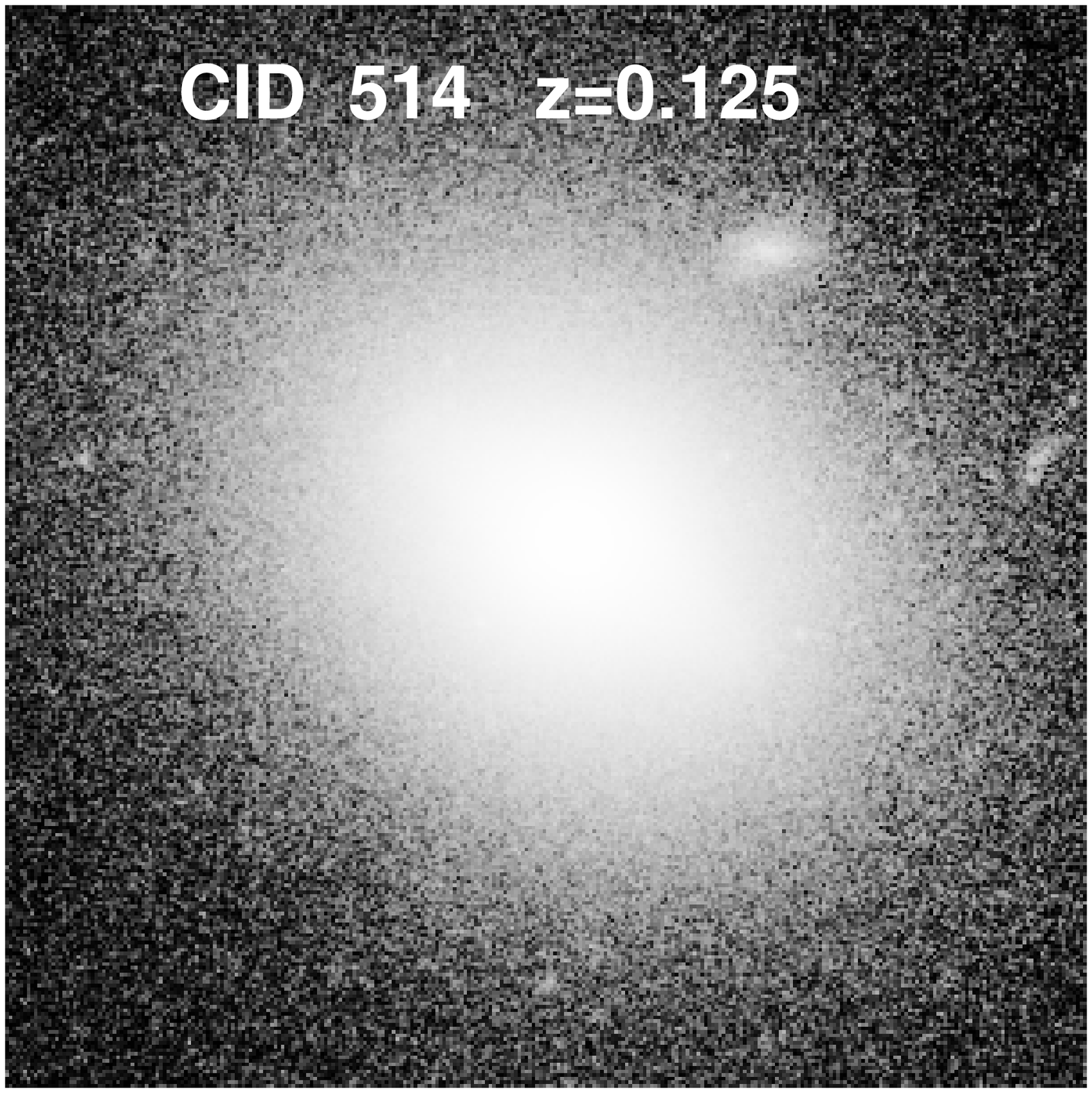}
\includegraphics[width=0.25\textwidth]{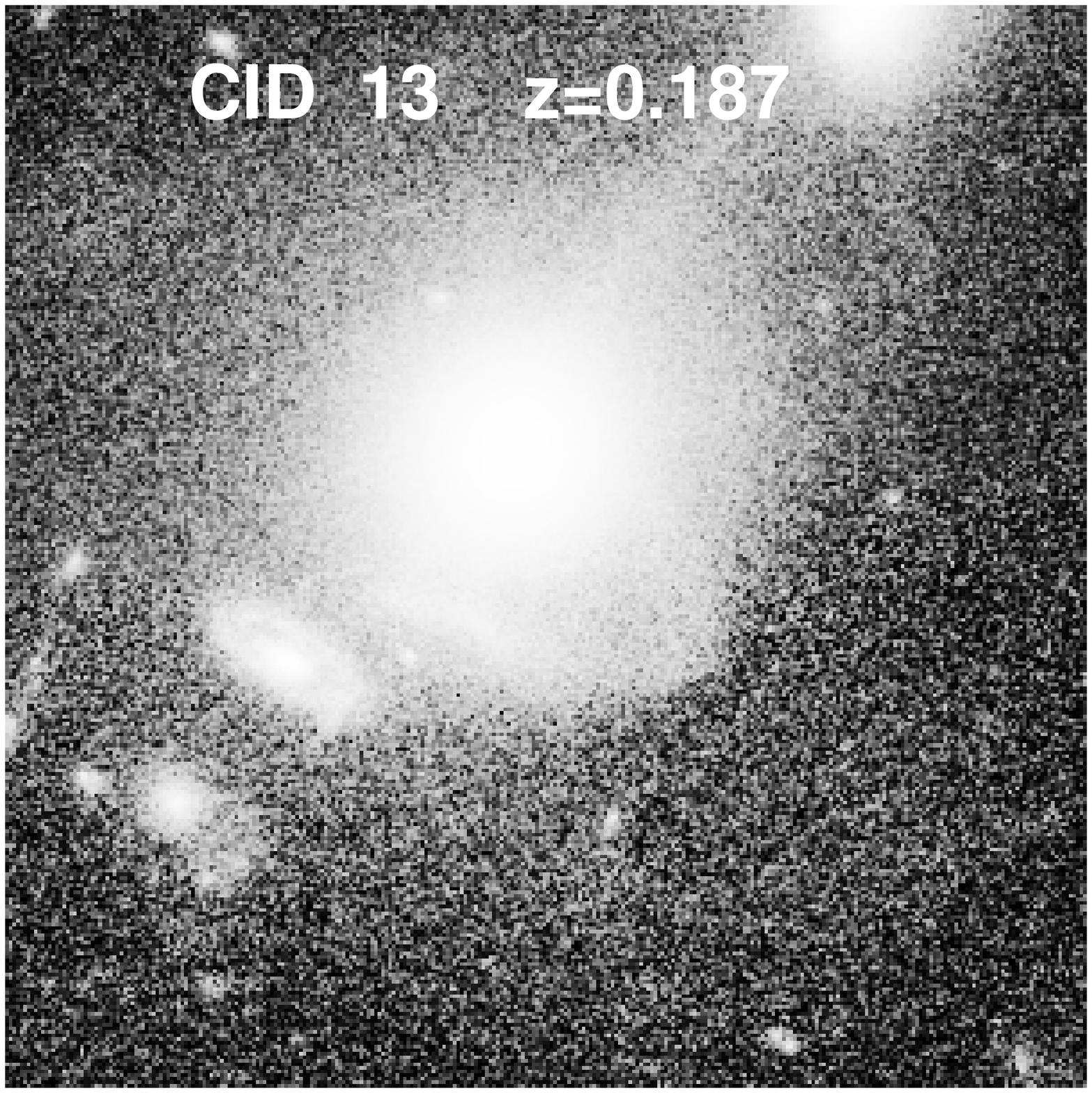}
\includegraphics[width=0.25\textwidth]{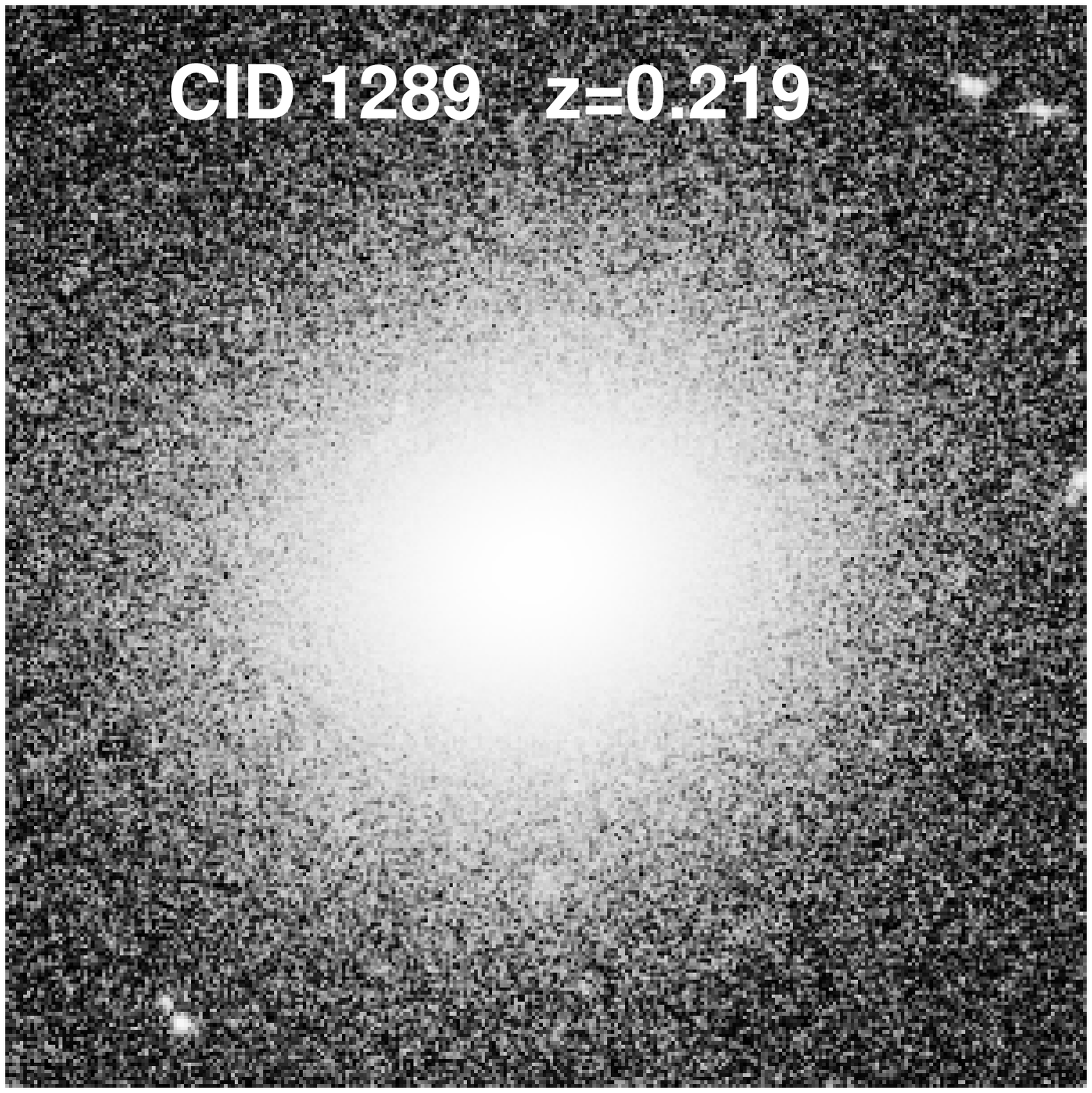}
\includegraphics[width=0.25\textwidth]{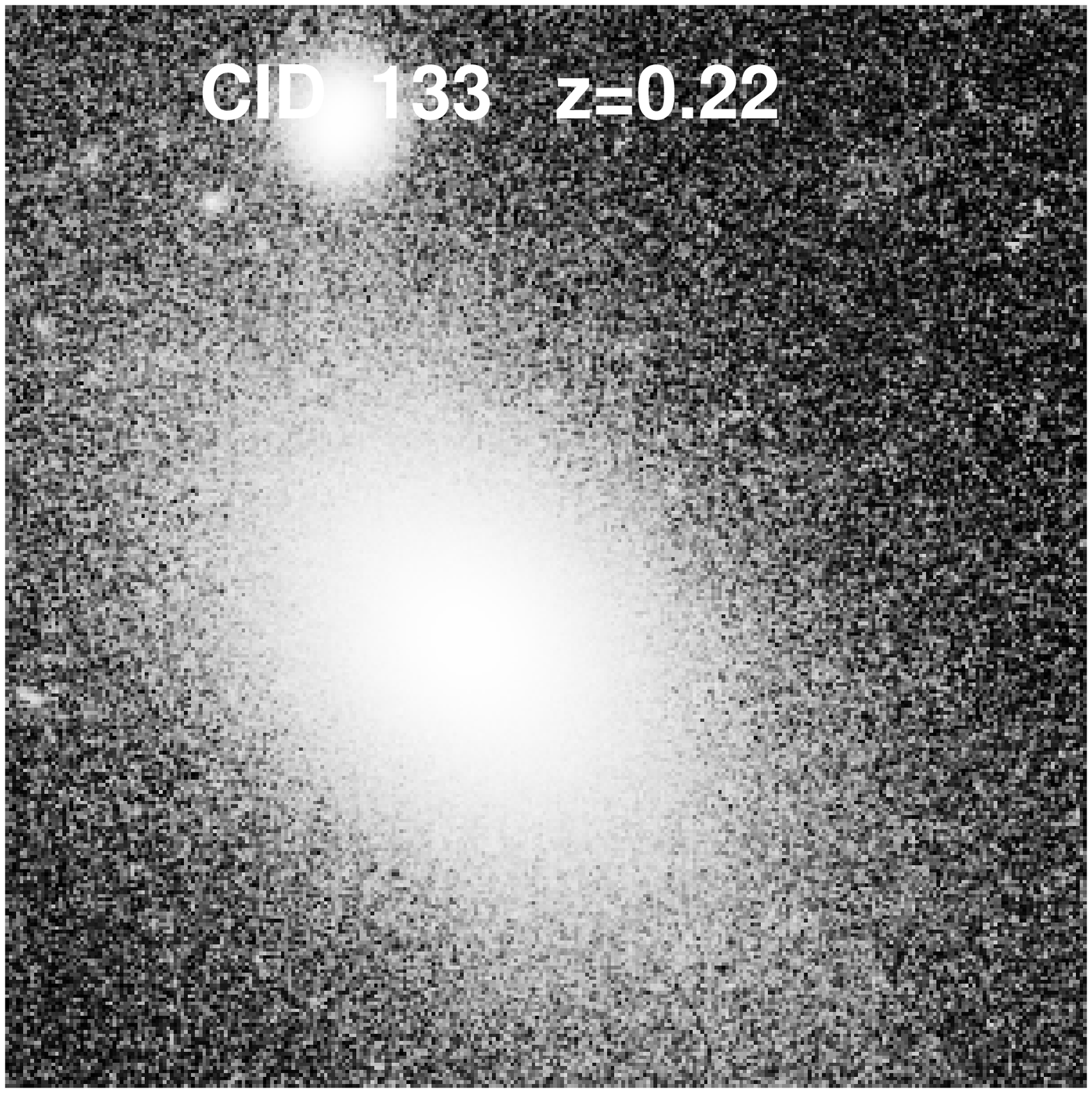}
\includegraphics[width=0.25\textwidth]{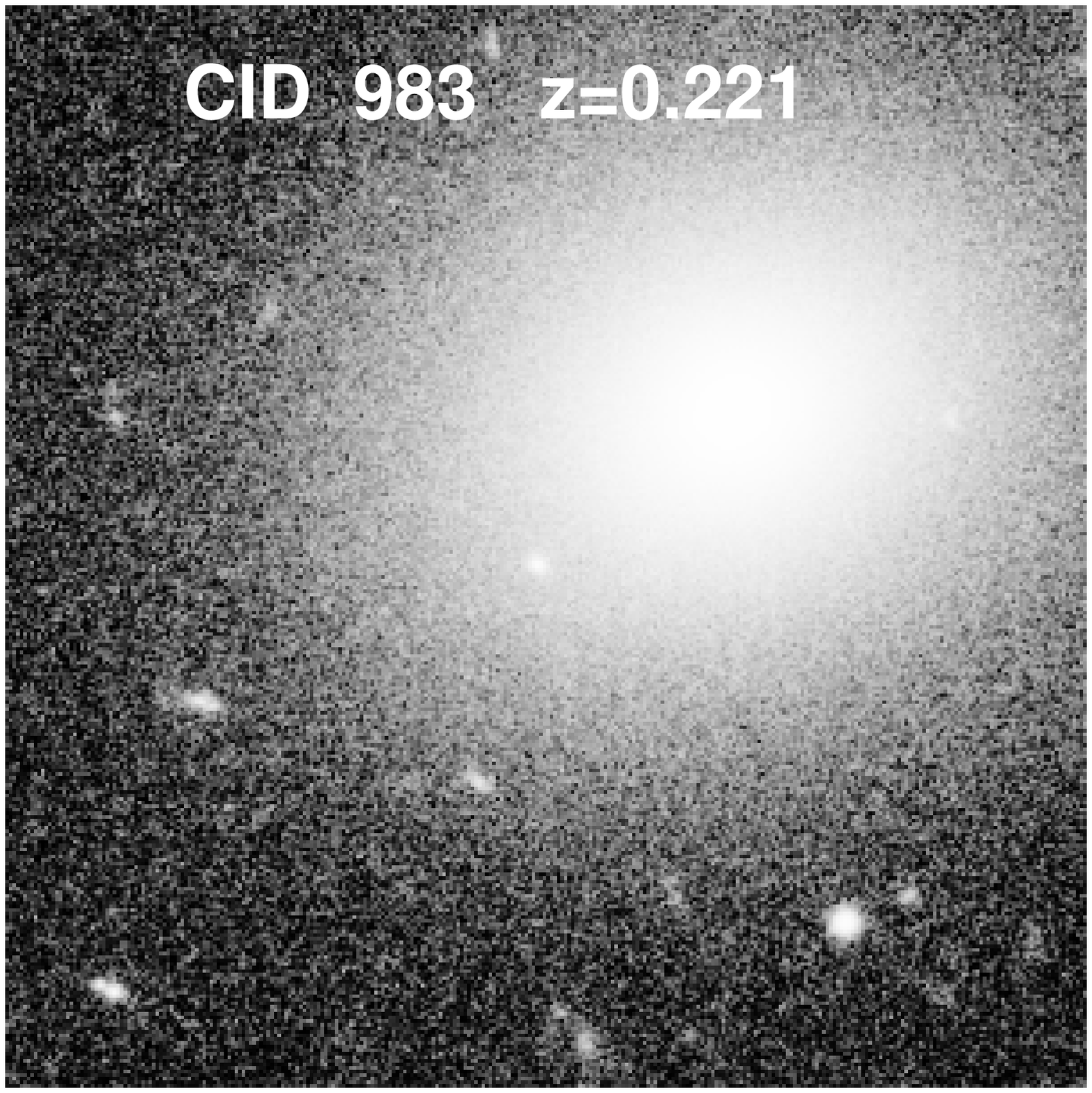}
\includegraphics[width=0.25\textwidth]{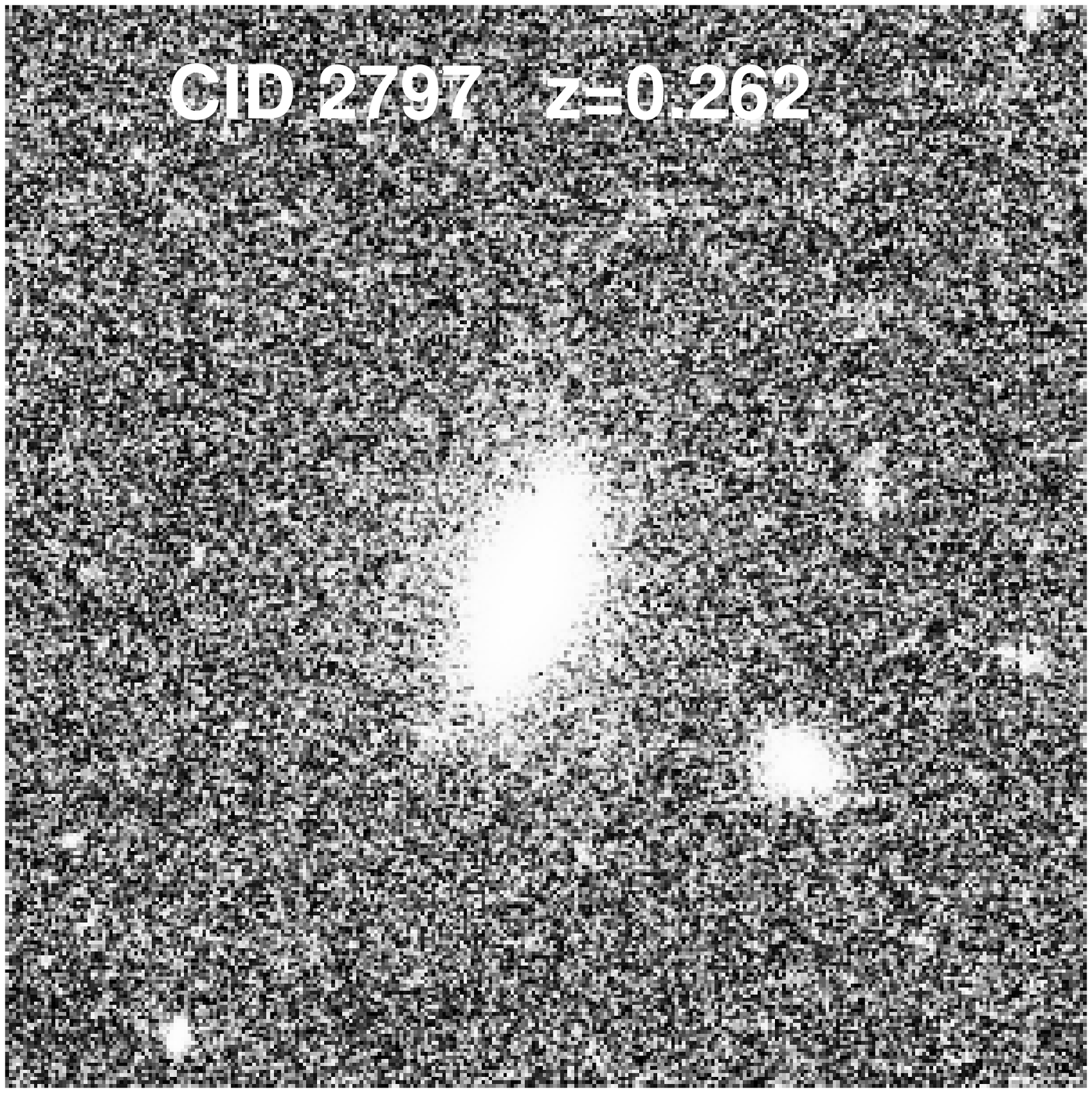}
\includegraphics[width=0.25\textwidth]{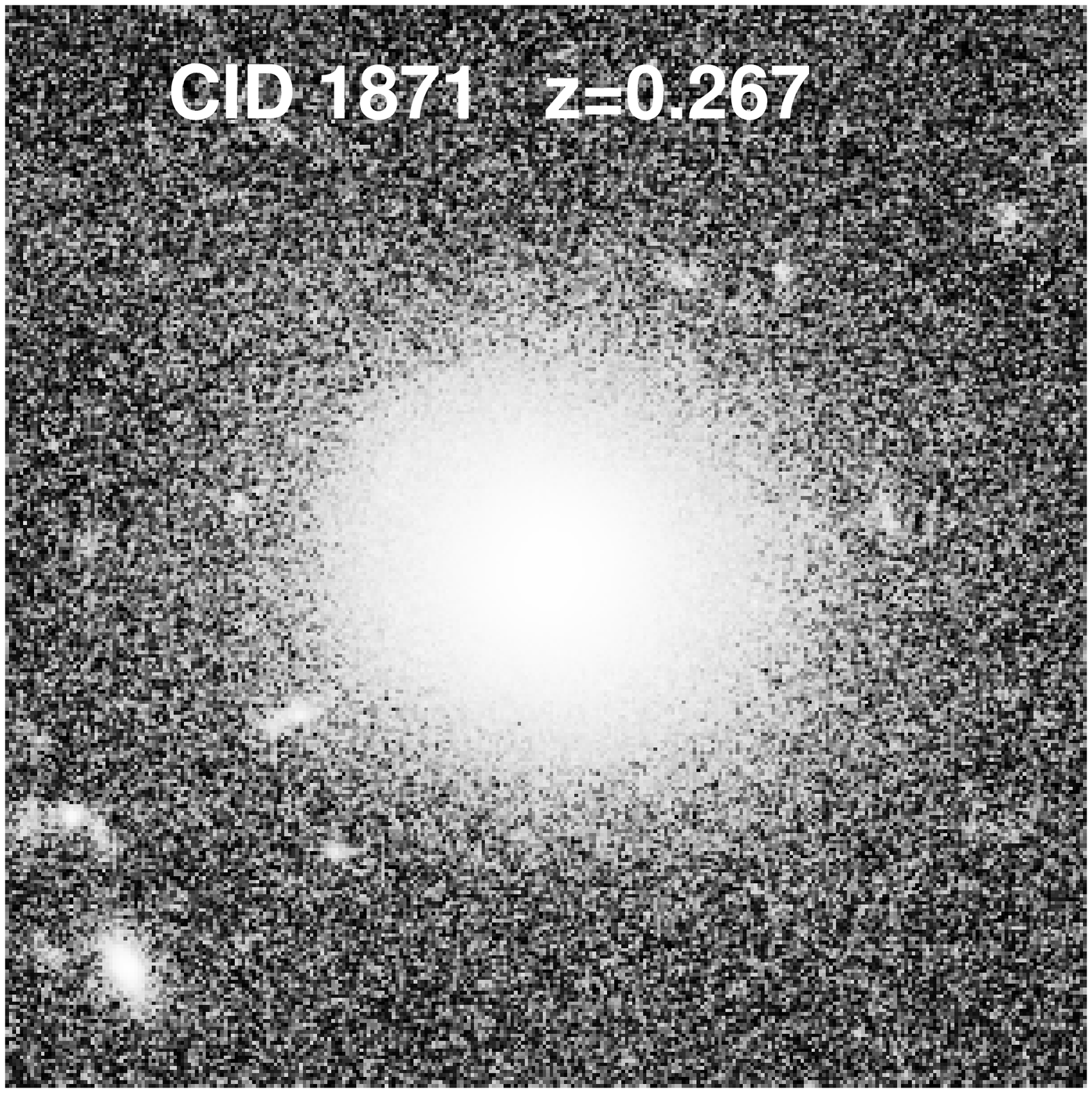}
\includegraphics[width=0.25\textwidth]{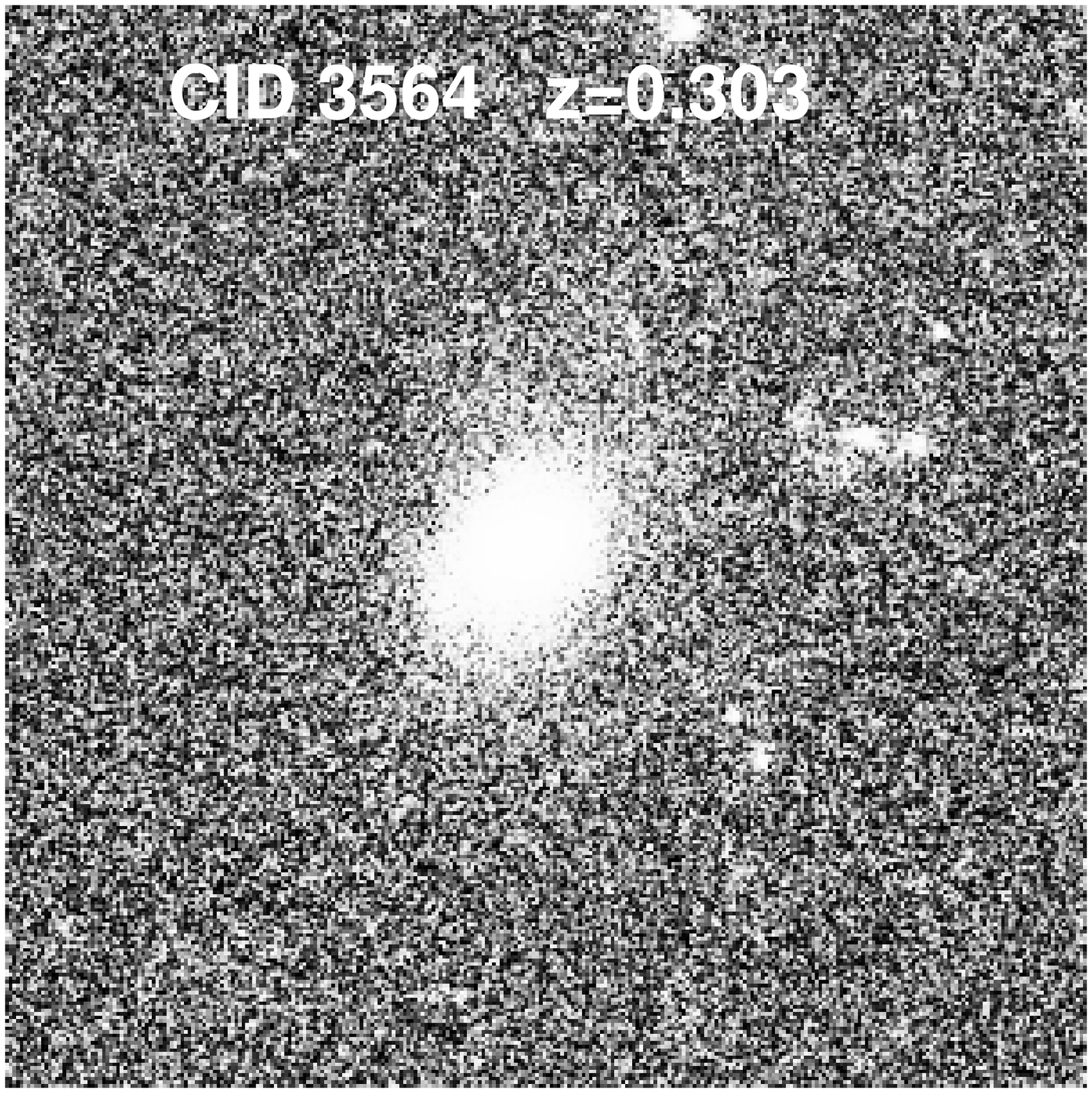}
\includegraphics[width=0.25\textwidth]{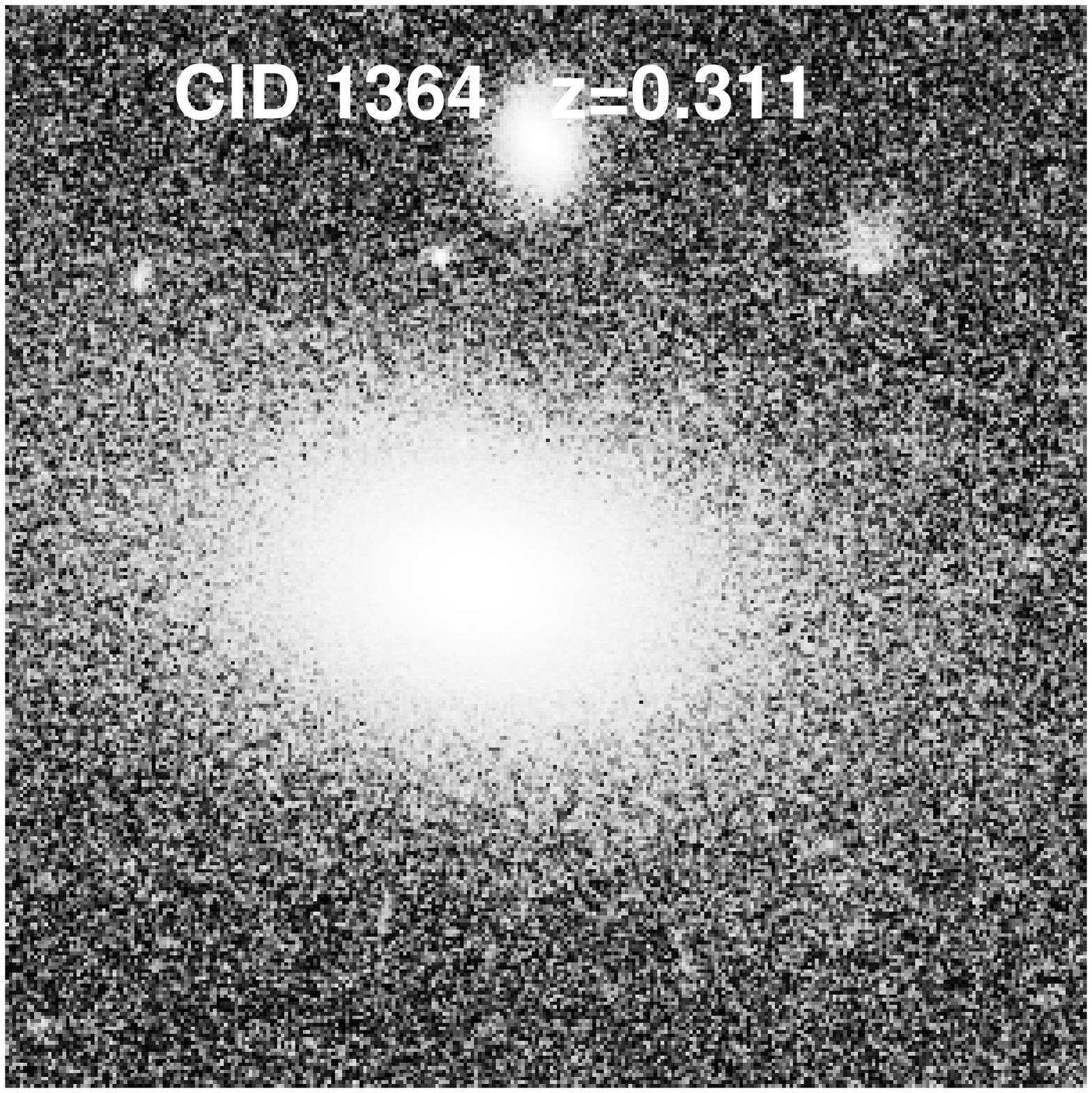}
\caption{\small HST ACS images (15$^{\prime\prime} \times$15$^{\prime\prime}$) of the X-ray ETGs in this paper sorted by increasing redshift.}
\label{fc1}
\end{figure}

\begin{figure}
\centering
\ContinuedFloat
\includegraphics[width=0.25\textwidth]{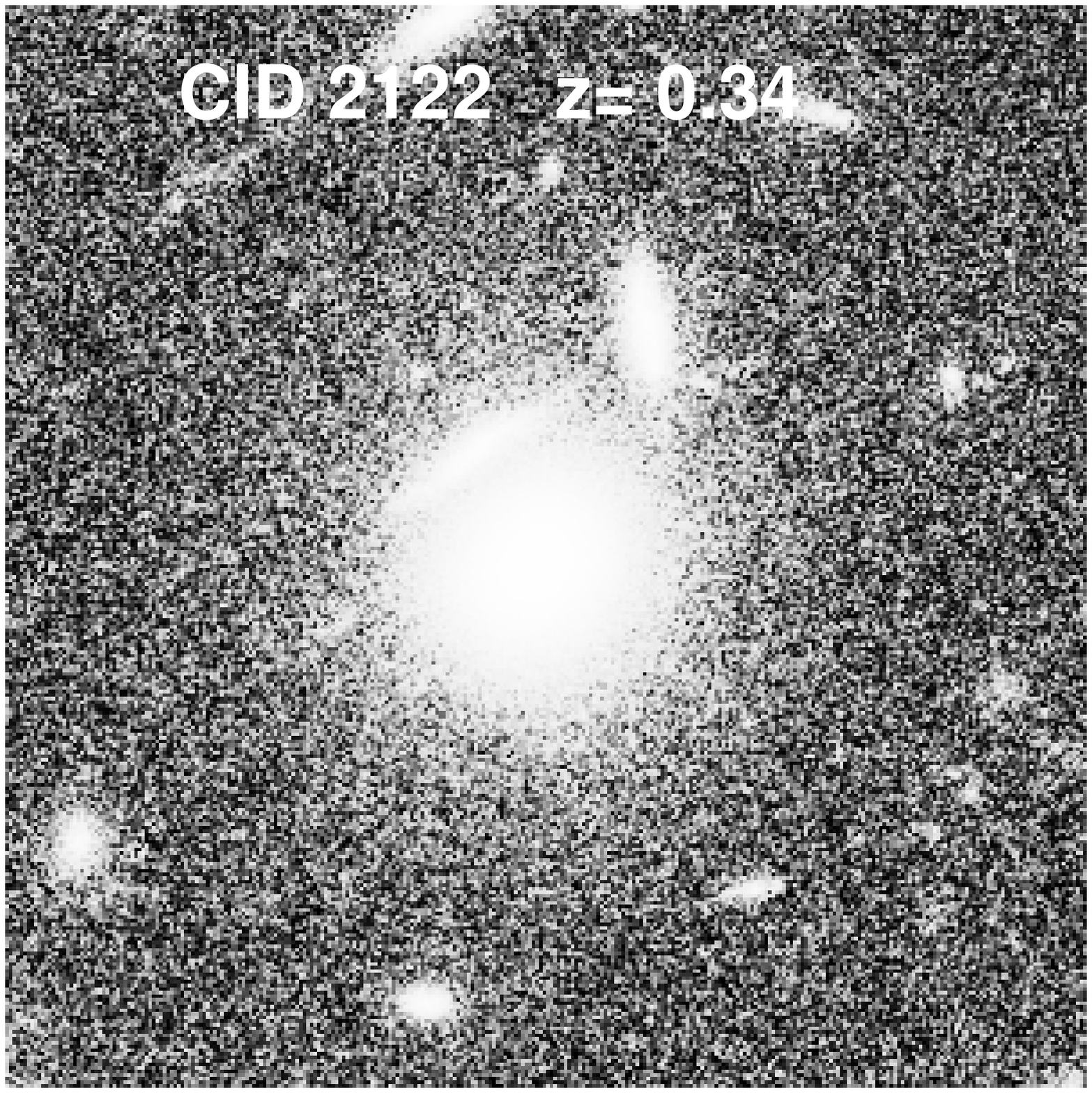}
\includegraphics[width=0.25\textwidth]{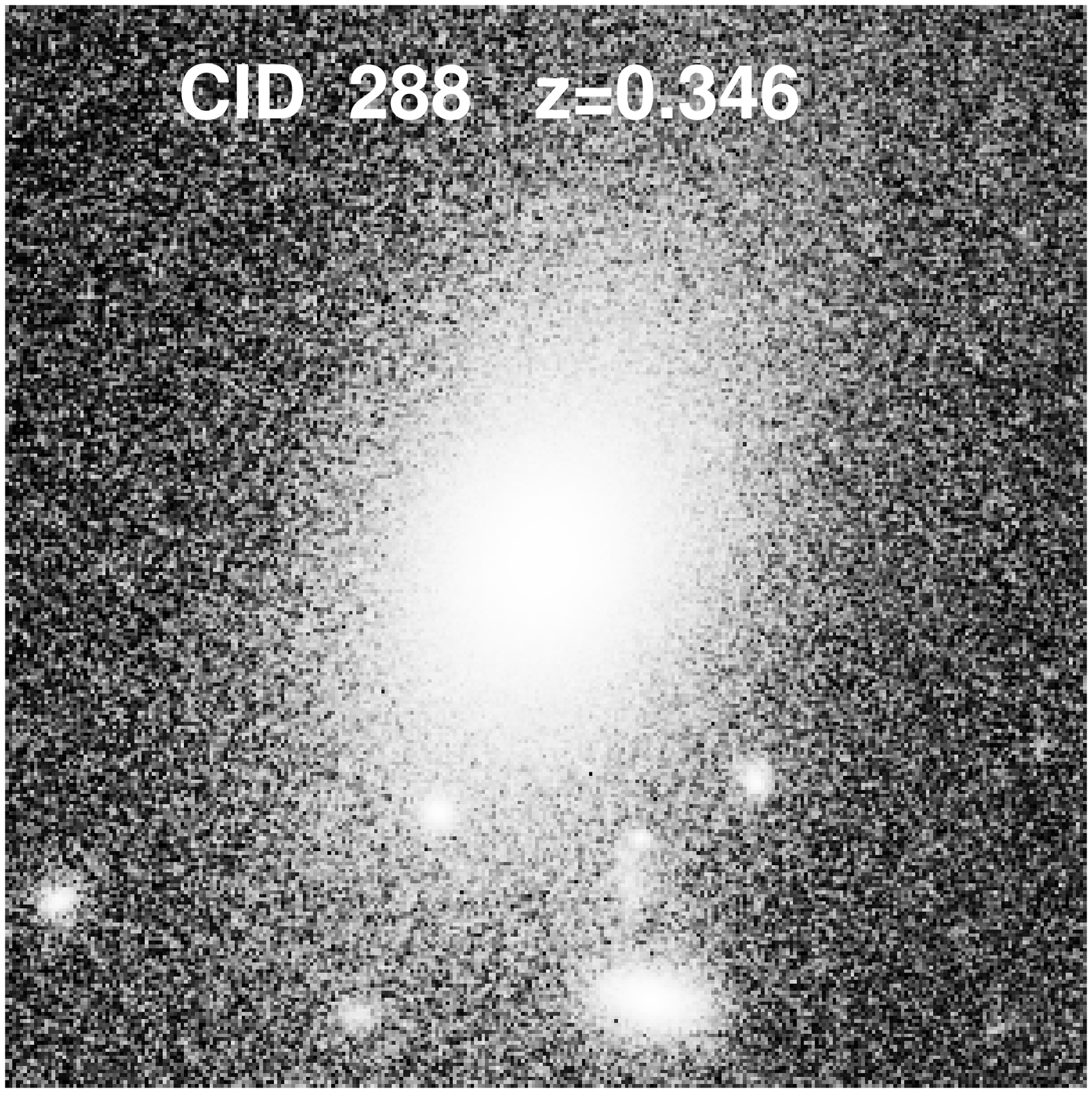}
\includegraphics[width=0.25\textwidth]{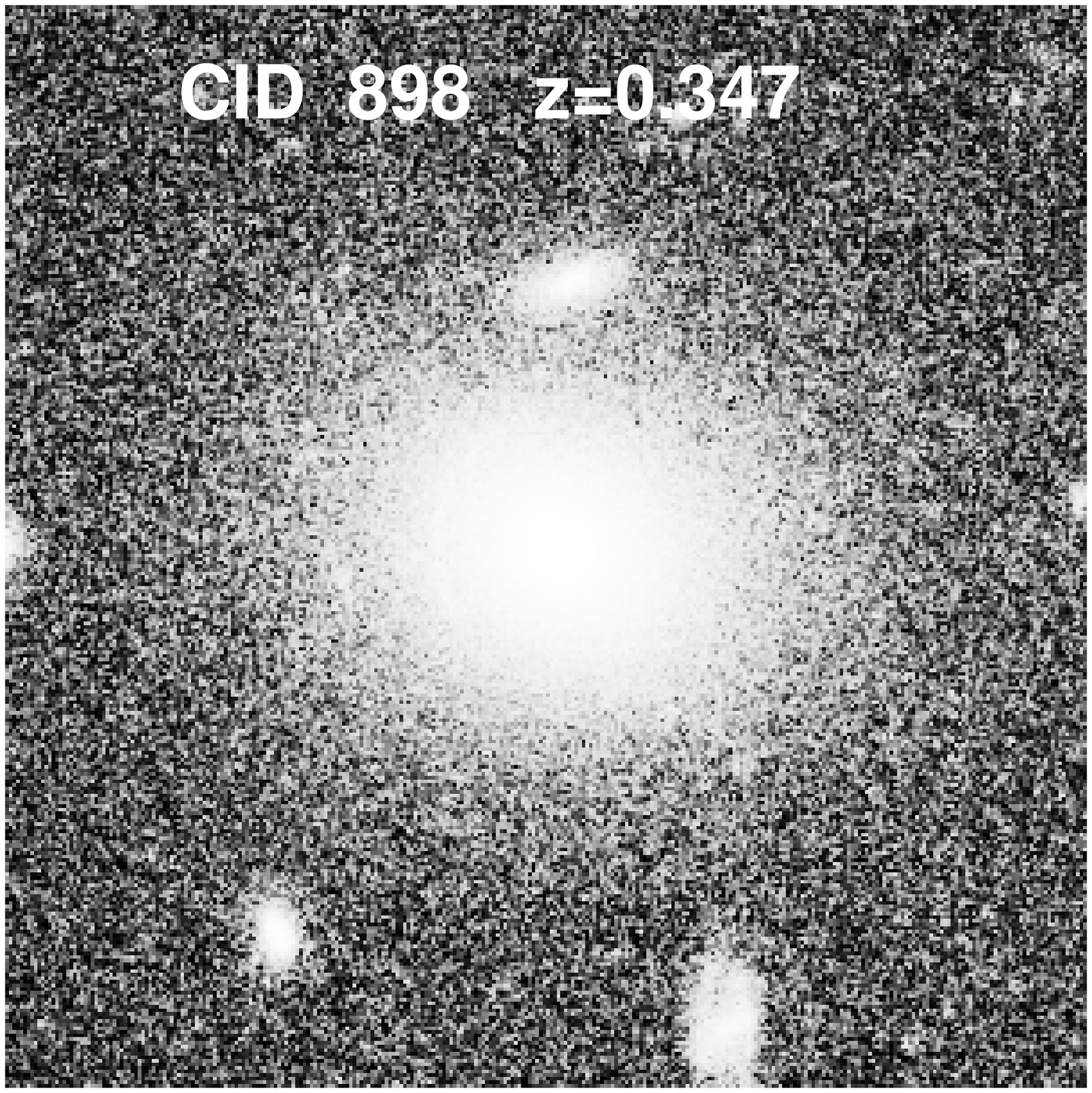}
\includegraphics[width=0.25\textwidth]{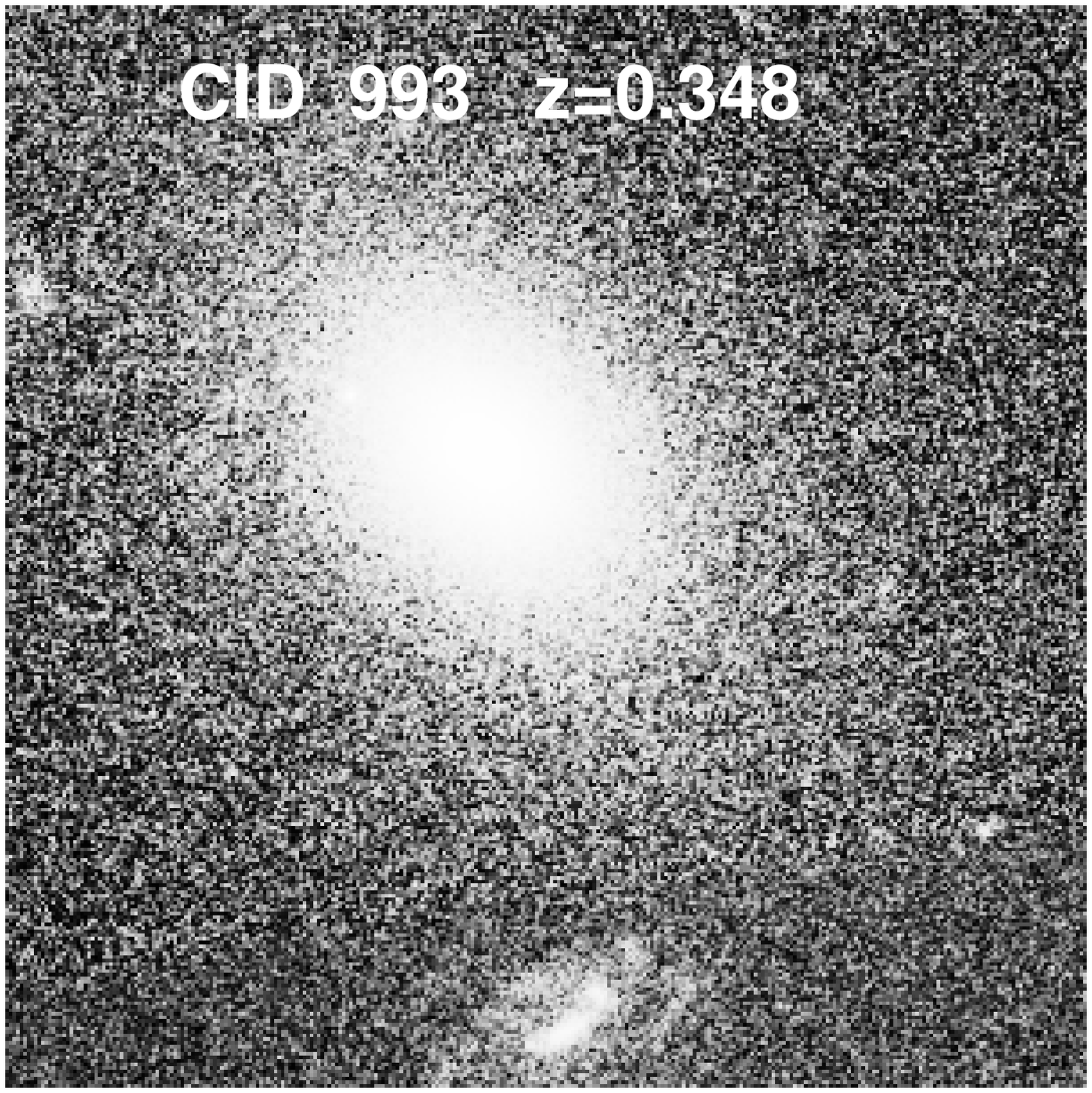}
\includegraphics[width=0.25\textwidth]{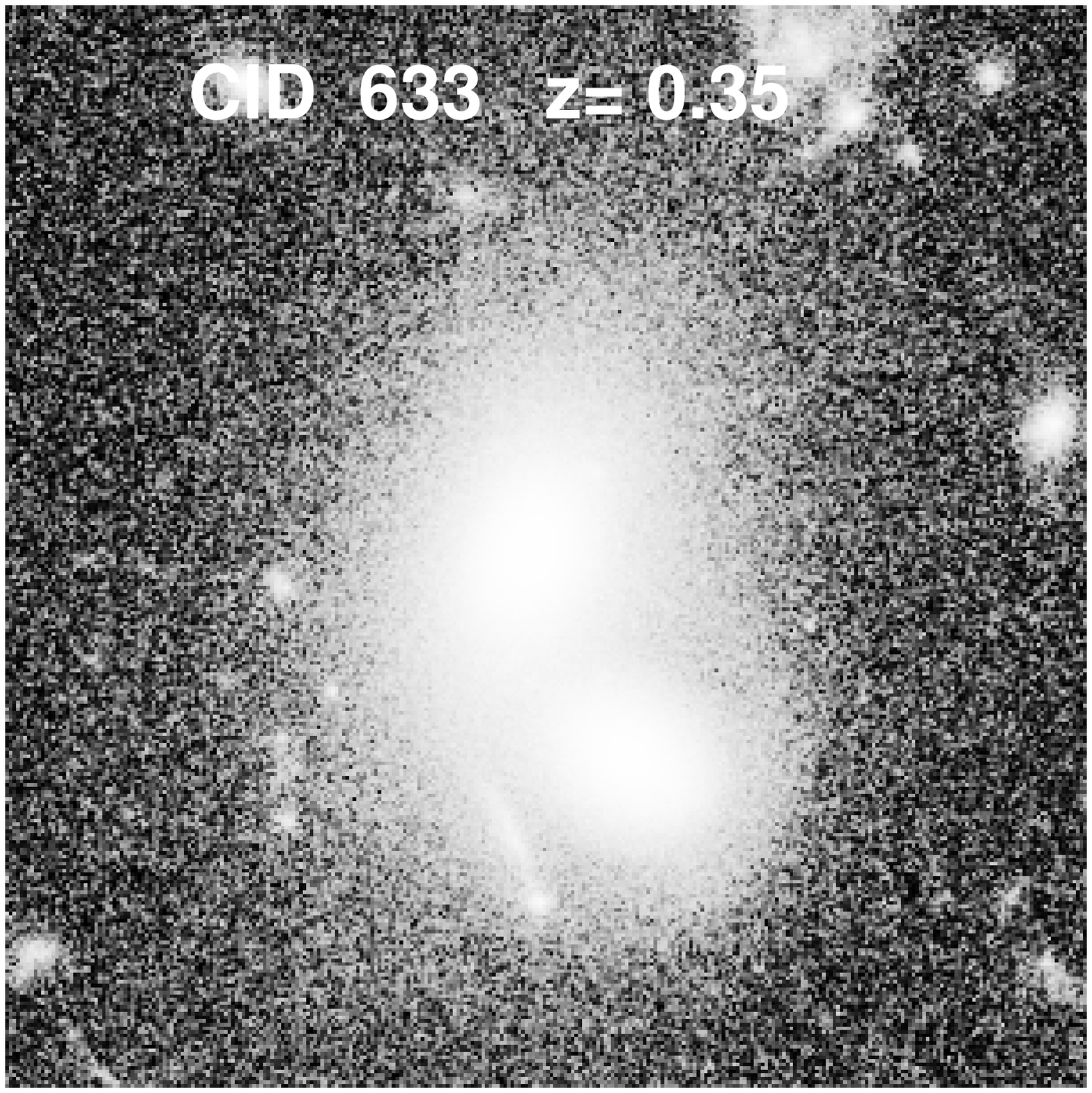}
\includegraphics[width=0.25\textwidth]{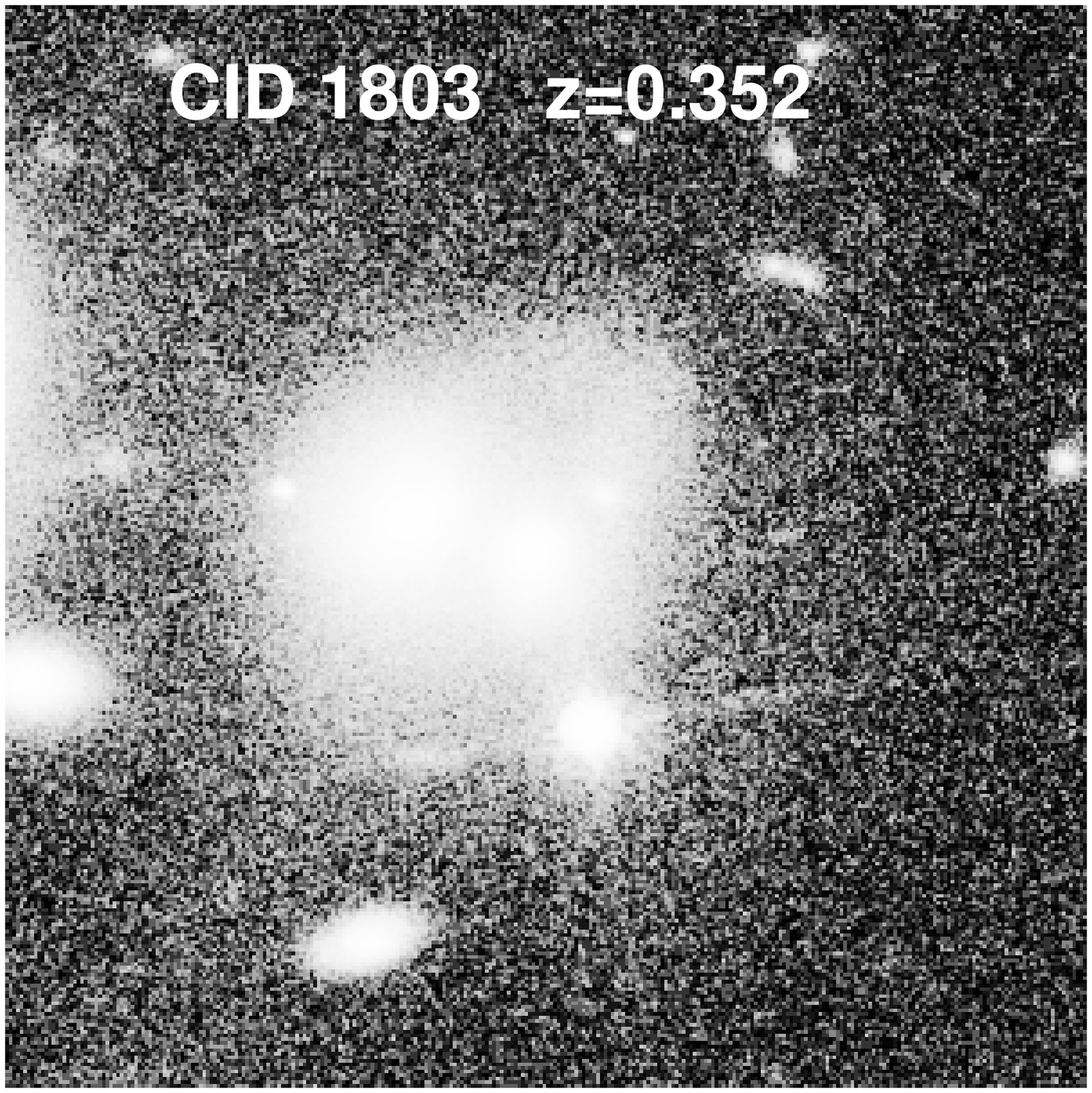}
\includegraphics[width=0.25\textwidth]{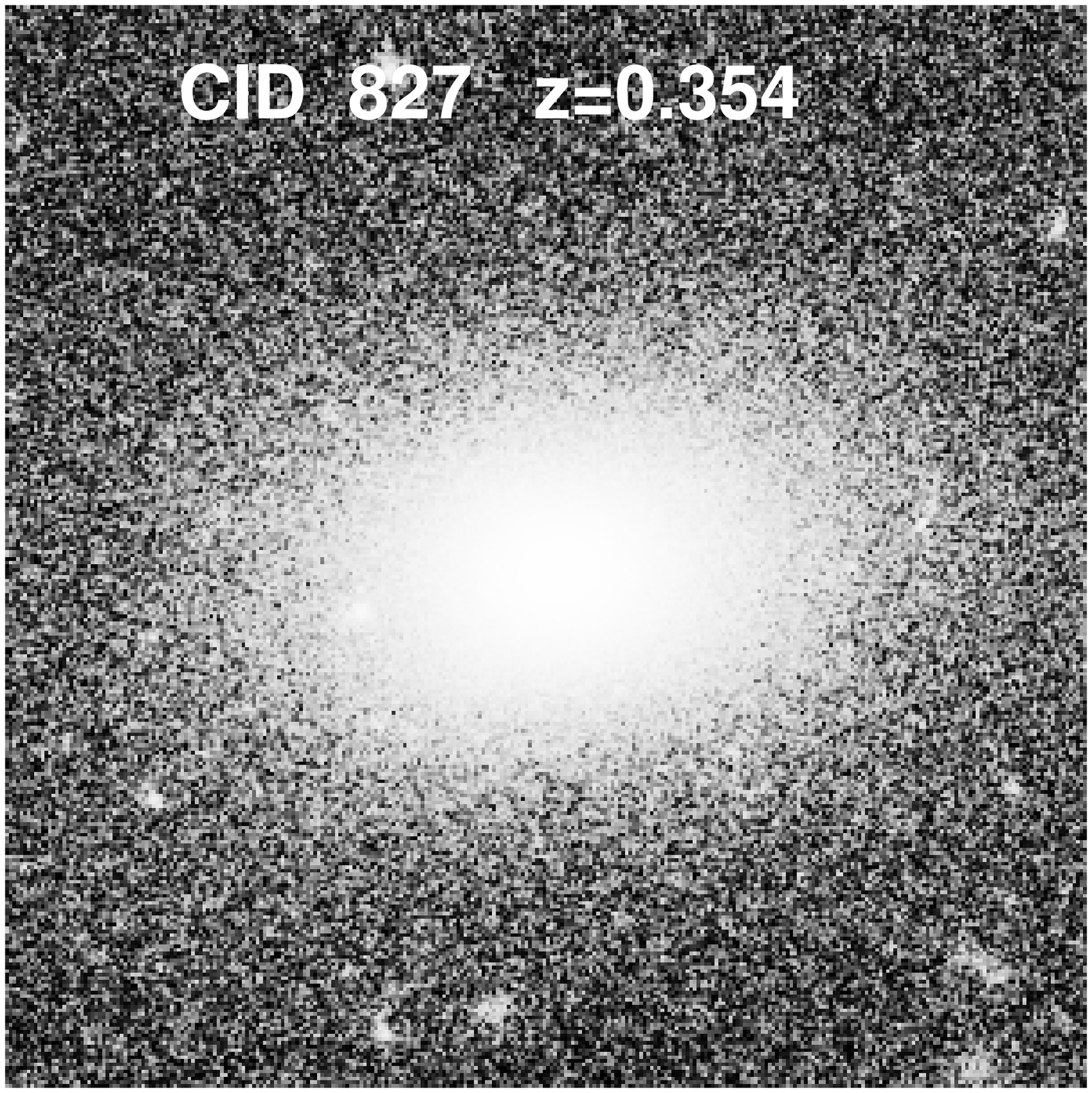}
\includegraphics[width=0.25\textwidth]{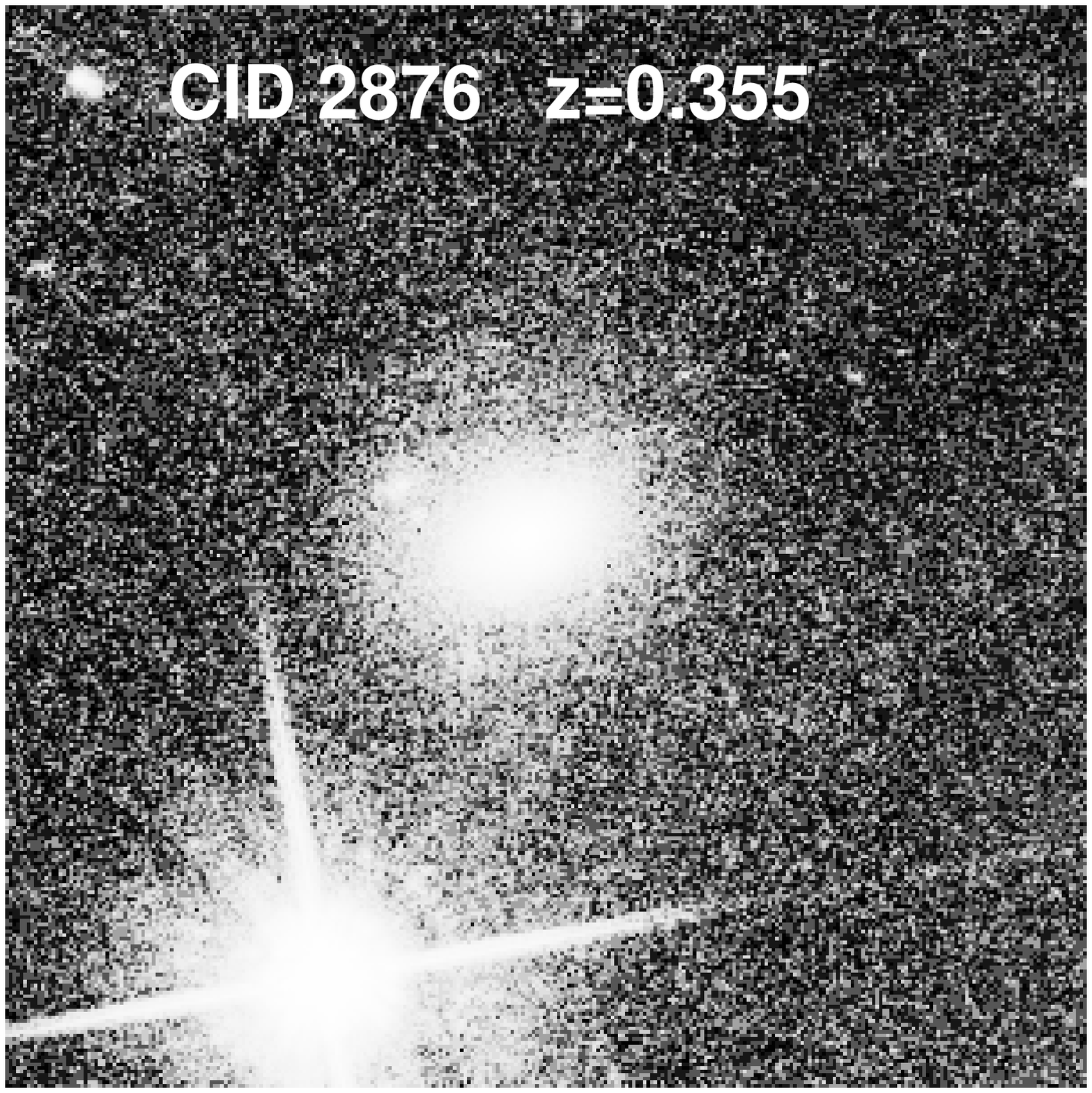}
\includegraphics[width=0.25\textwidth]{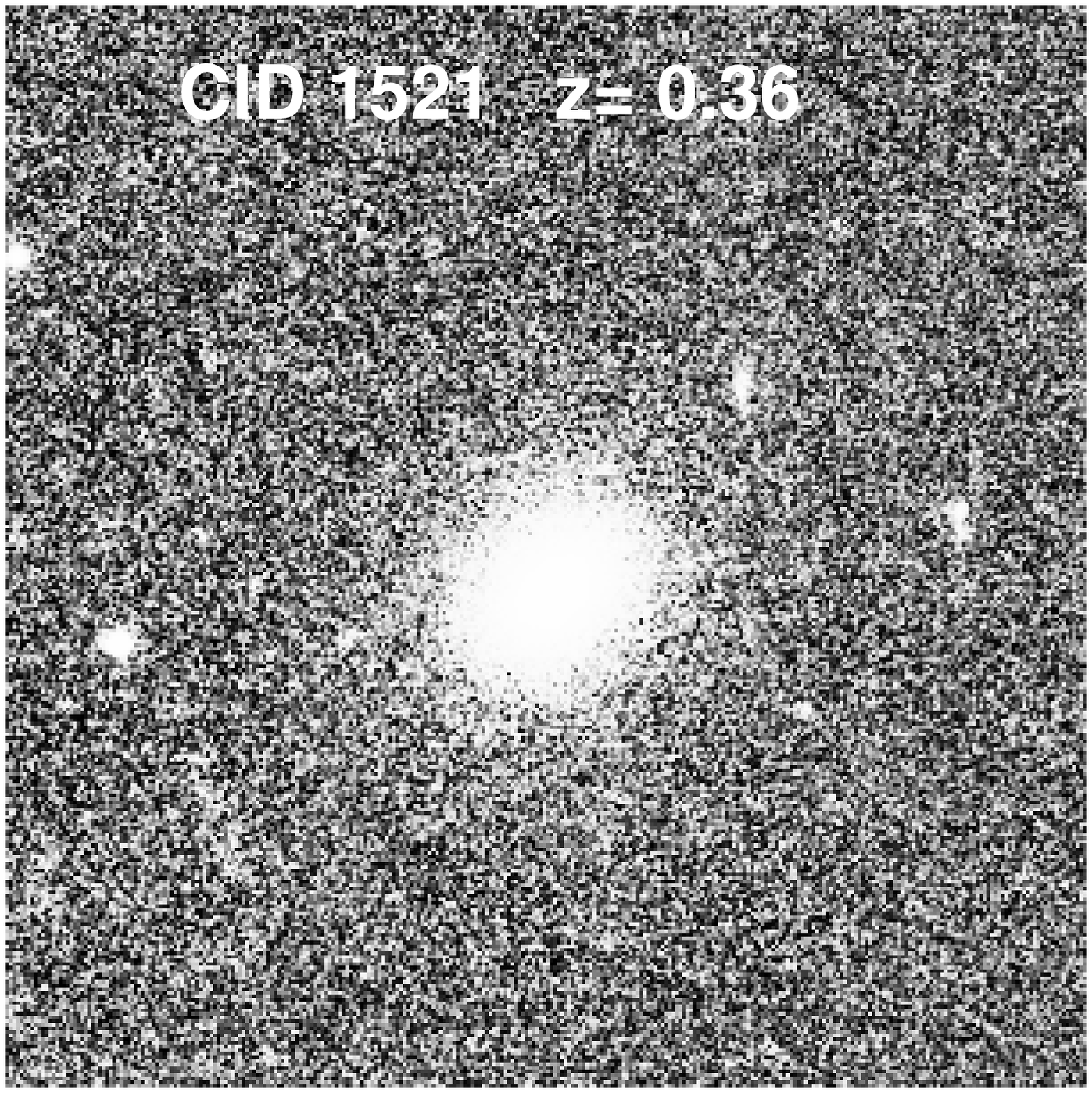}
\includegraphics[width=0.25\textwidth]{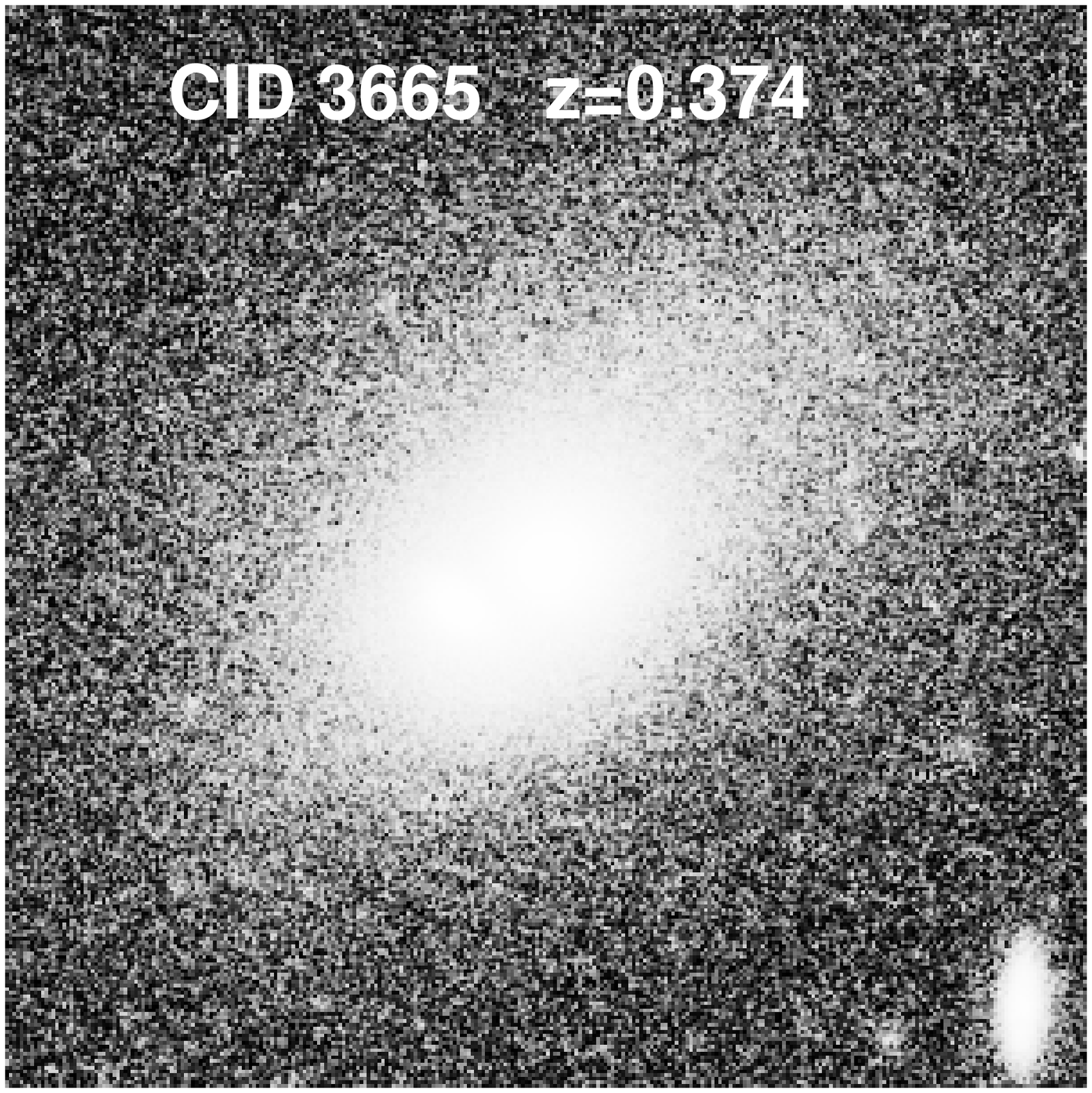}
\includegraphics[width=0.25\textwidth]{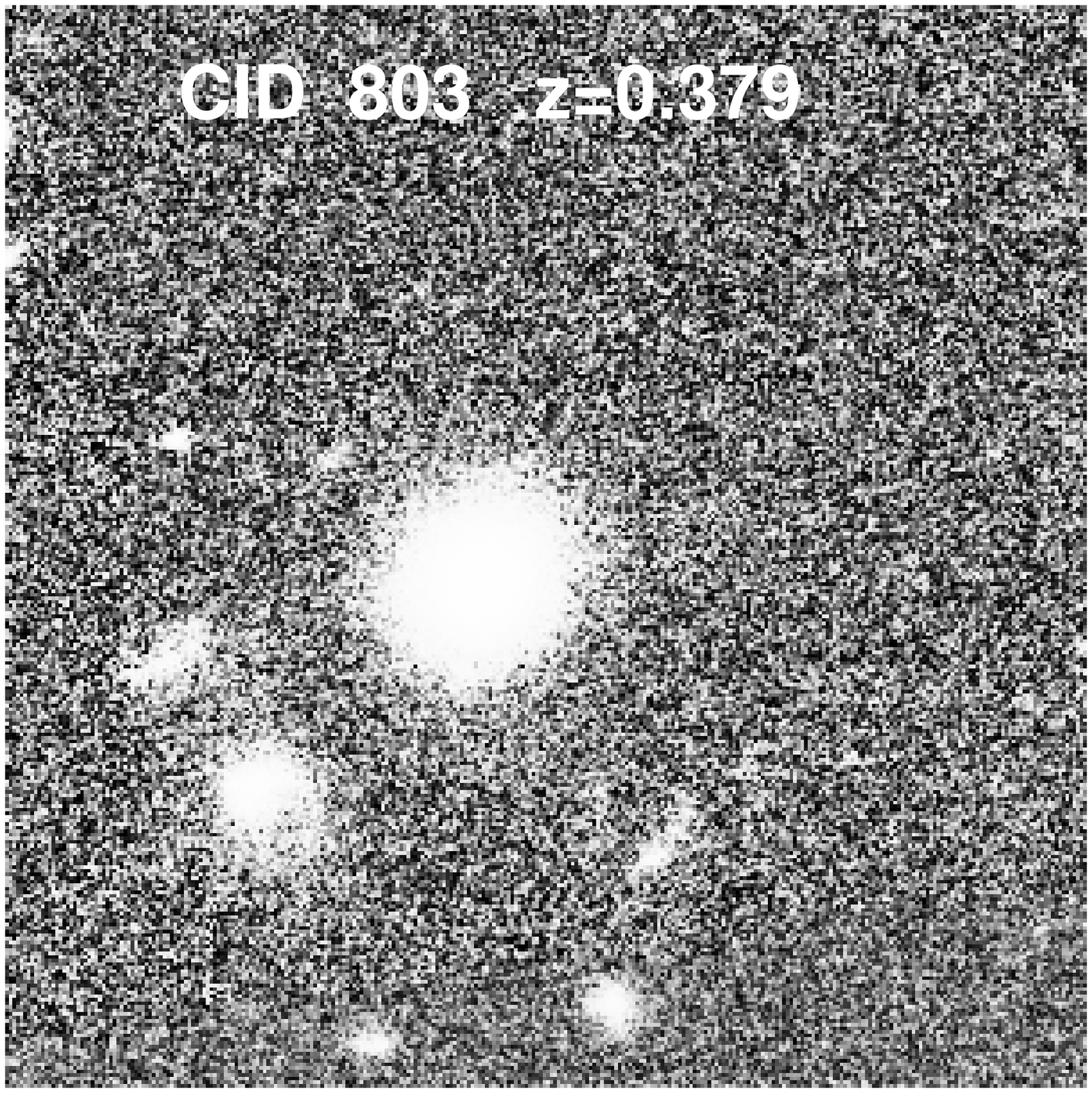}
\includegraphics[width=0.25\textwidth]{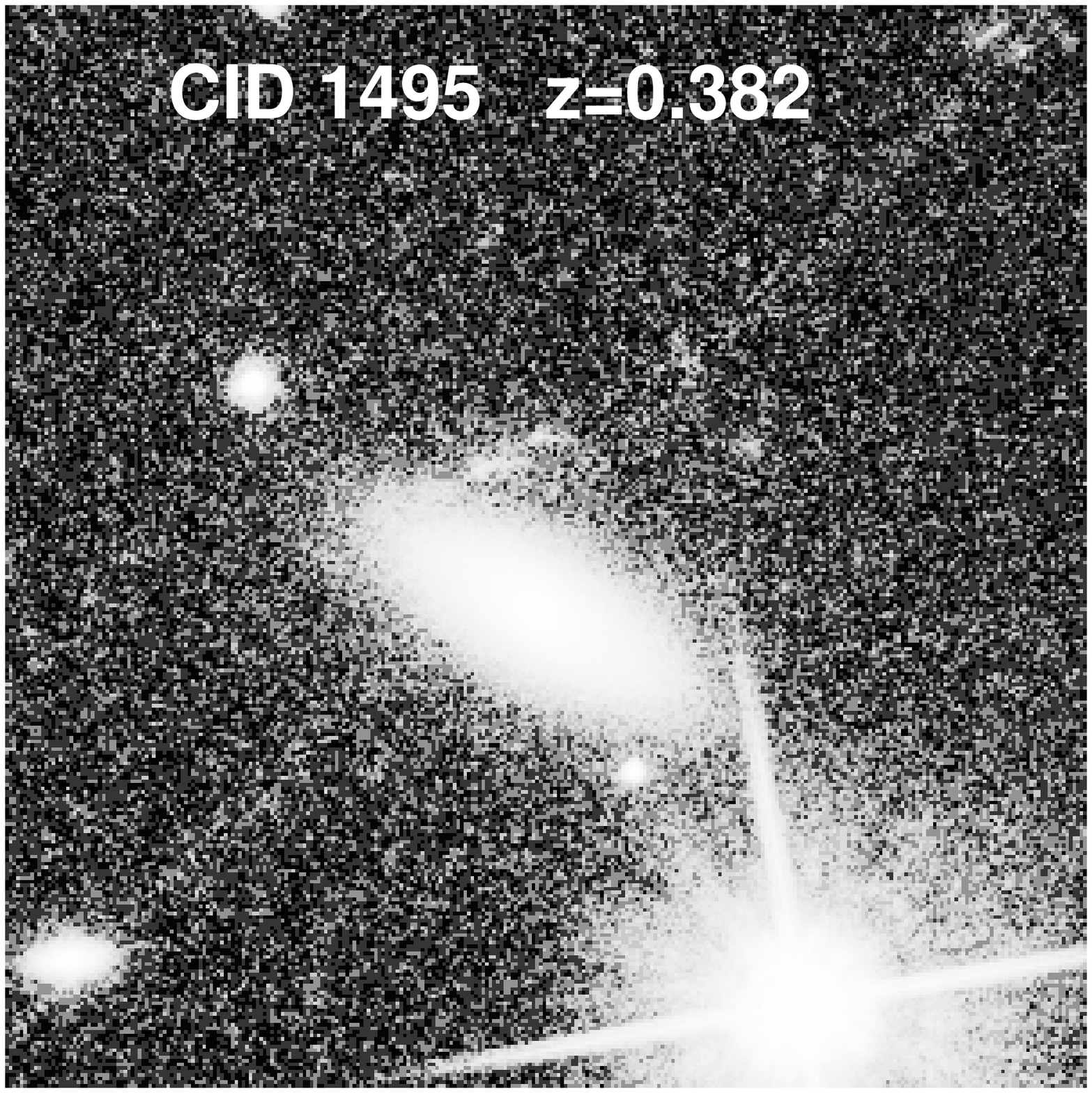}
\includegraphics[width=0.25\textwidth]{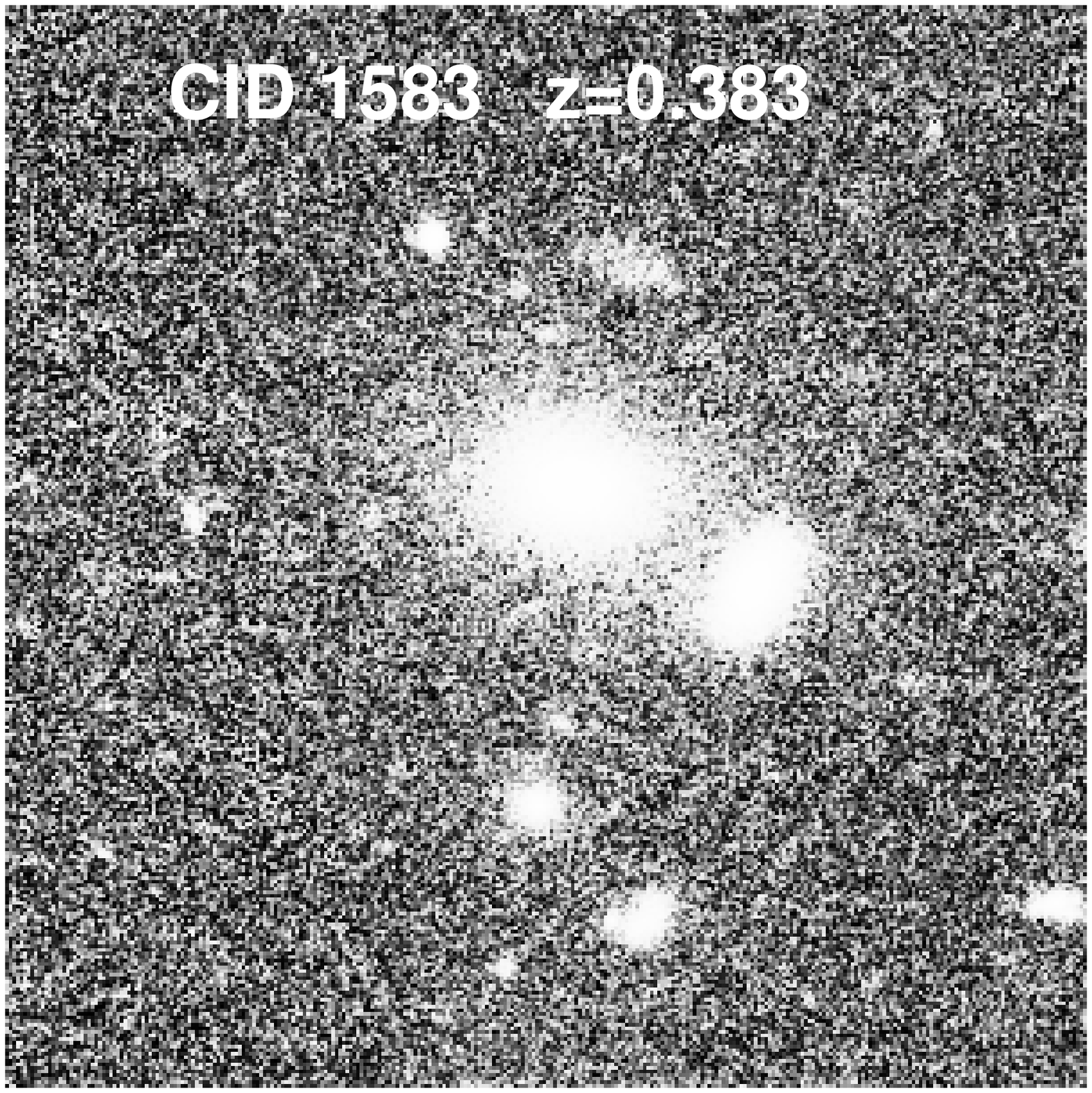}
\includegraphics[width=0.25\textwidth]{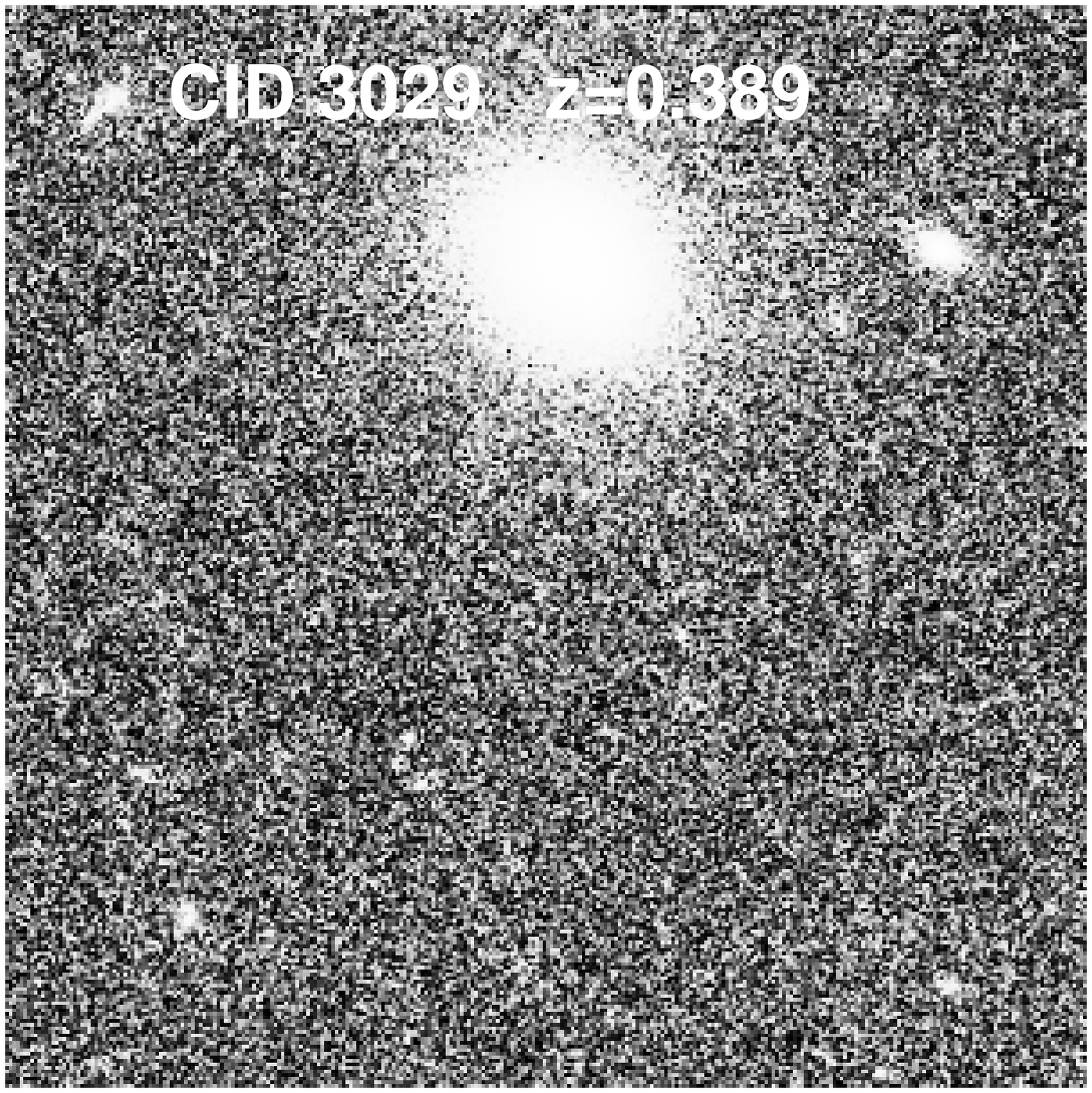}
\includegraphics[width=0.25\textwidth]{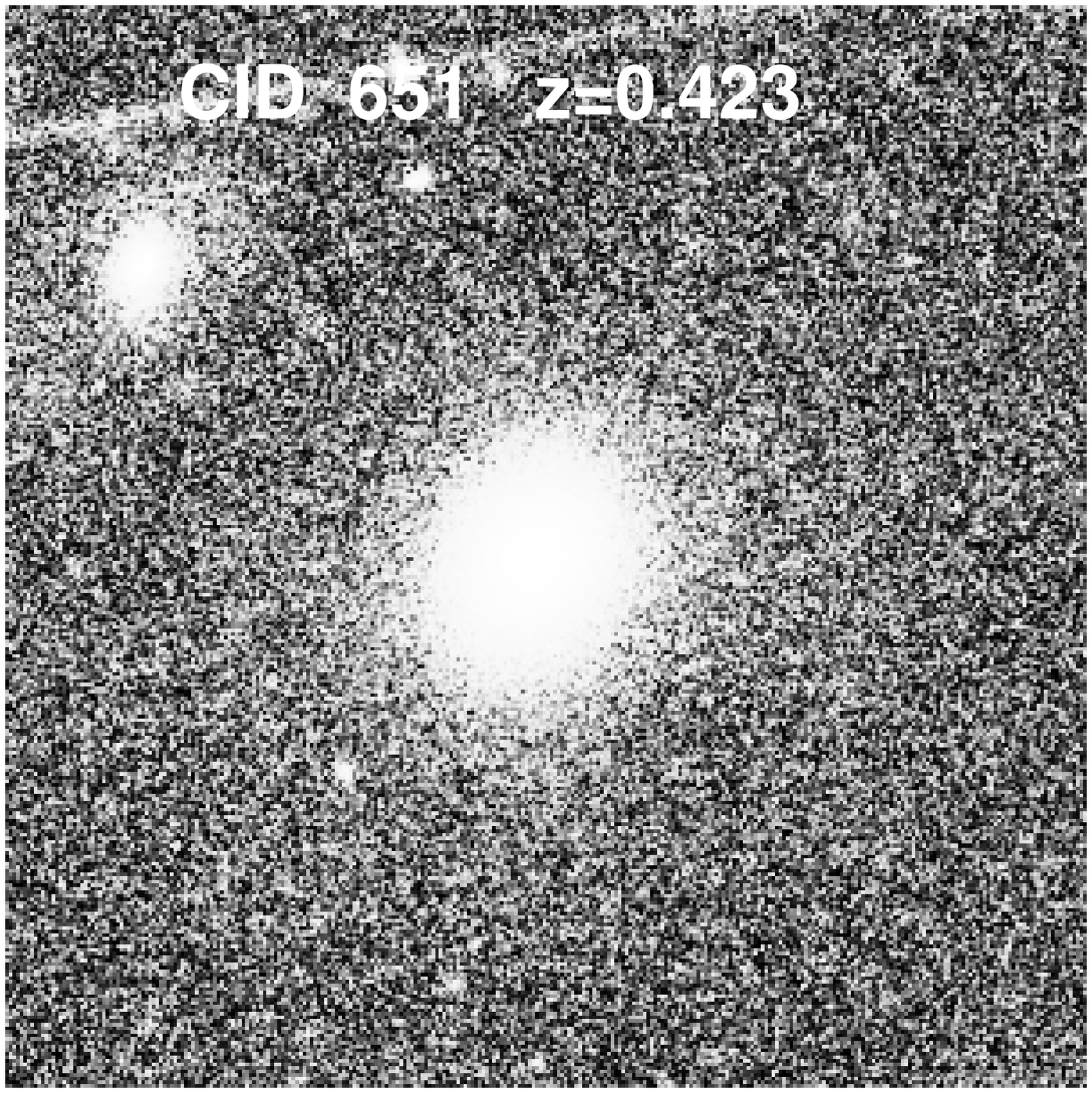}
\caption{\small HST ACS images (15$^{\prime\prime} \times$15$^{\prime\prime}$) of the X-ray ETGs in this paper sorted by increasing redshift.}
\label{fc1}
\end{figure}

\begin{figure}
\centering
\ContinuedFloat
\includegraphics[width=0.25\textwidth]{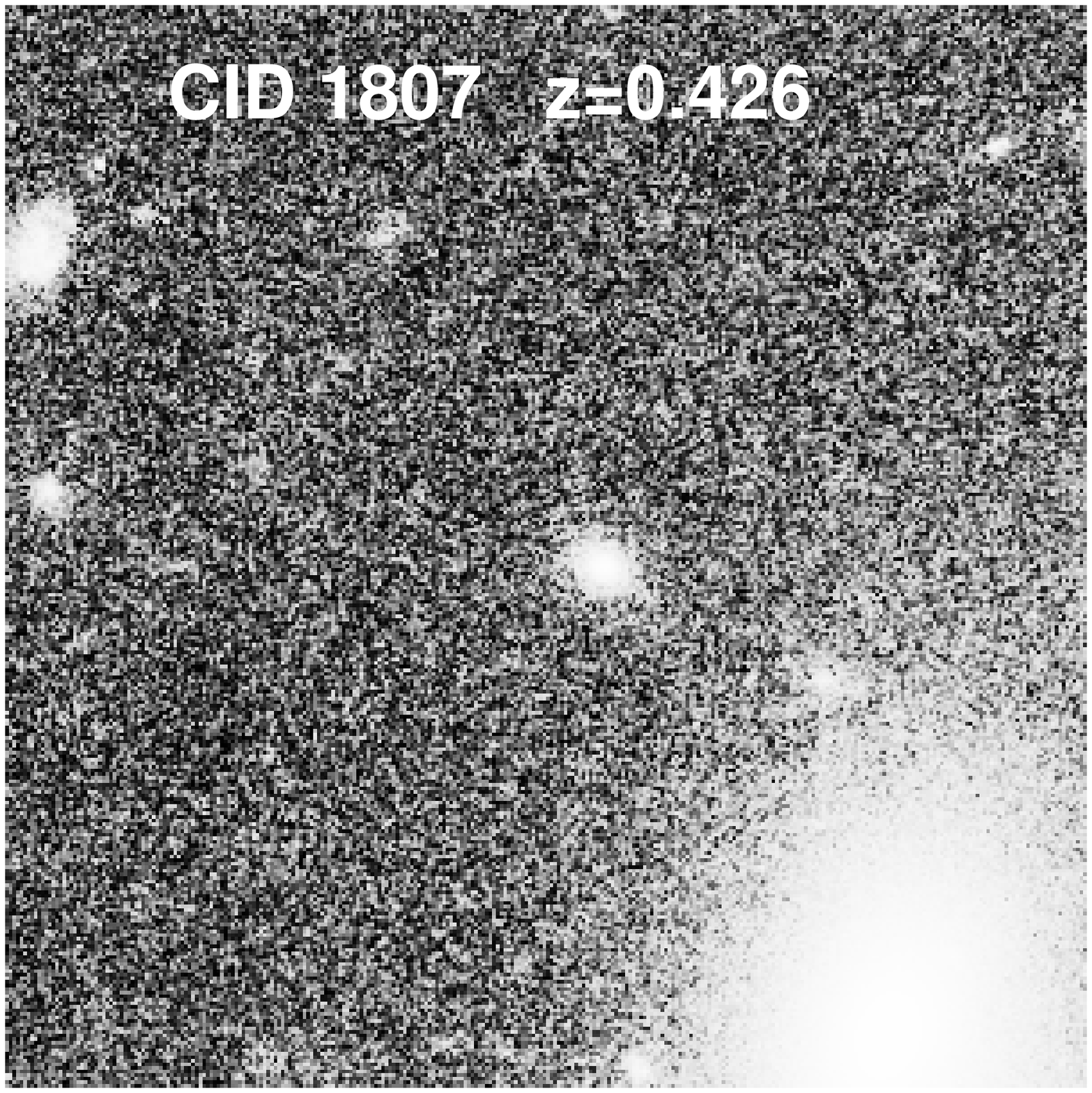}
\includegraphics[width=0.25\textwidth]{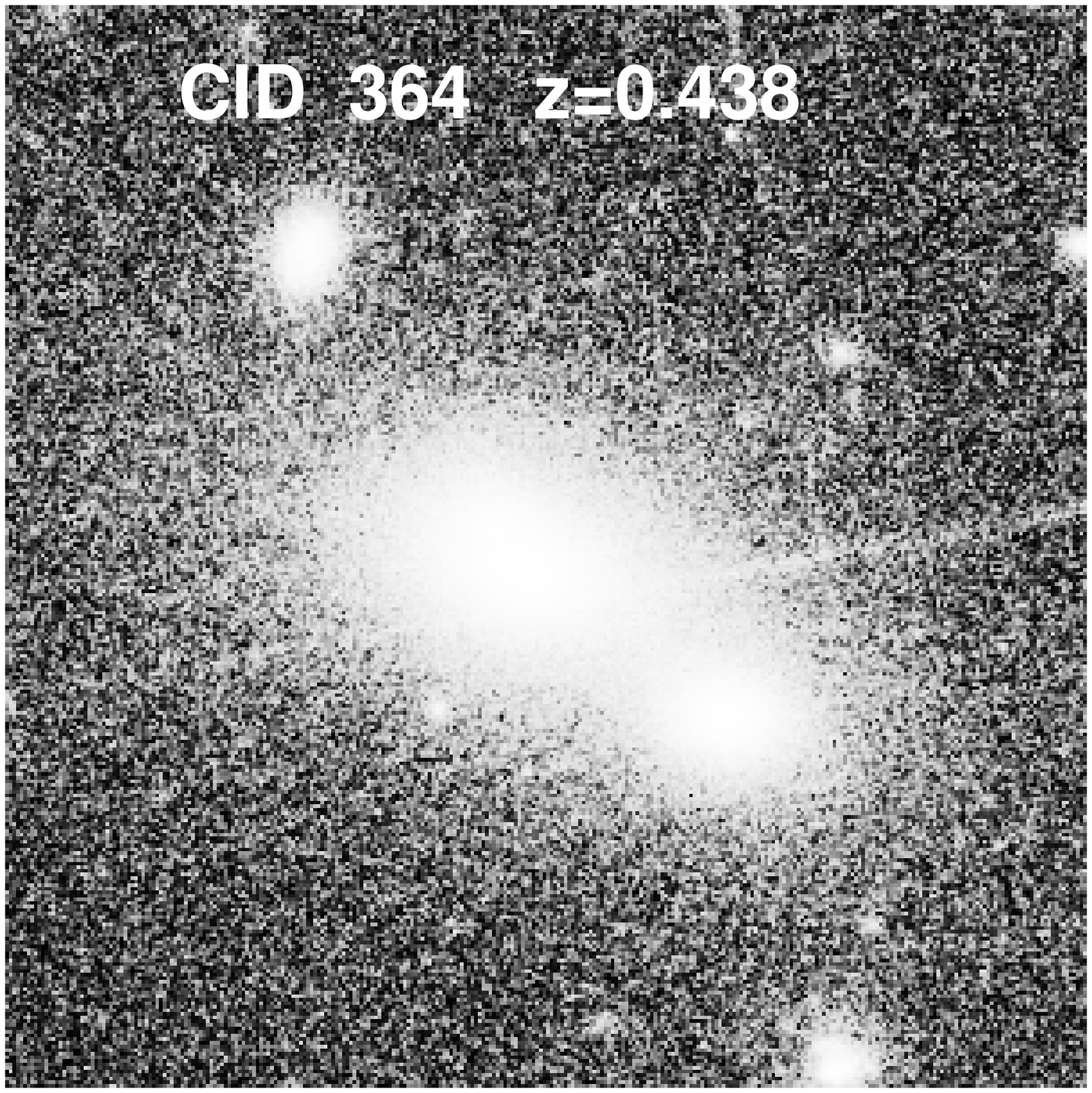}
\includegraphics[width=0.25\textwidth]{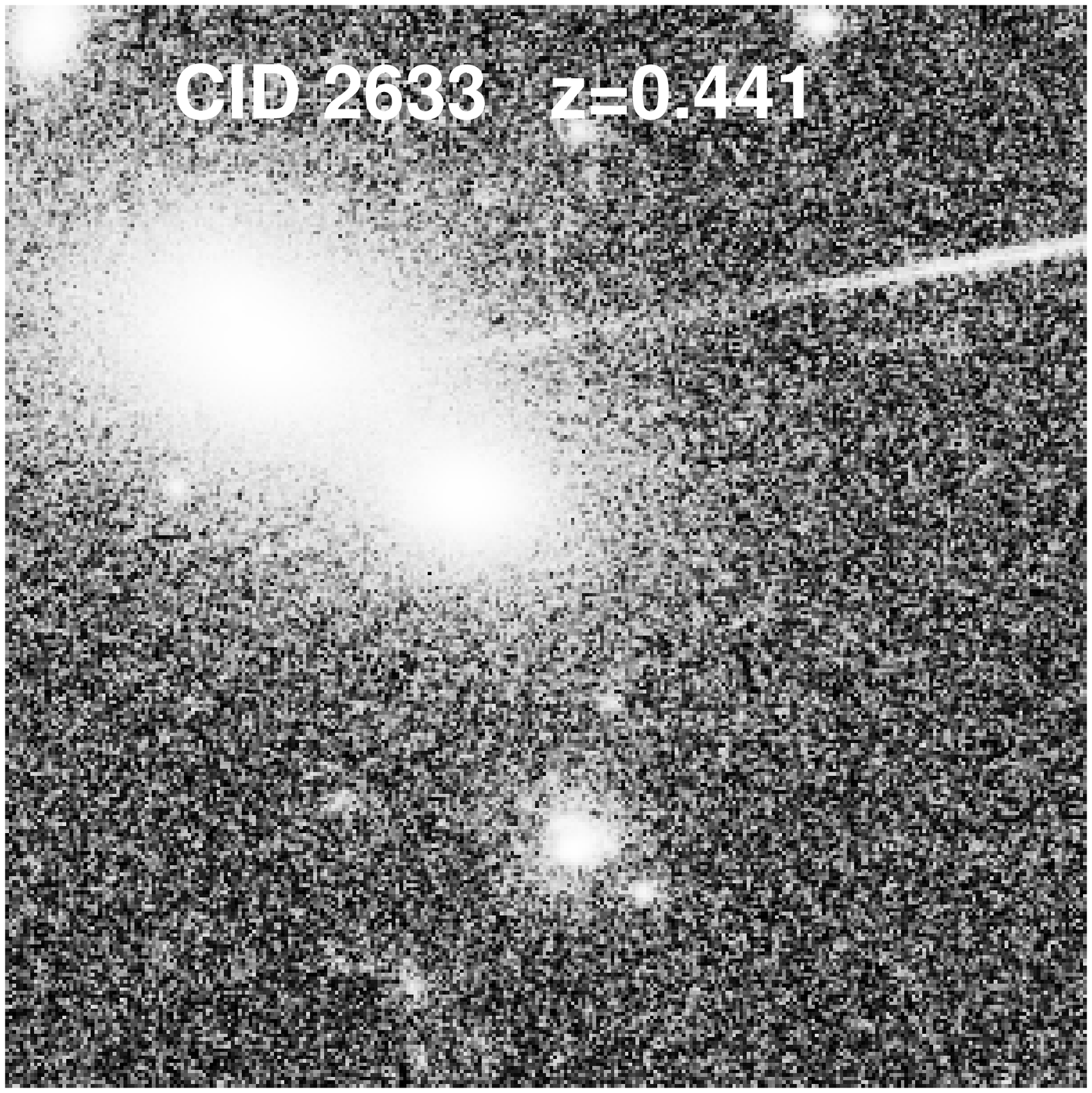}
\includegraphics[width=0.25\textwidth]{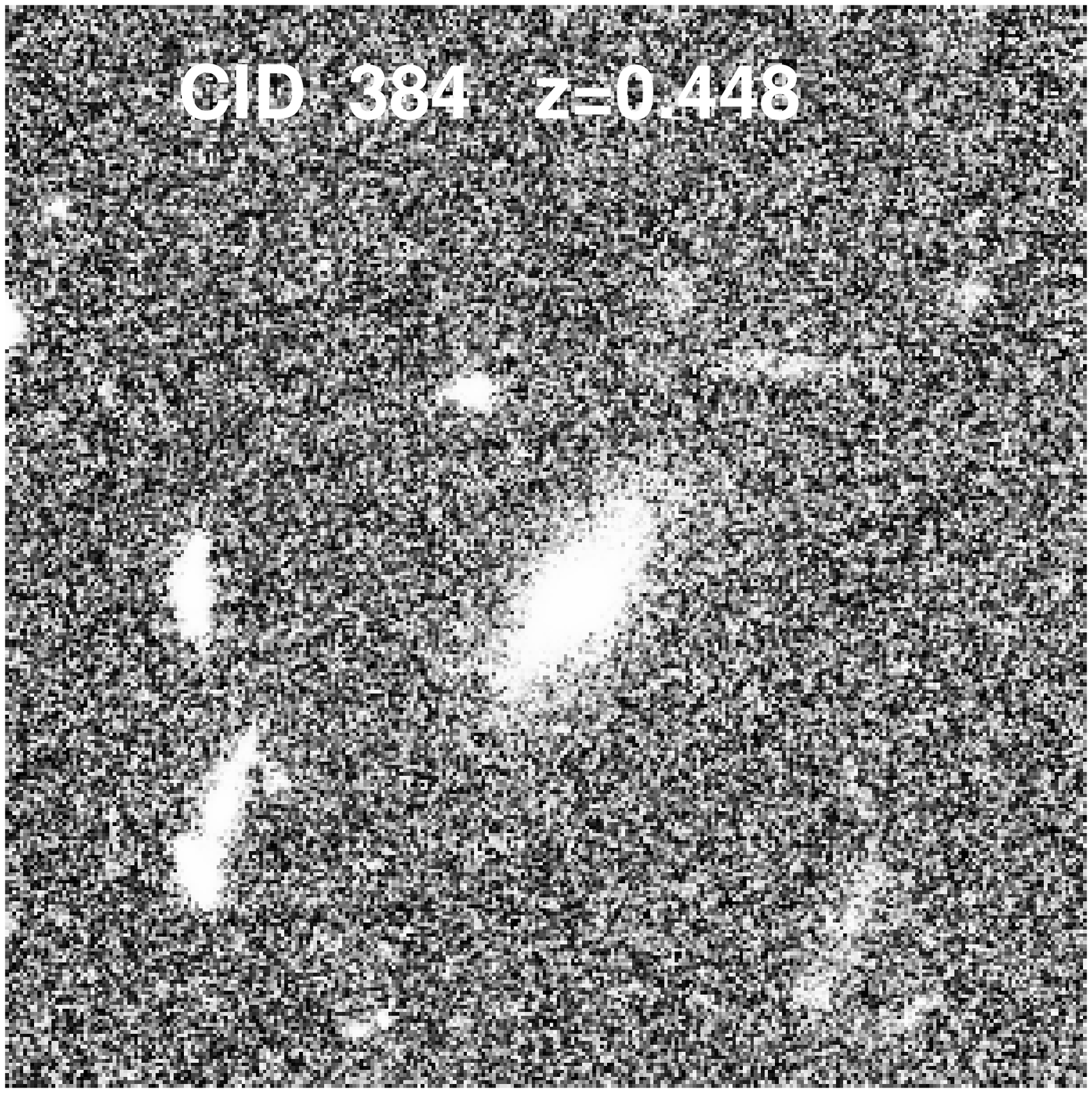}
\includegraphics[width=0.25\textwidth]{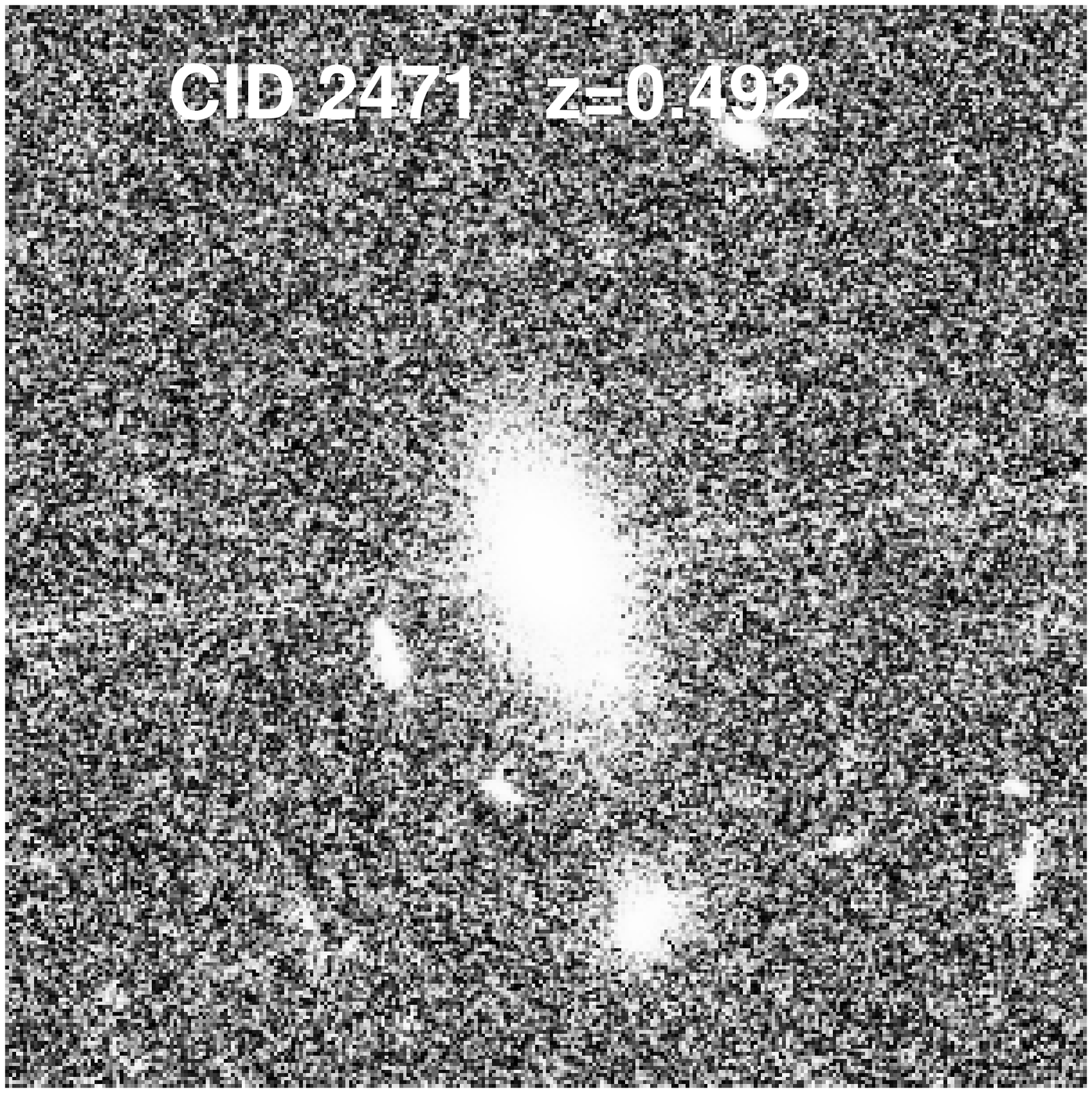}
\includegraphics[width=0.25\textwidth]{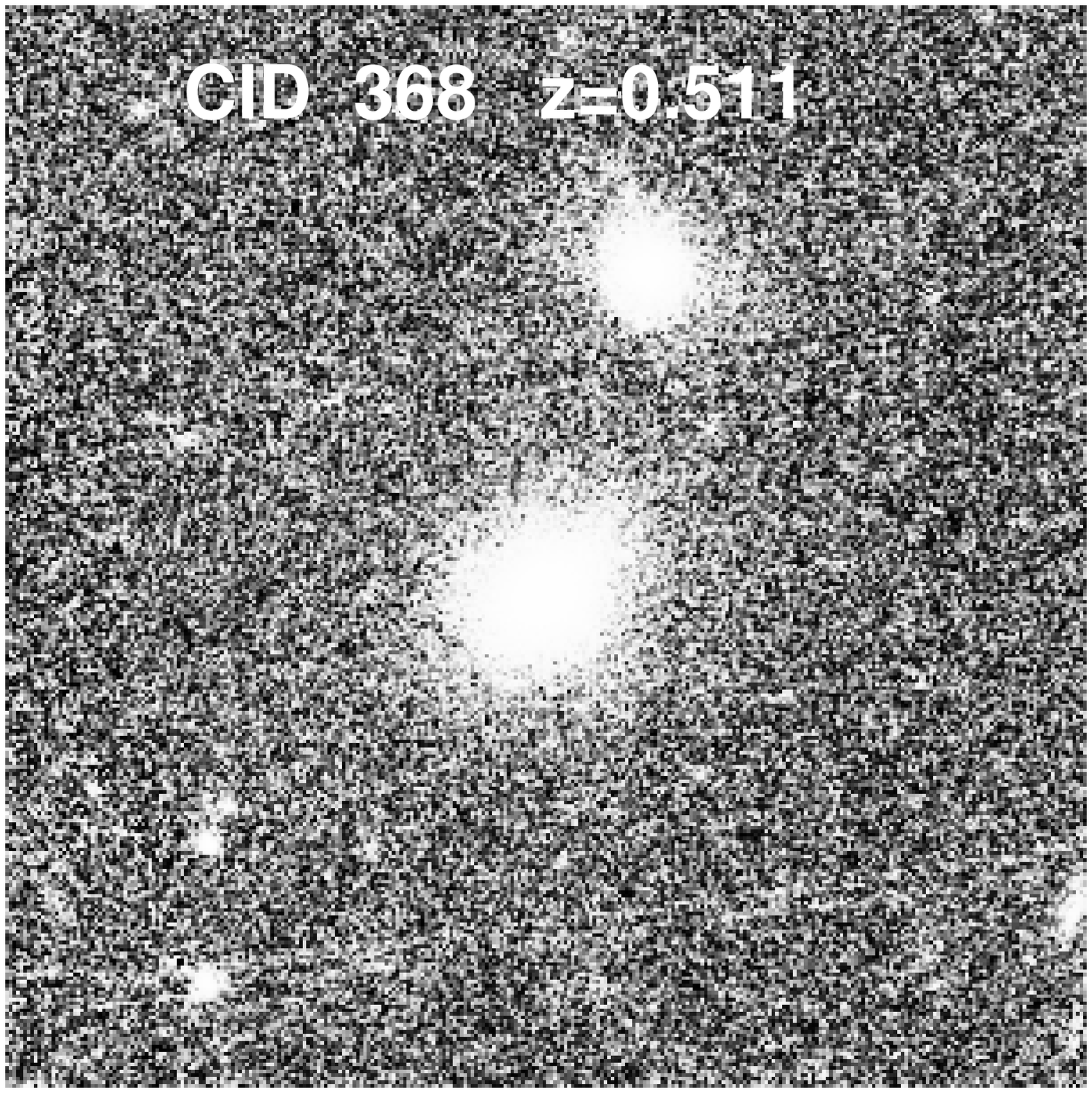}
\includegraphics[width=0.25\textwidth]{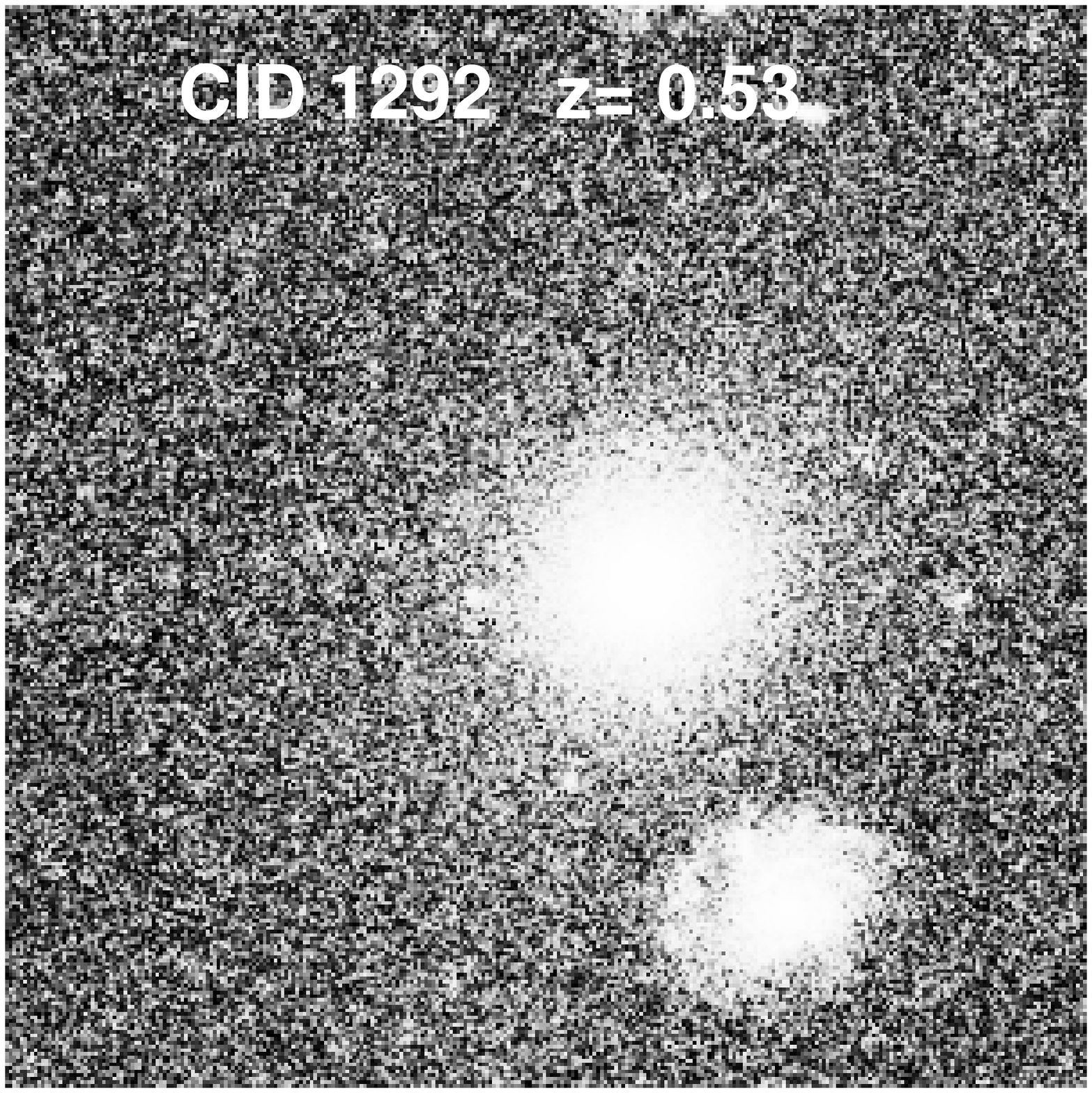}
\includegraphics[width=0.25\textwidth]{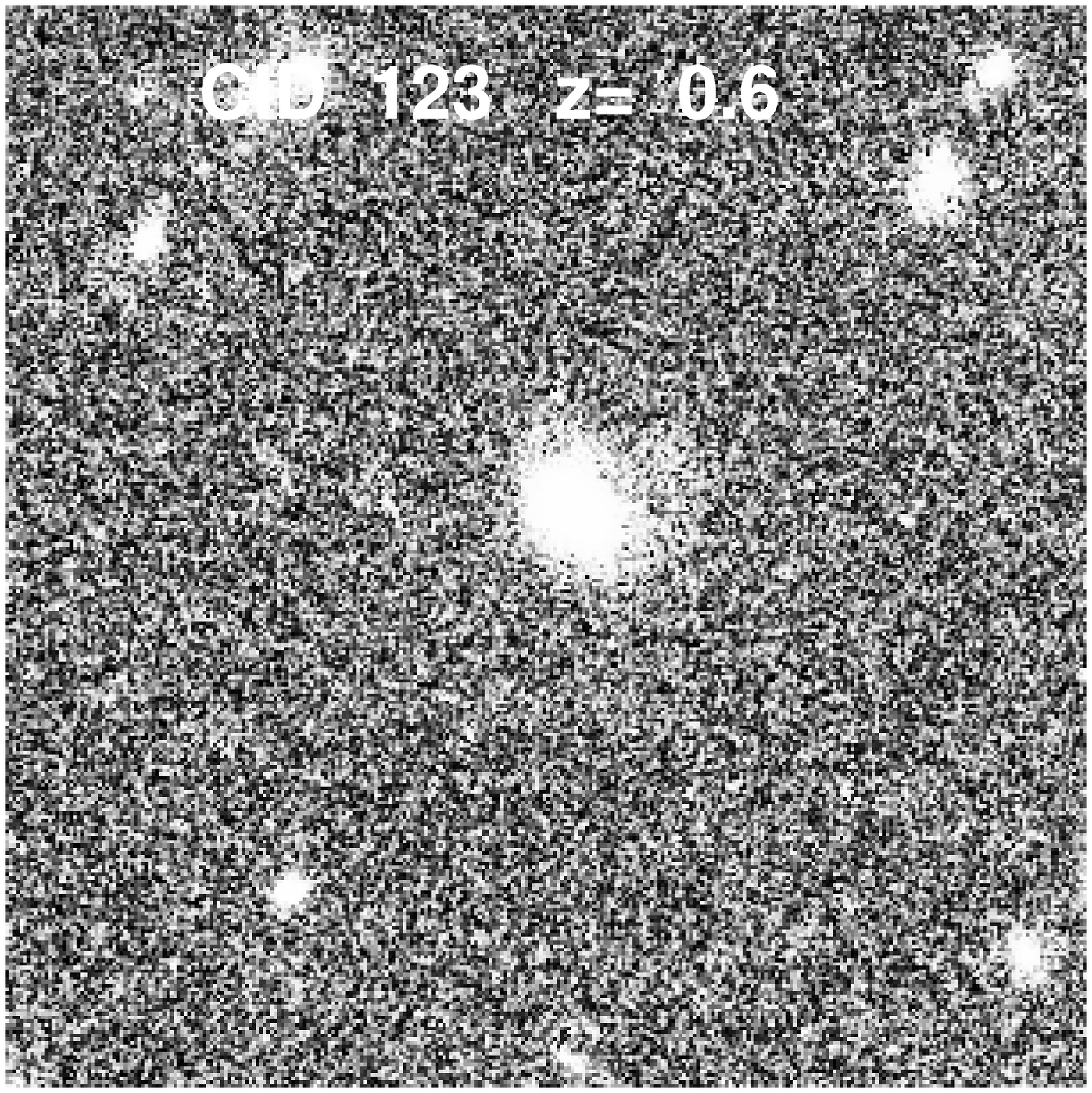}
\includegraphics[width=0.25\textwidth]{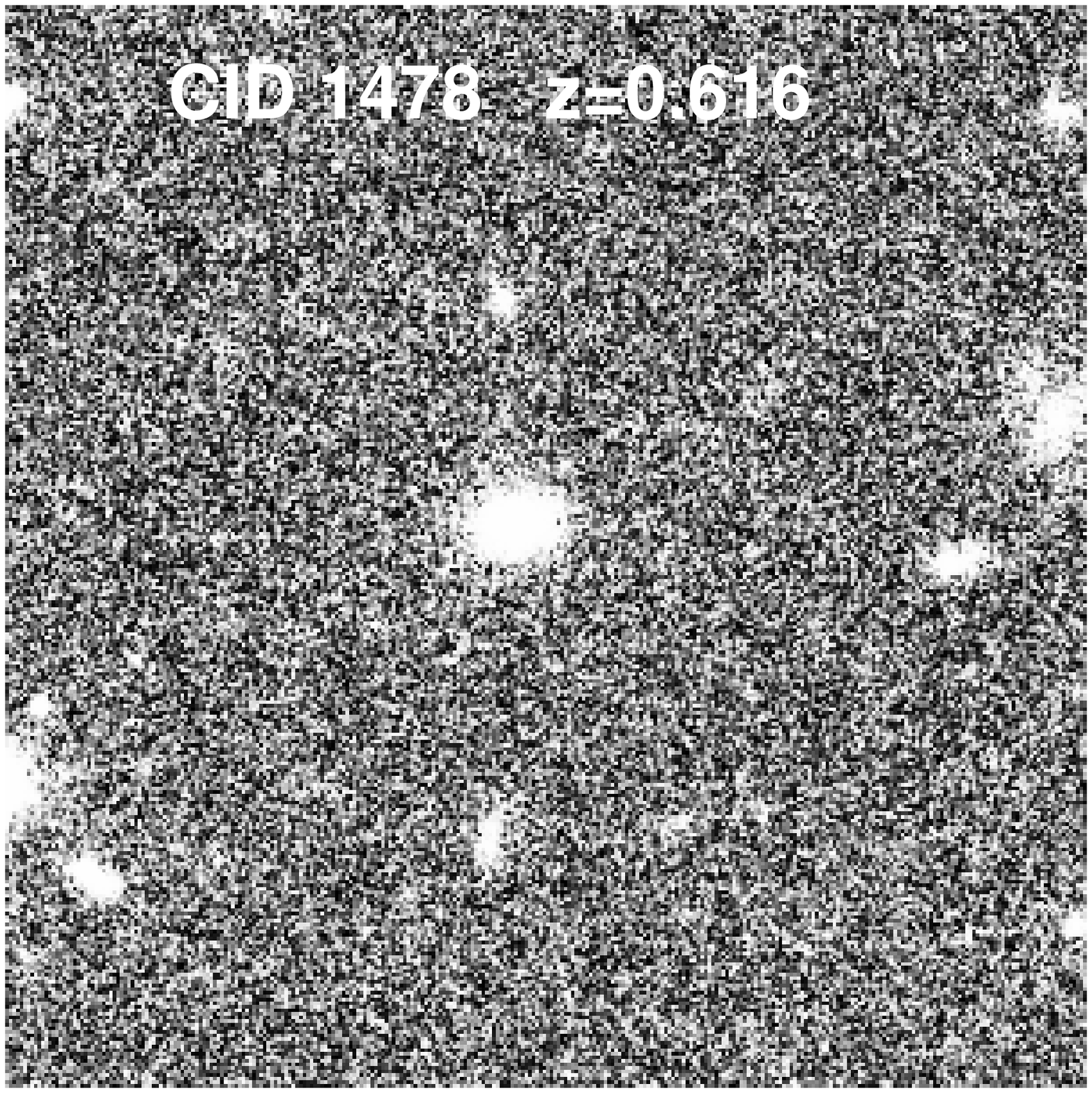}
\includegraphics[width=0.25\textwidth]{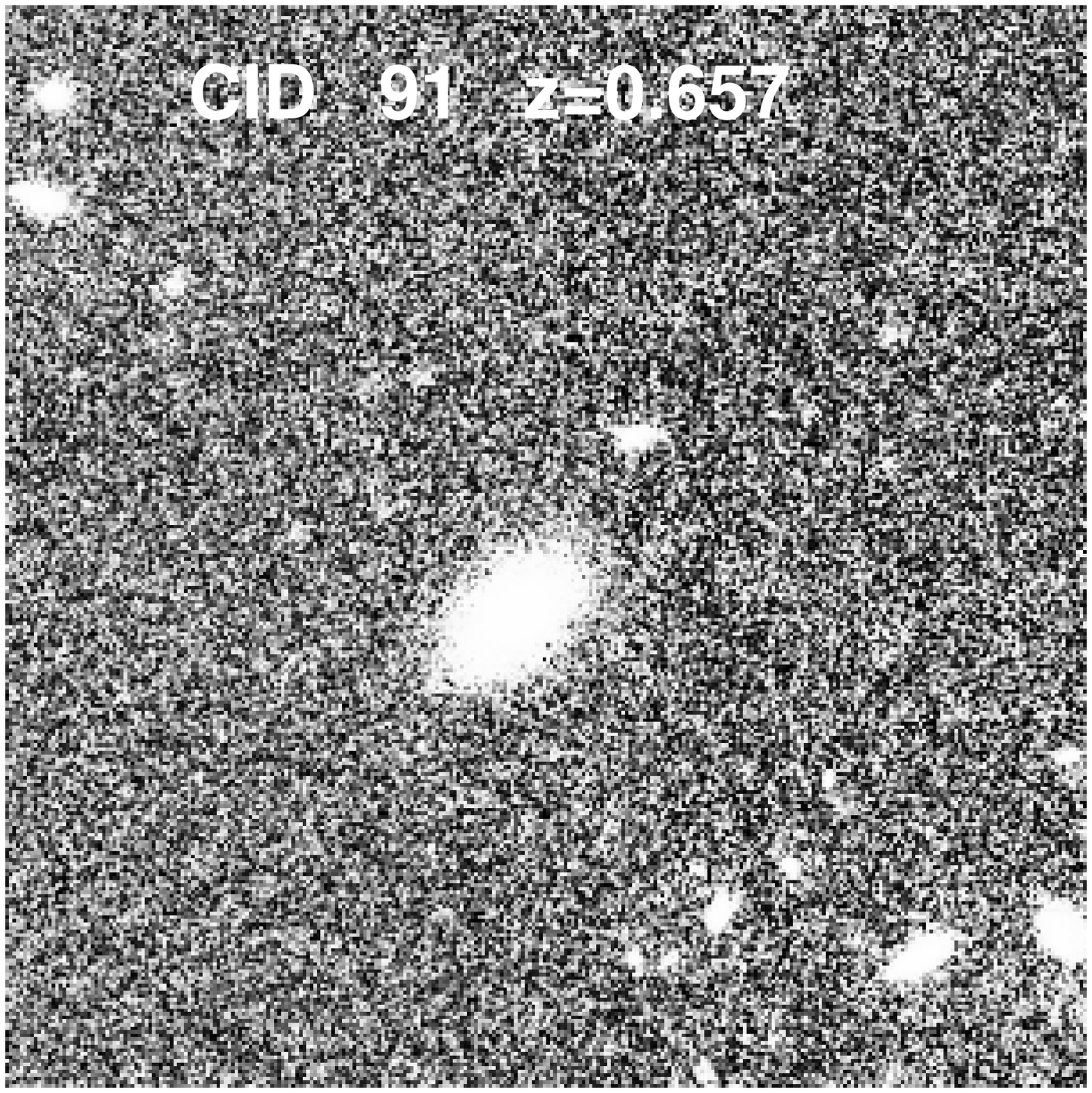}
\includegraphics[width=0.25\textwidth]{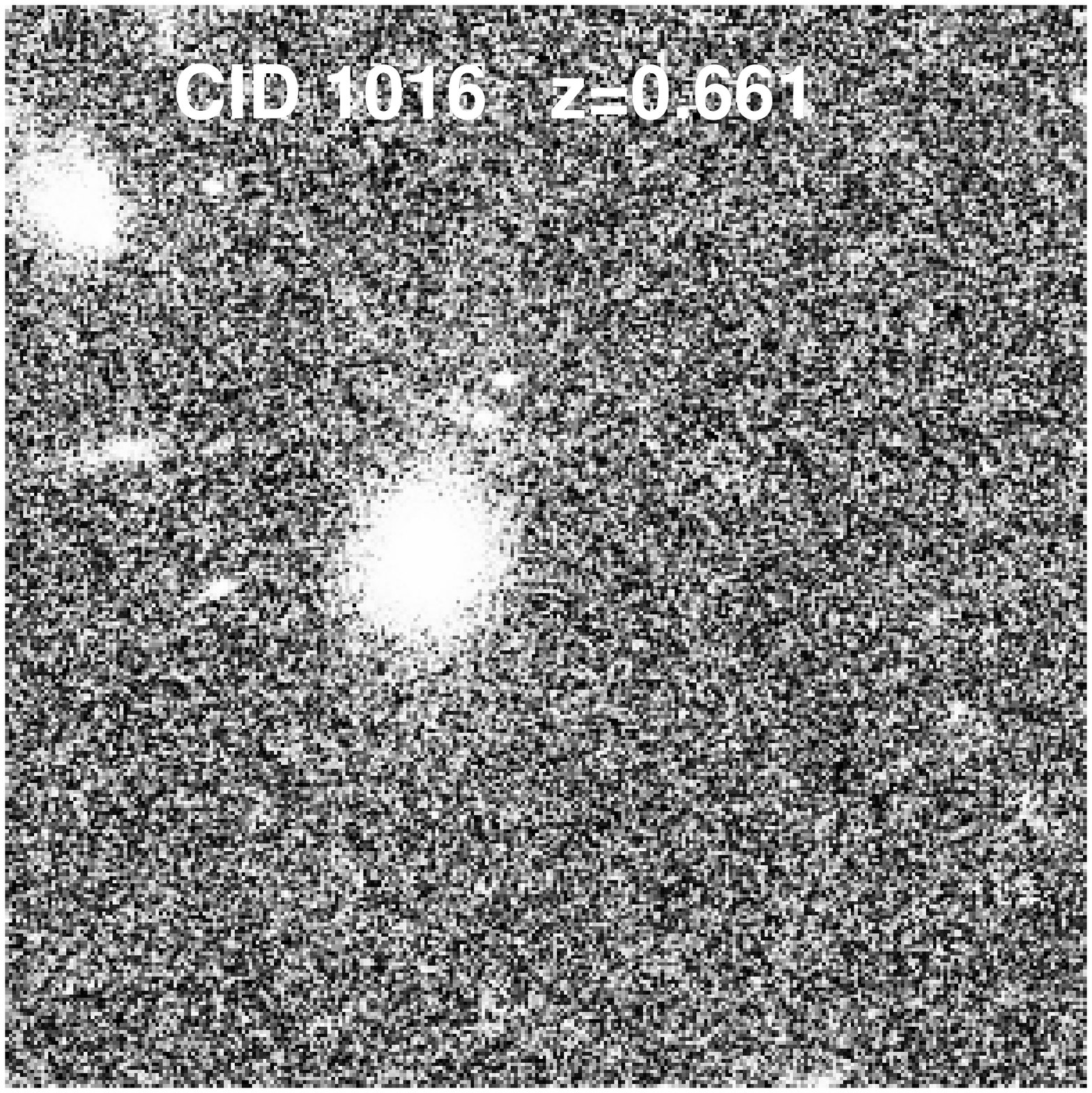}
\includegraphics[width=0.25\textwidth]{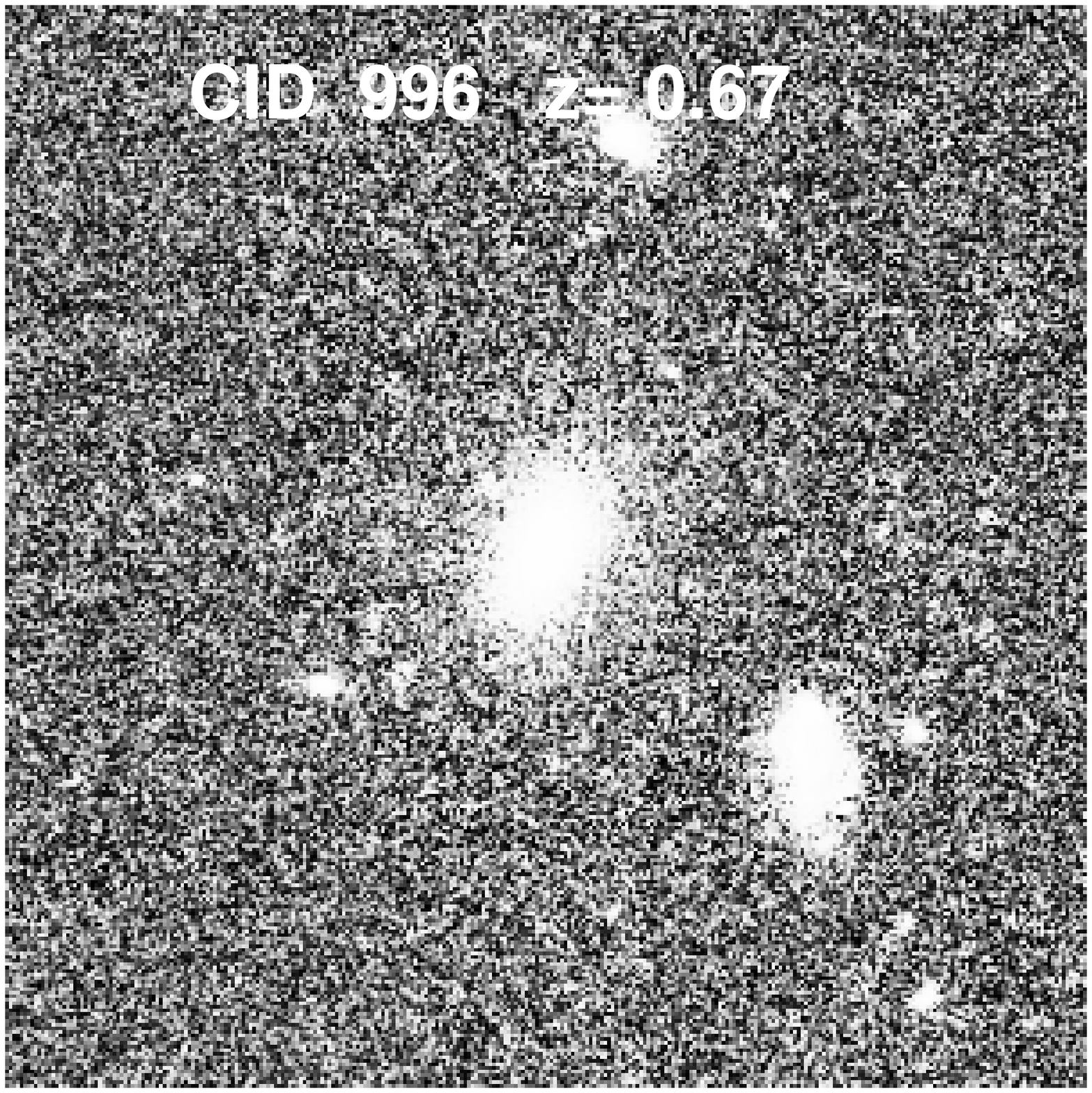}
\includegraphics[width=0.25\textwidth]{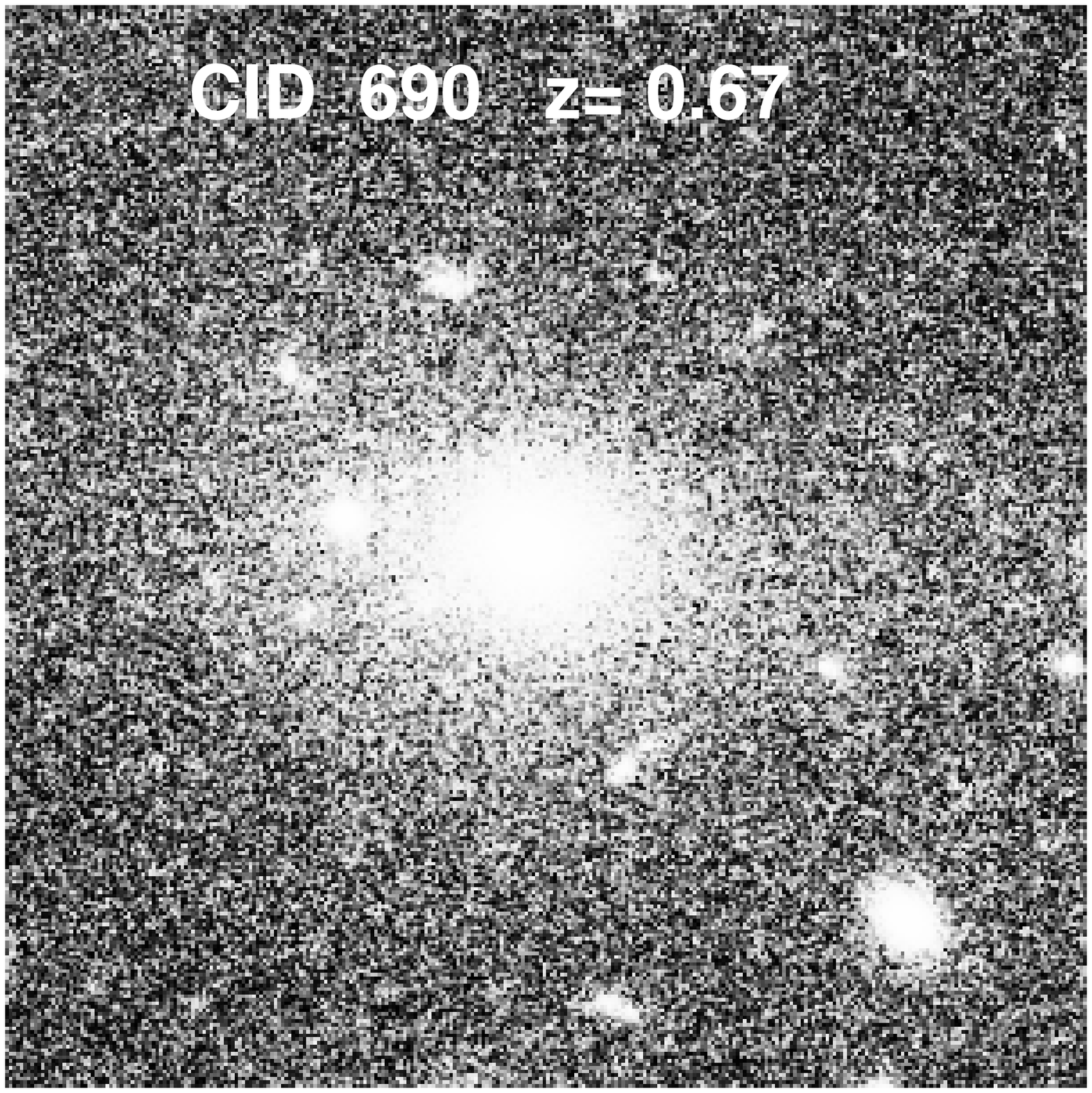}
\includegraphics[width=0.25\textwidth]{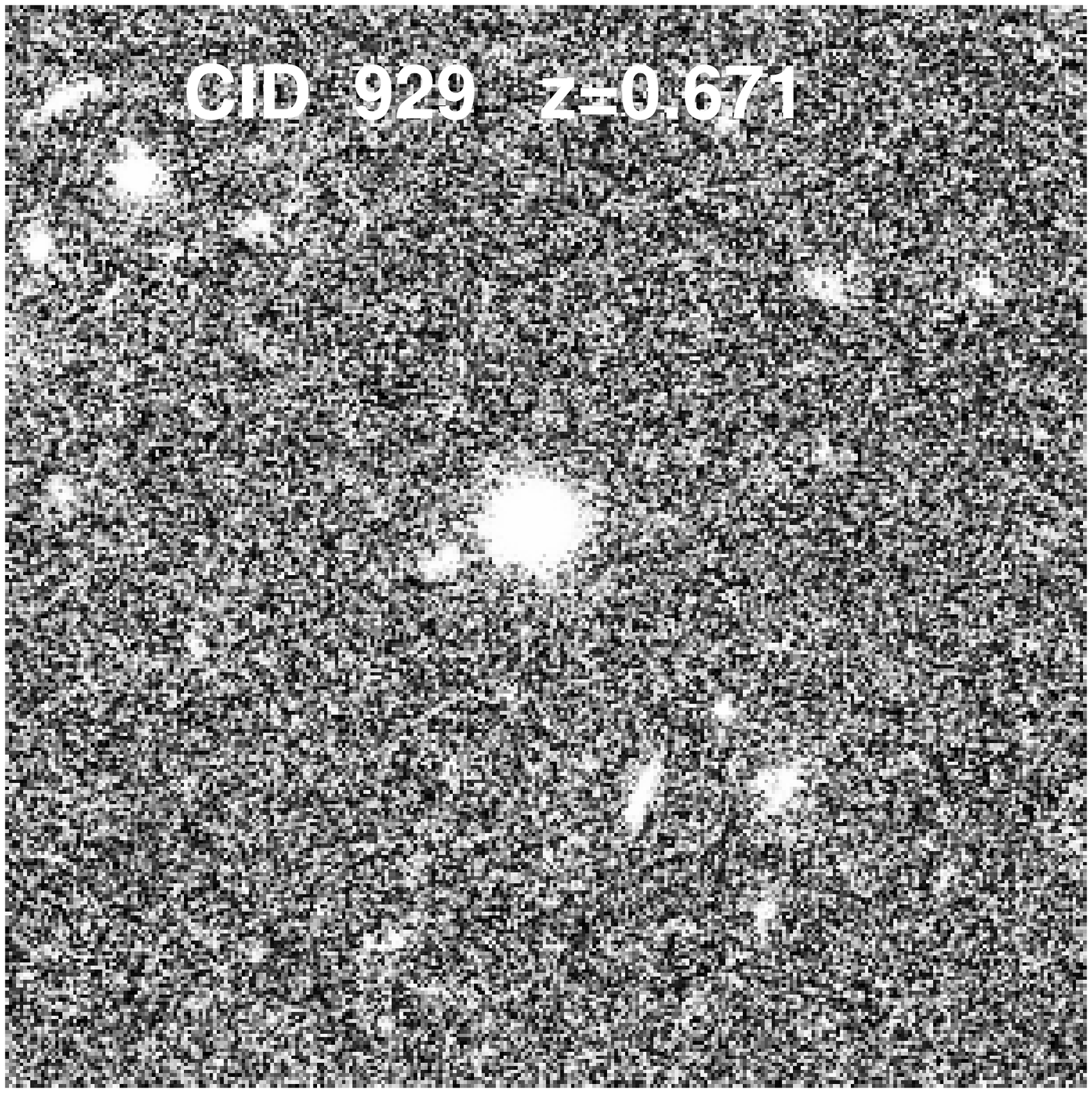}
\includegraphics[width=0.25\textwidth]{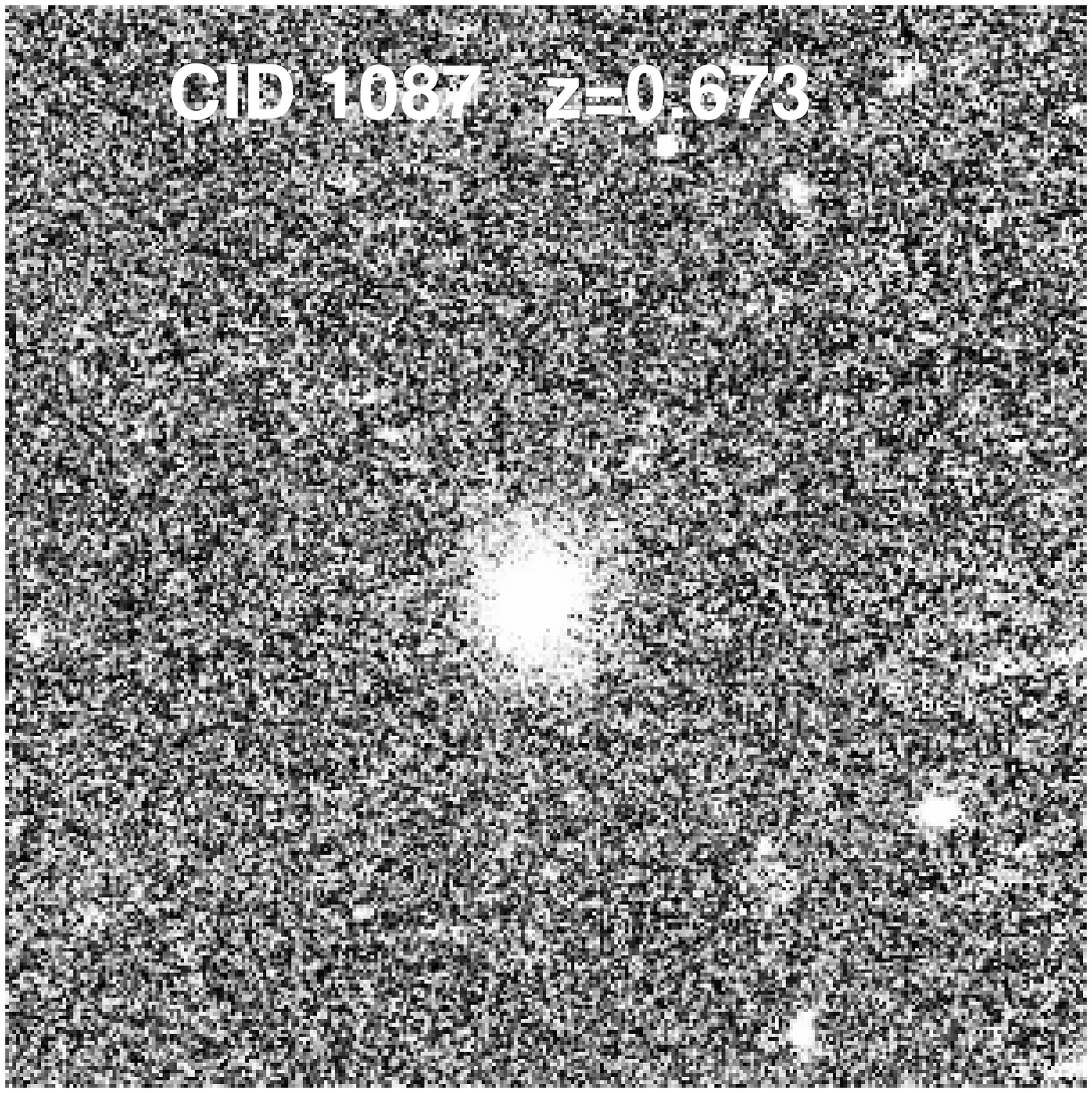}
\caption{\small HST ACS images (15$^{\prime\prime} \times$15$^{\prime\prime}$) of the X-ray ETGs in this paper sorted by increasing redshift.}
\label{fc1}
\end{figure}

\begin{figure}
\centering
\ContinuedFloat
\includegraphics[width=0.25\textwidth]{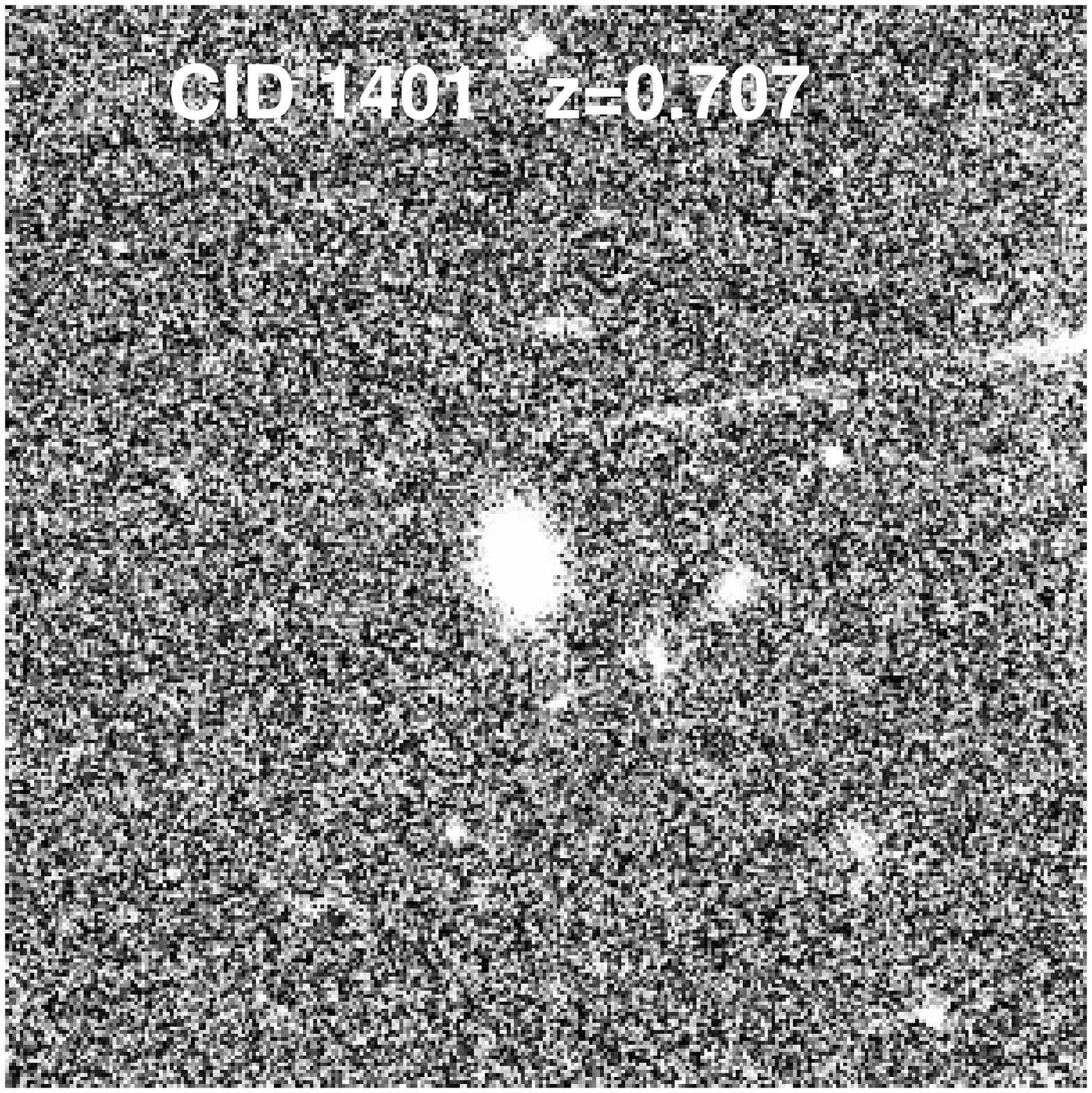}
\includegraphics[width=0.25\textwidth]{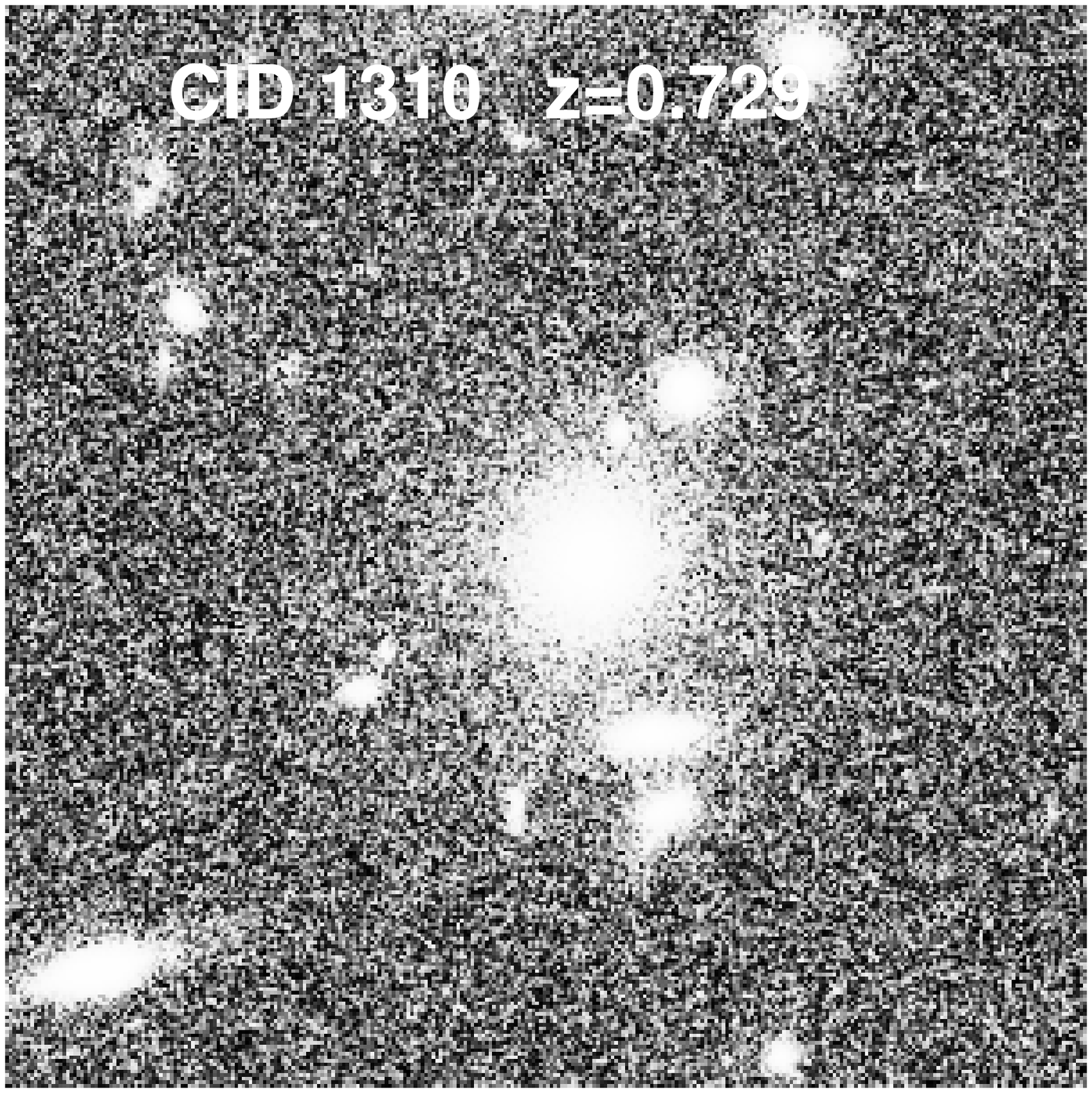}
\includegraphics[width=0.25\textwidth]{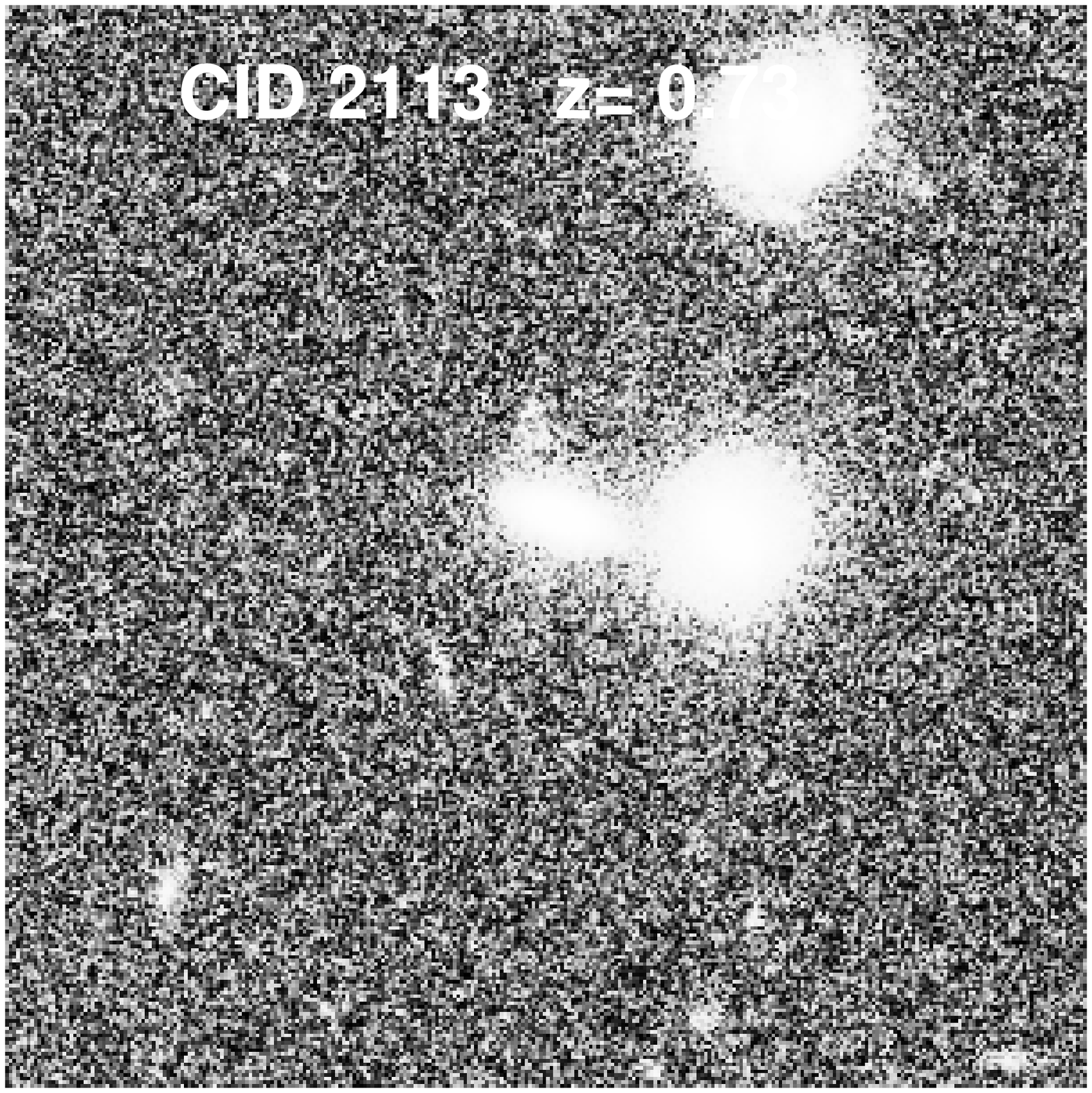}
\includegraphics[width=0.25\textwidth]{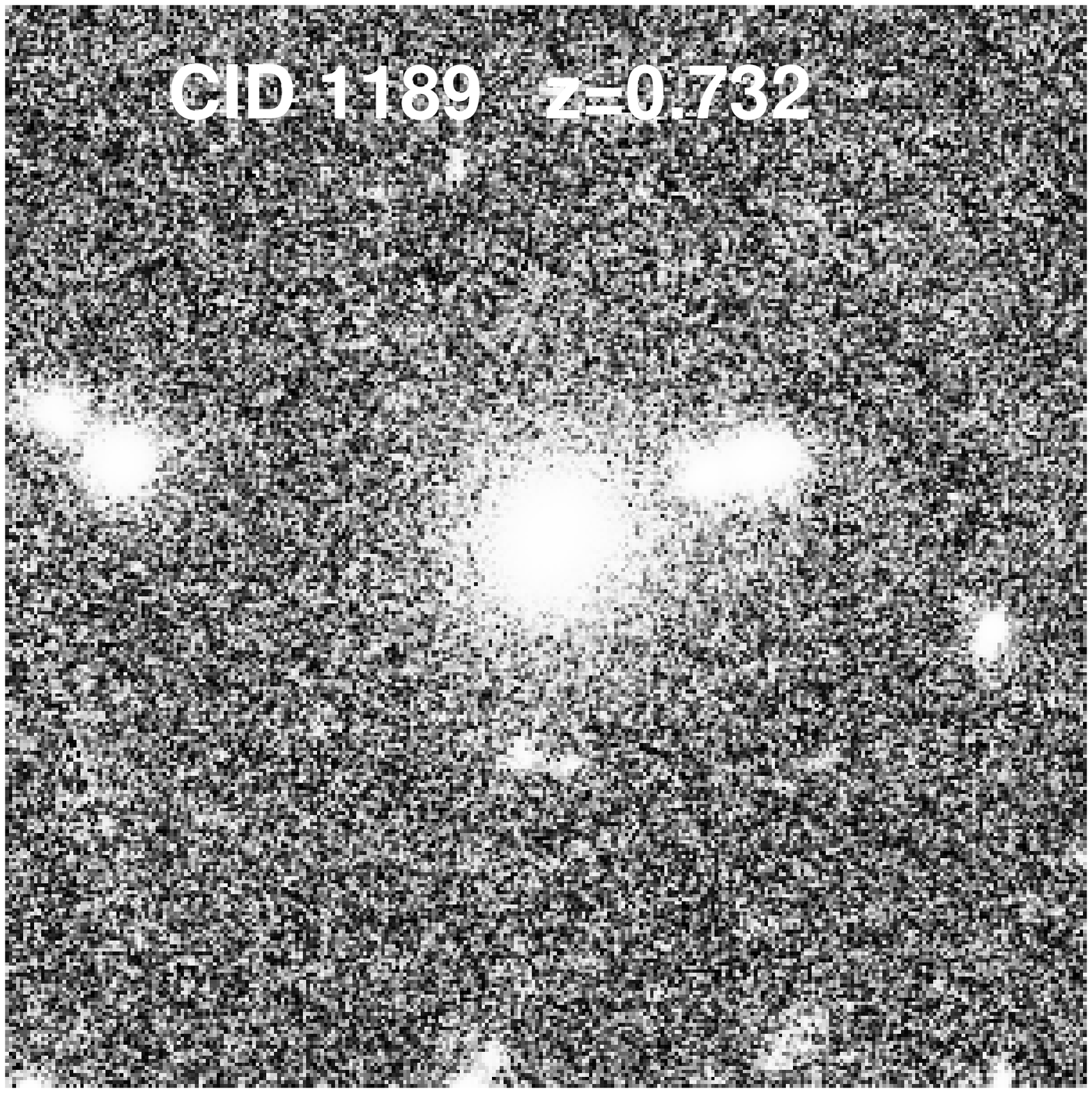}
\includegraphics[width=0.25\textwidth]{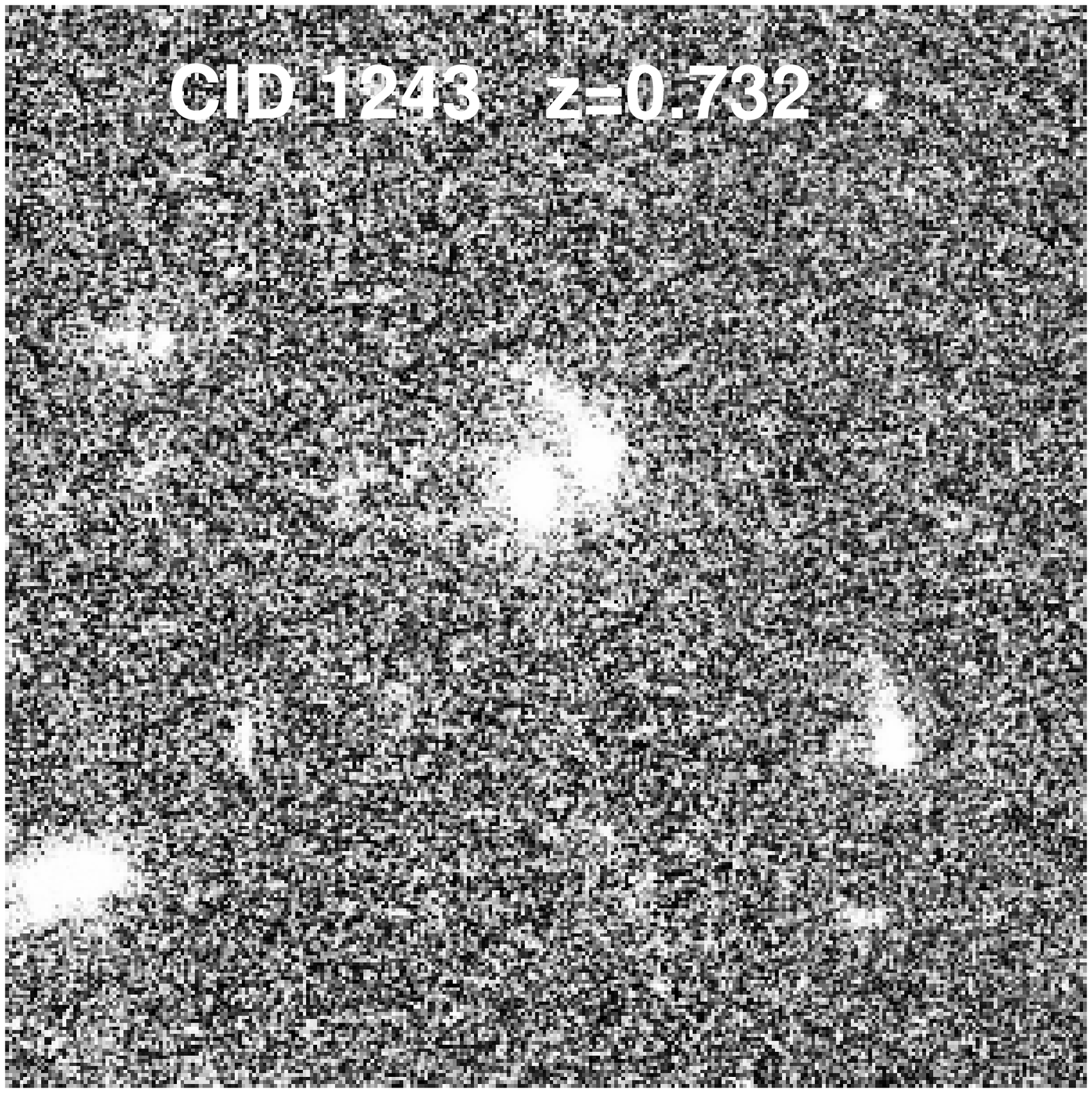}
\includegraphics[width=0.25\textwidth]{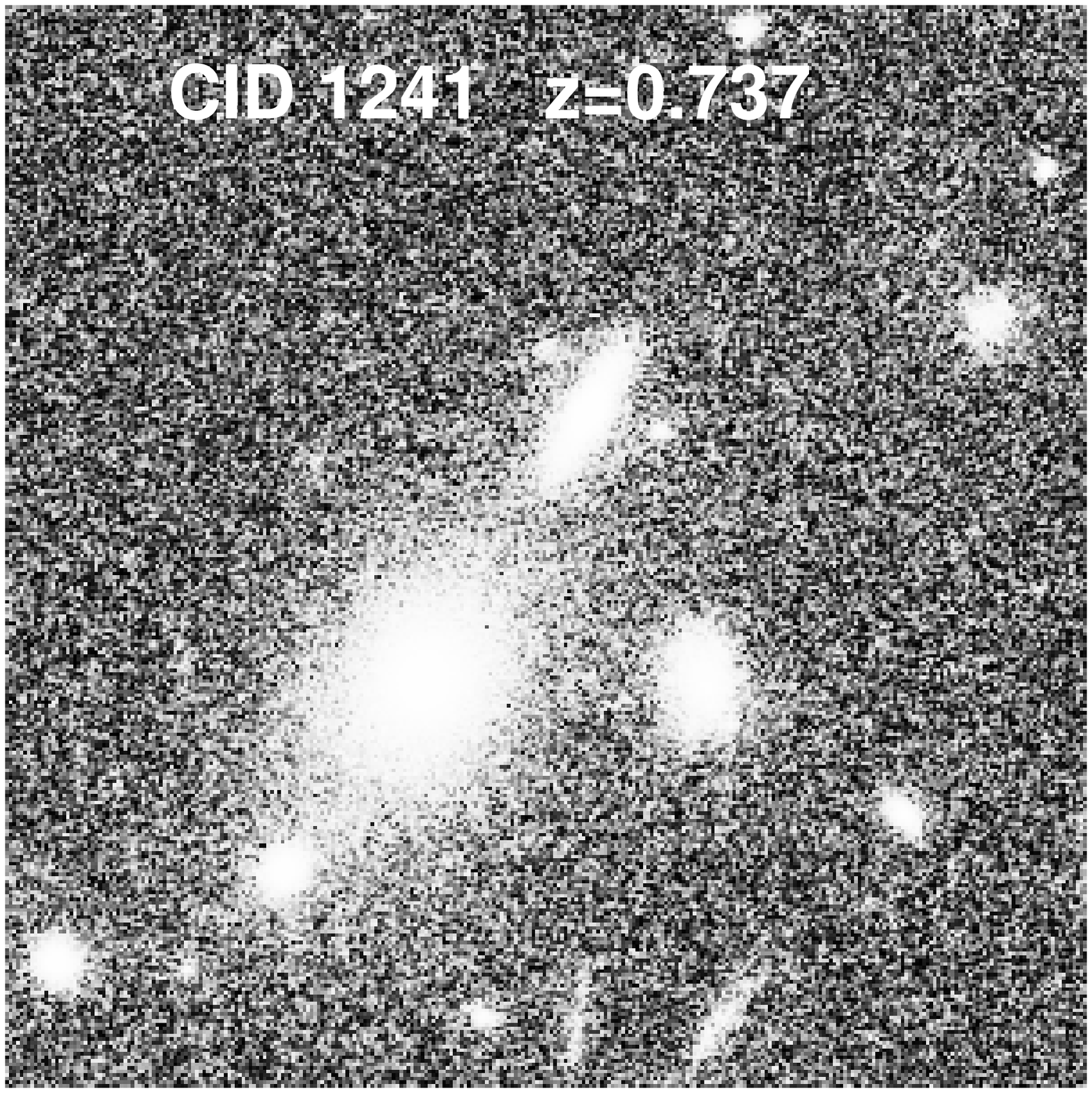}
\includegraphics[width=0.25\textwidth]{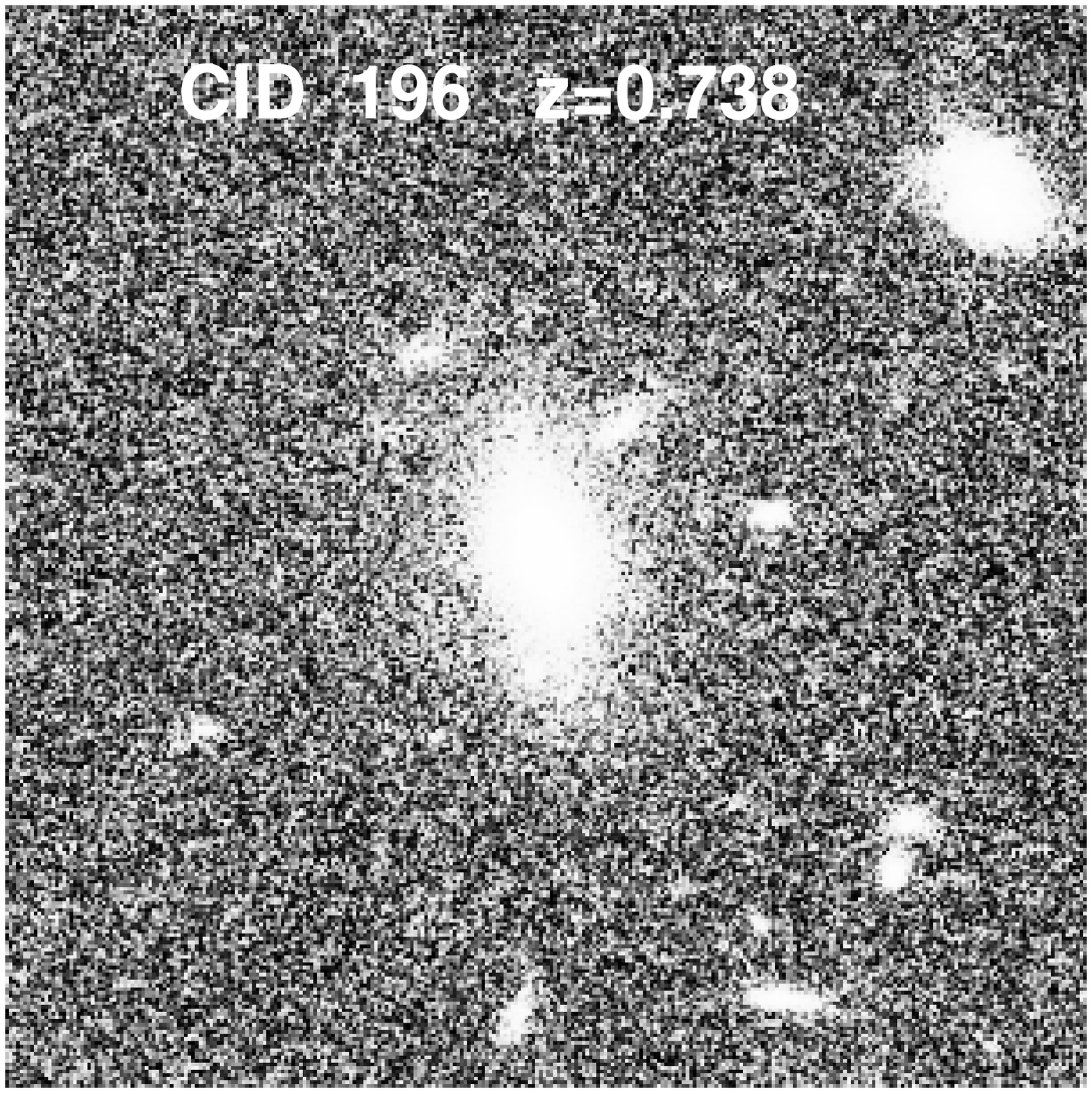}
\includegraphics[width=0.25\textwidth]{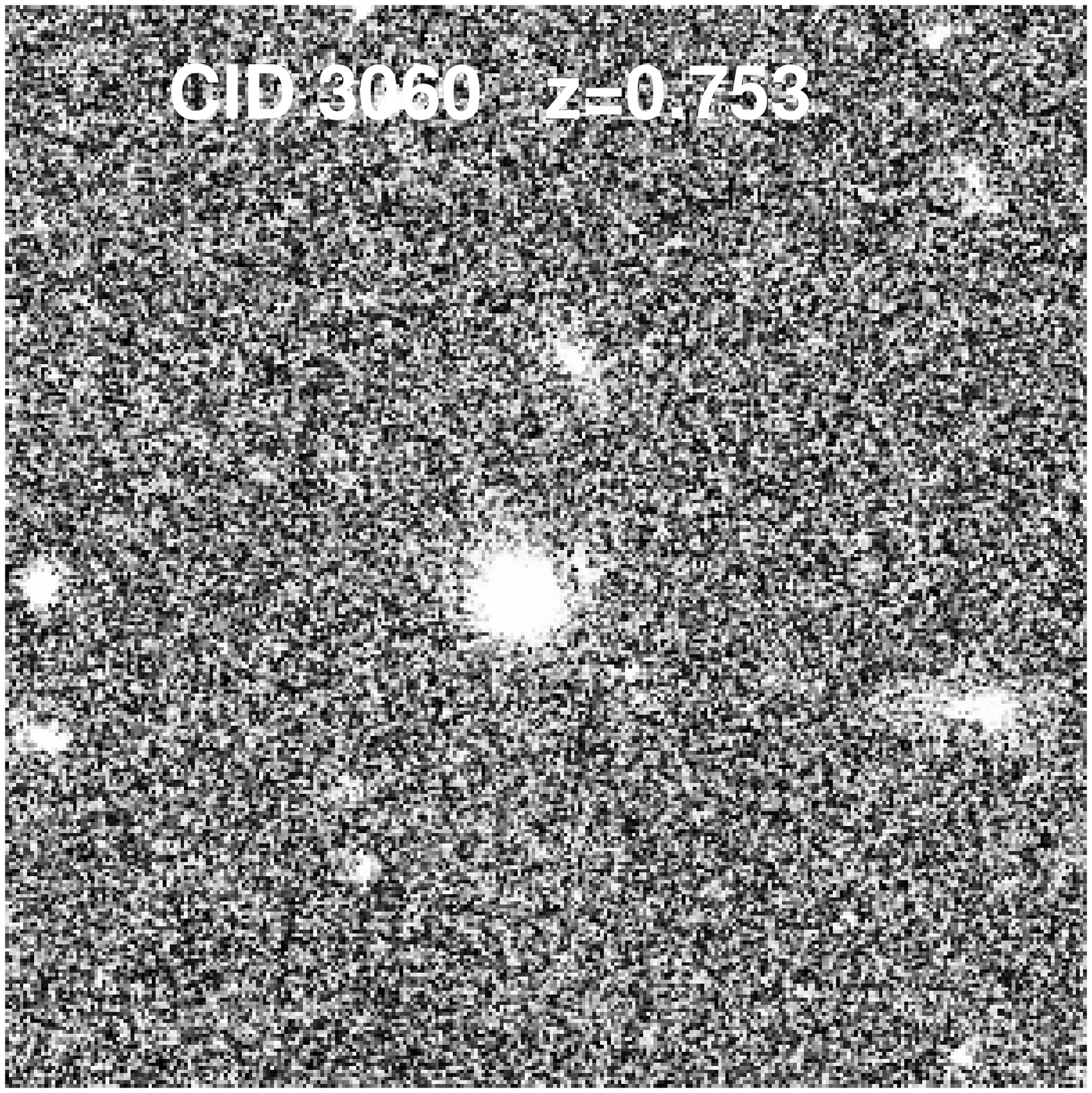}
\includegraphics[width=0.25\textwidth]{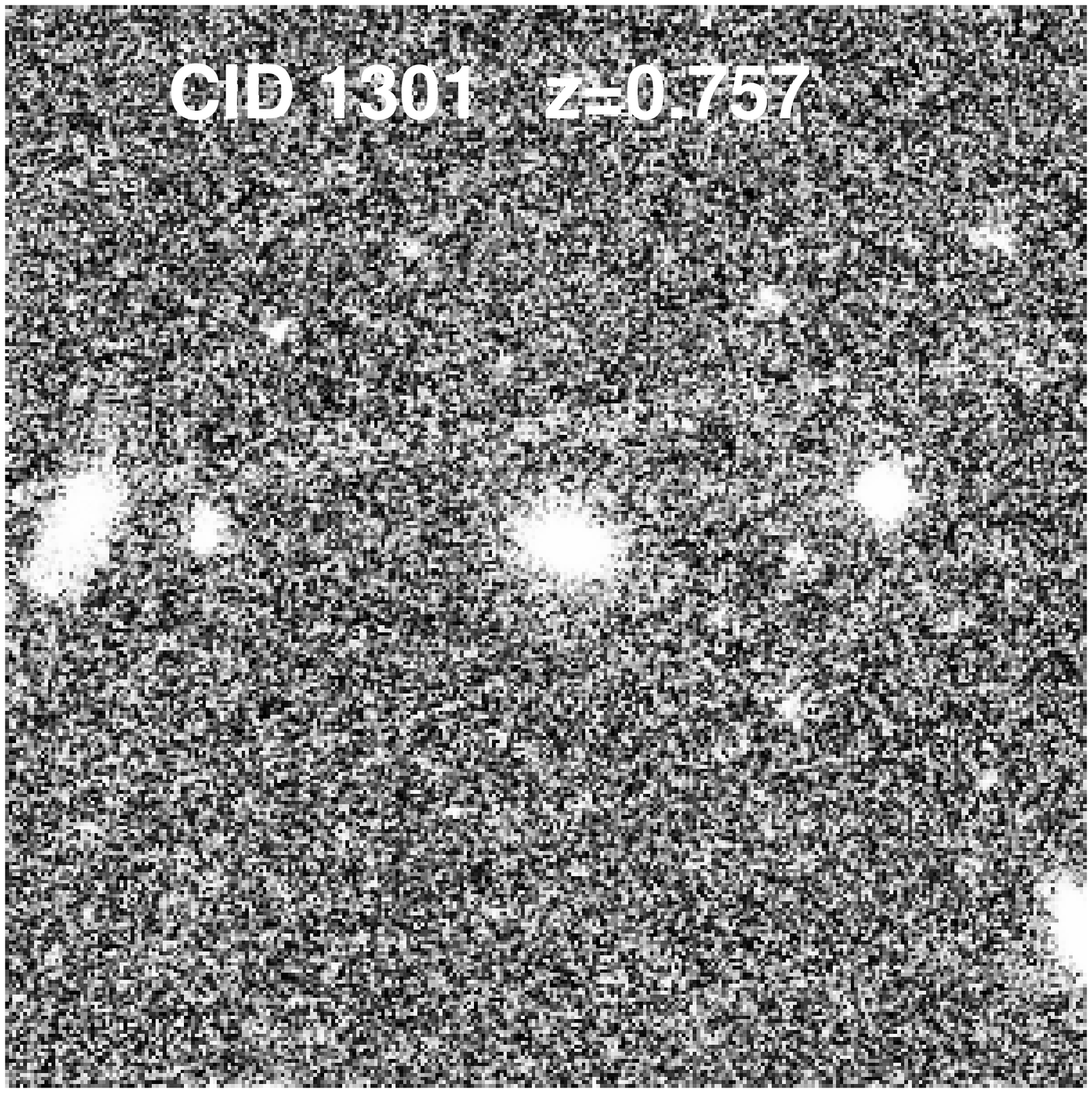}
\includegraphics[width=0.25\textwidth]{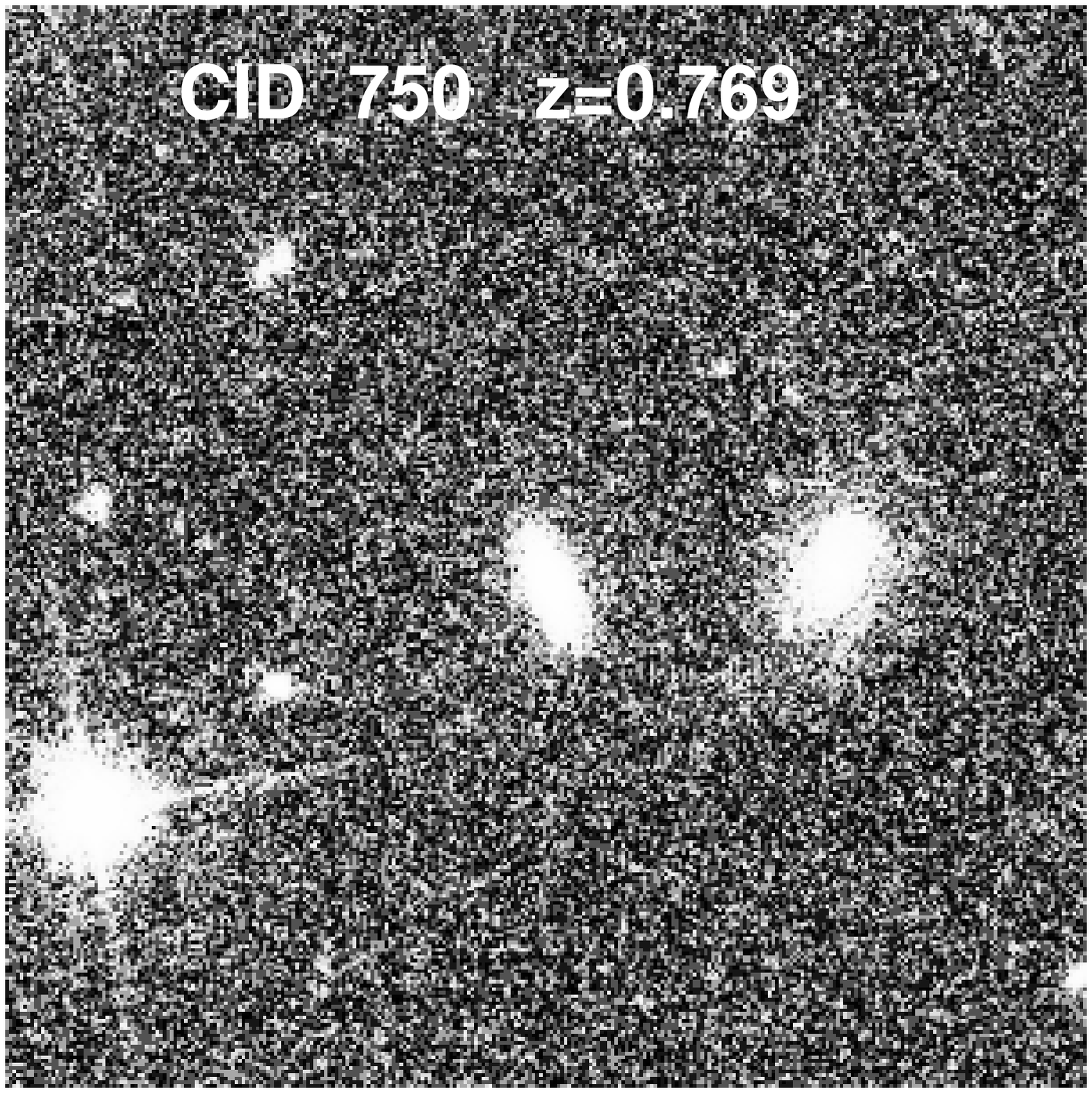}
\includegraphics[width=0.25\textwidth]{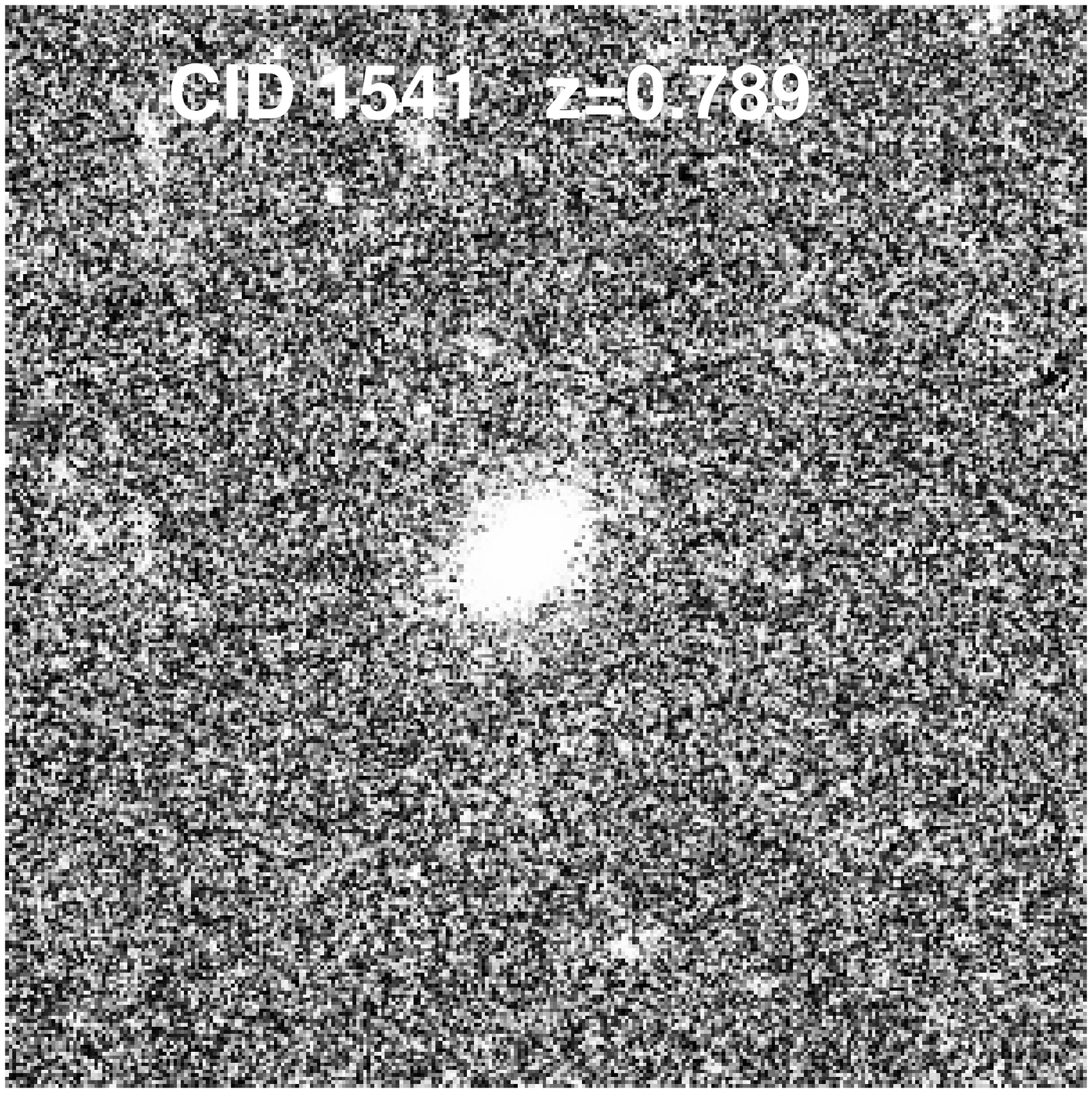}
\includegraphics[width=0.25\textwidth]{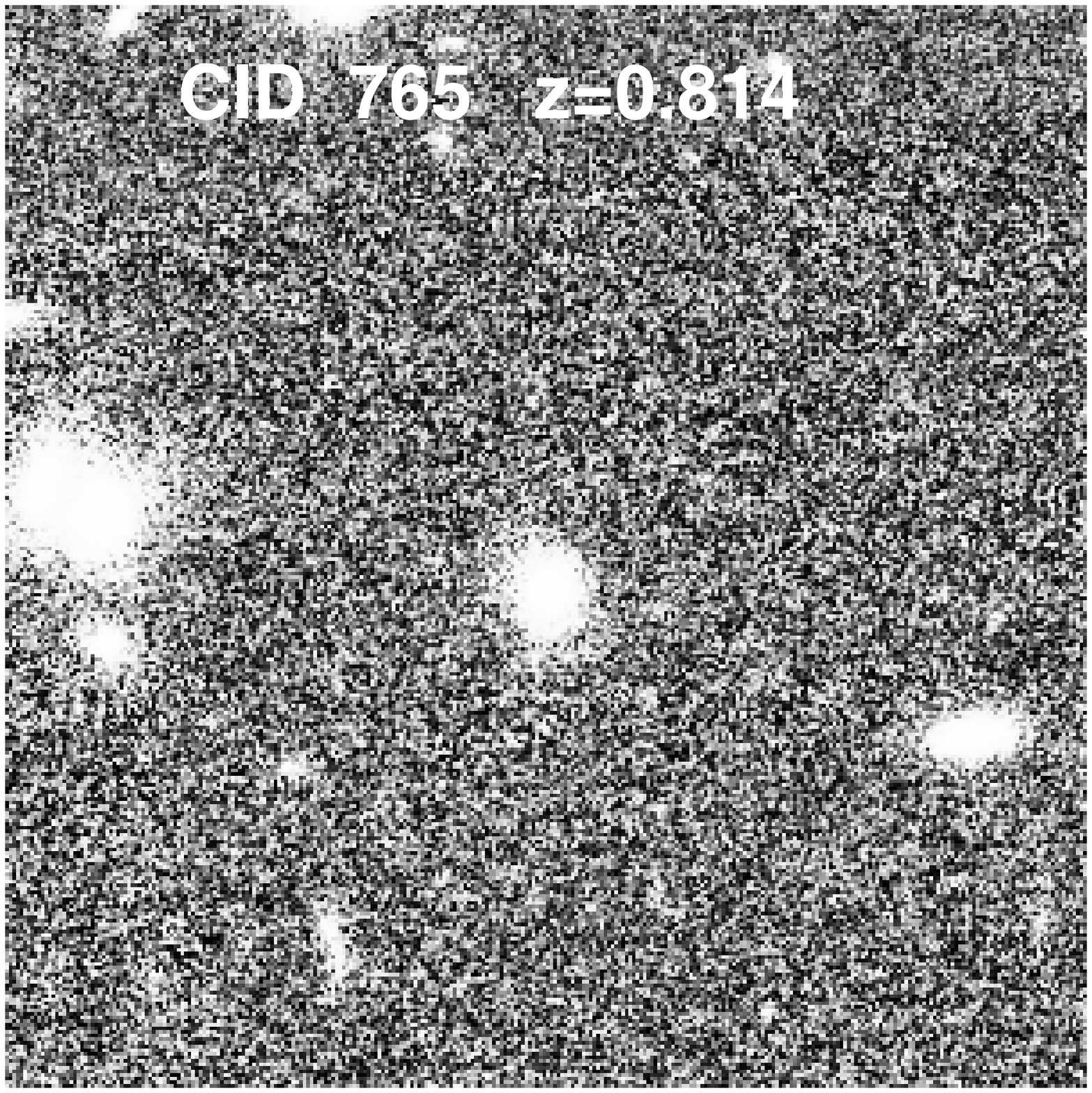}
\includegraphics[width=0.25\textwidth]{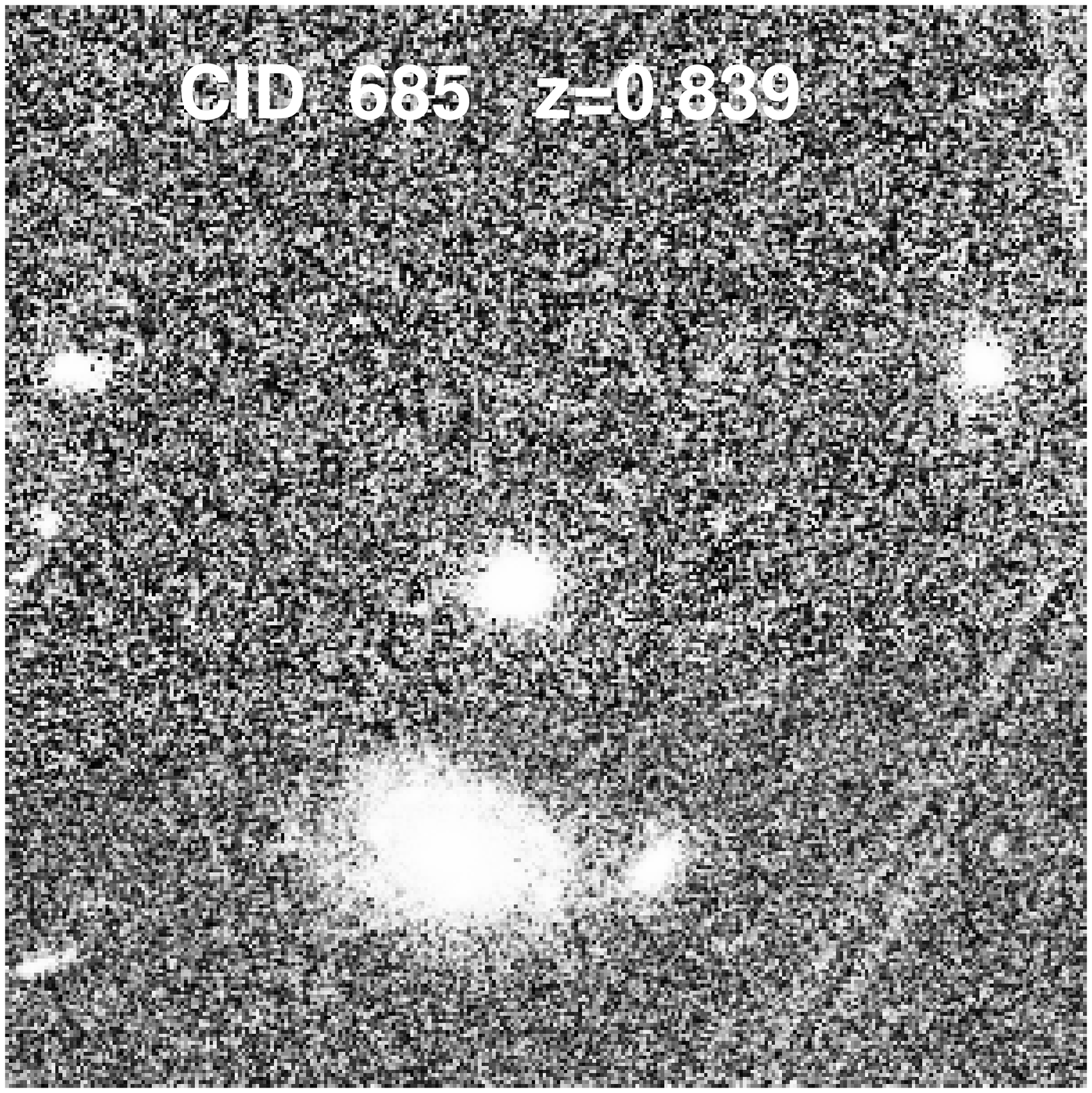}
\includegraphics[width=0.25\textwidth]{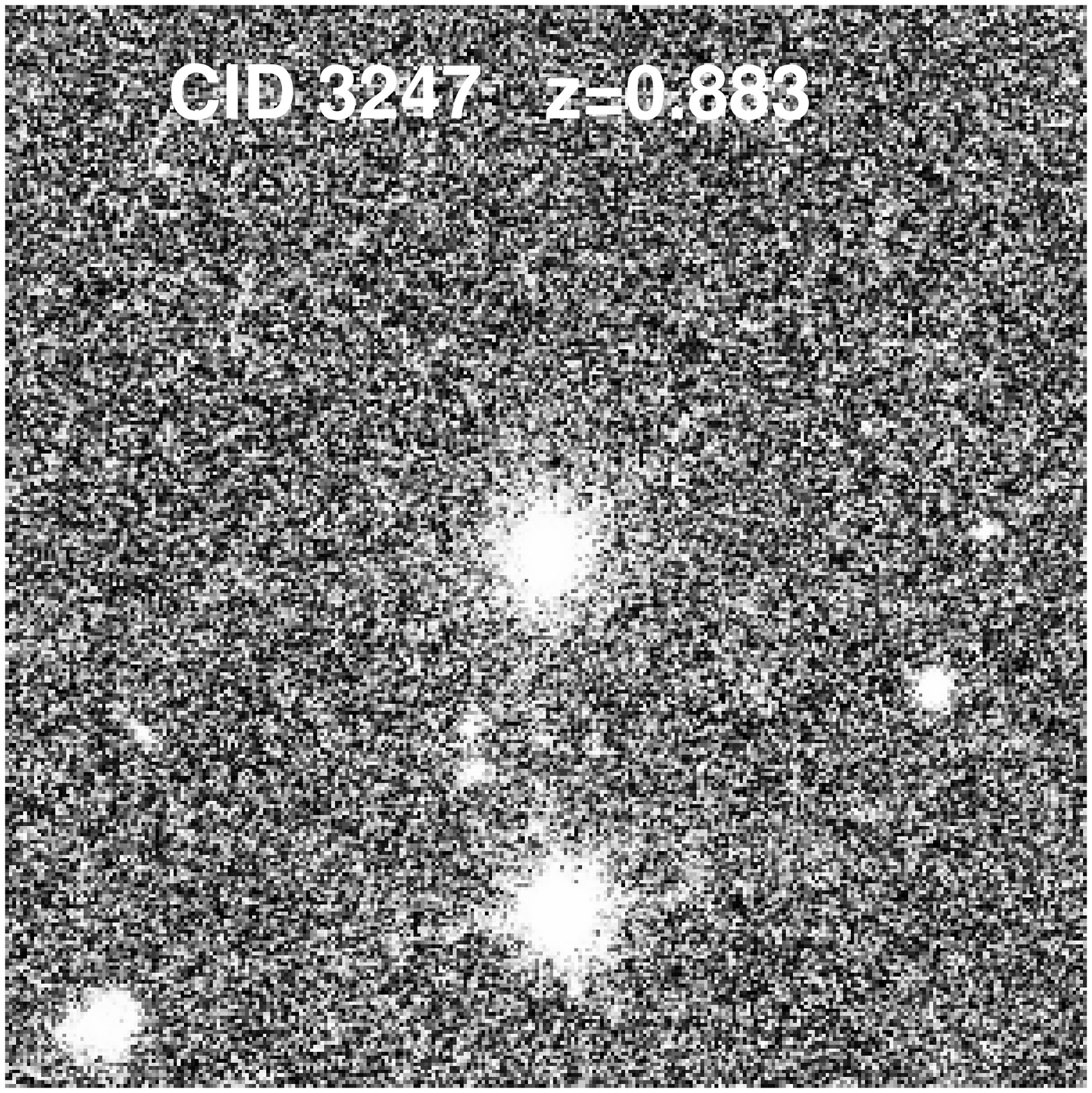}
\includegraphics[width=0.25\textwidth]{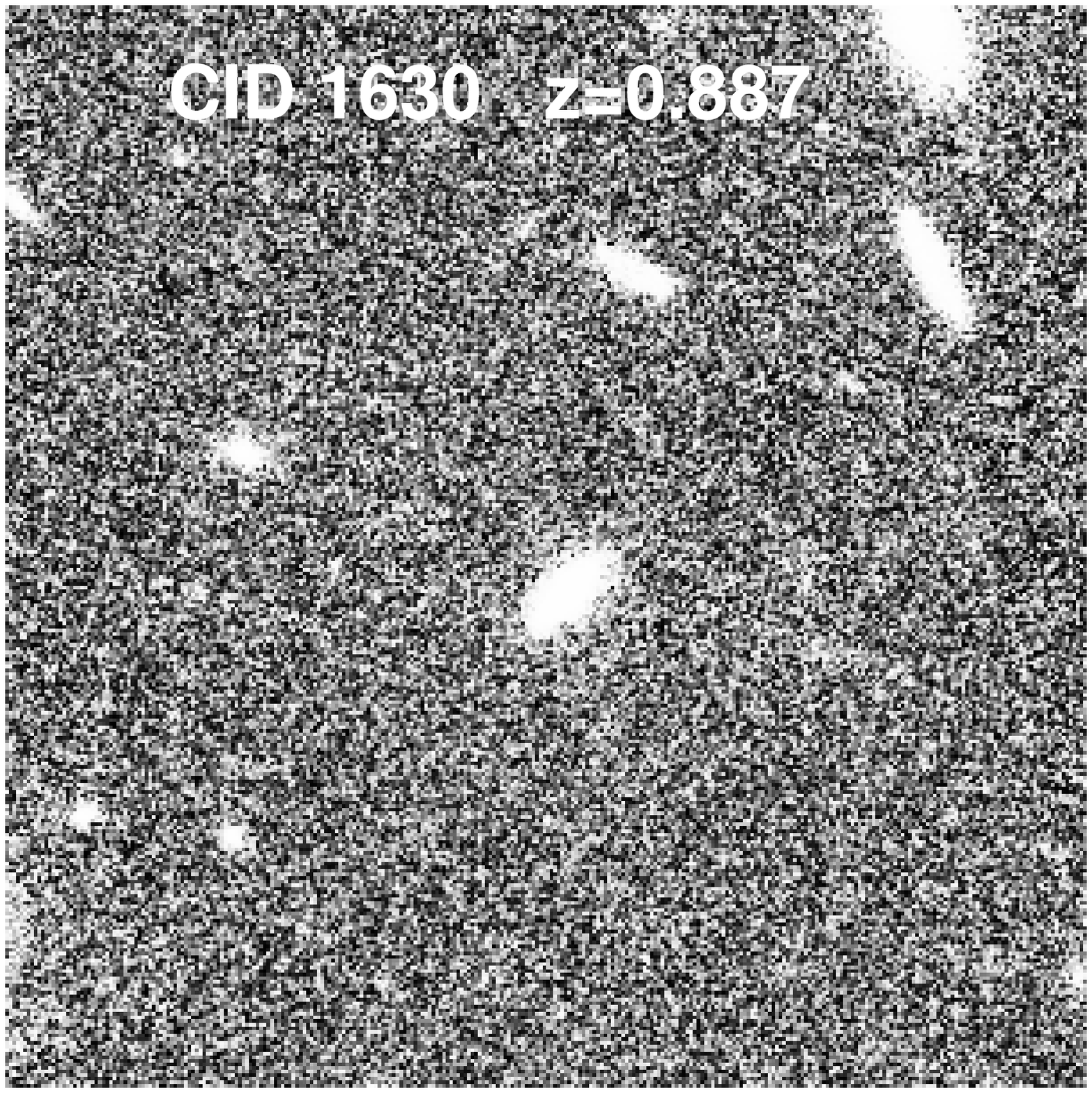}
\caption{\small HST ACS images (15$^{\prime\prime} \times$15$^{\prime\prime}$) of the X-ray ETGs in this paper sorted by increasing redshift.}
\label{fc1}
\end{figure}

\begin{figure}
\centering
\ContinuedFloat
\includegraphics[width=0.25\textwidth]{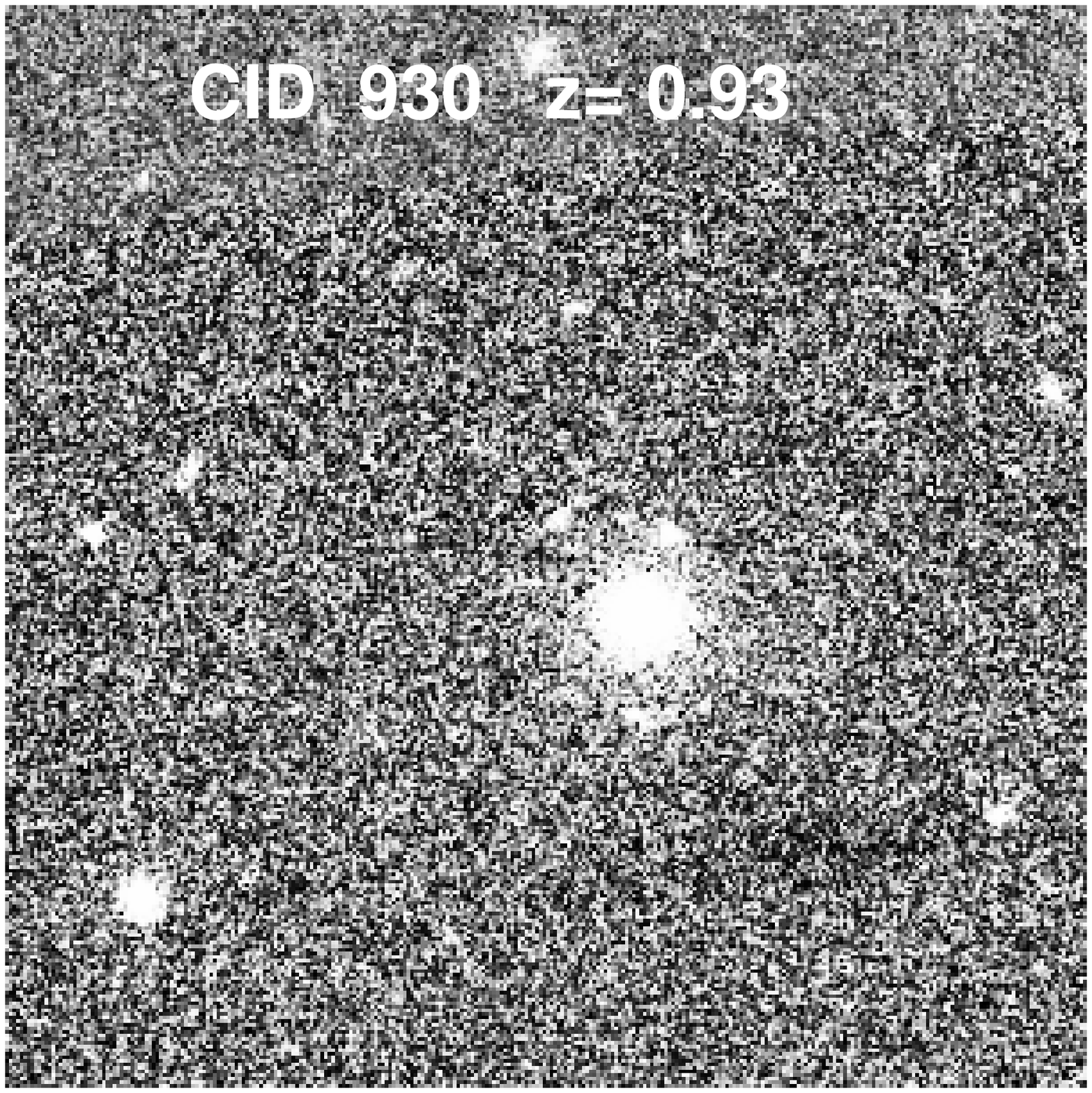}
\includegraphics[width=0.25\textwidth]{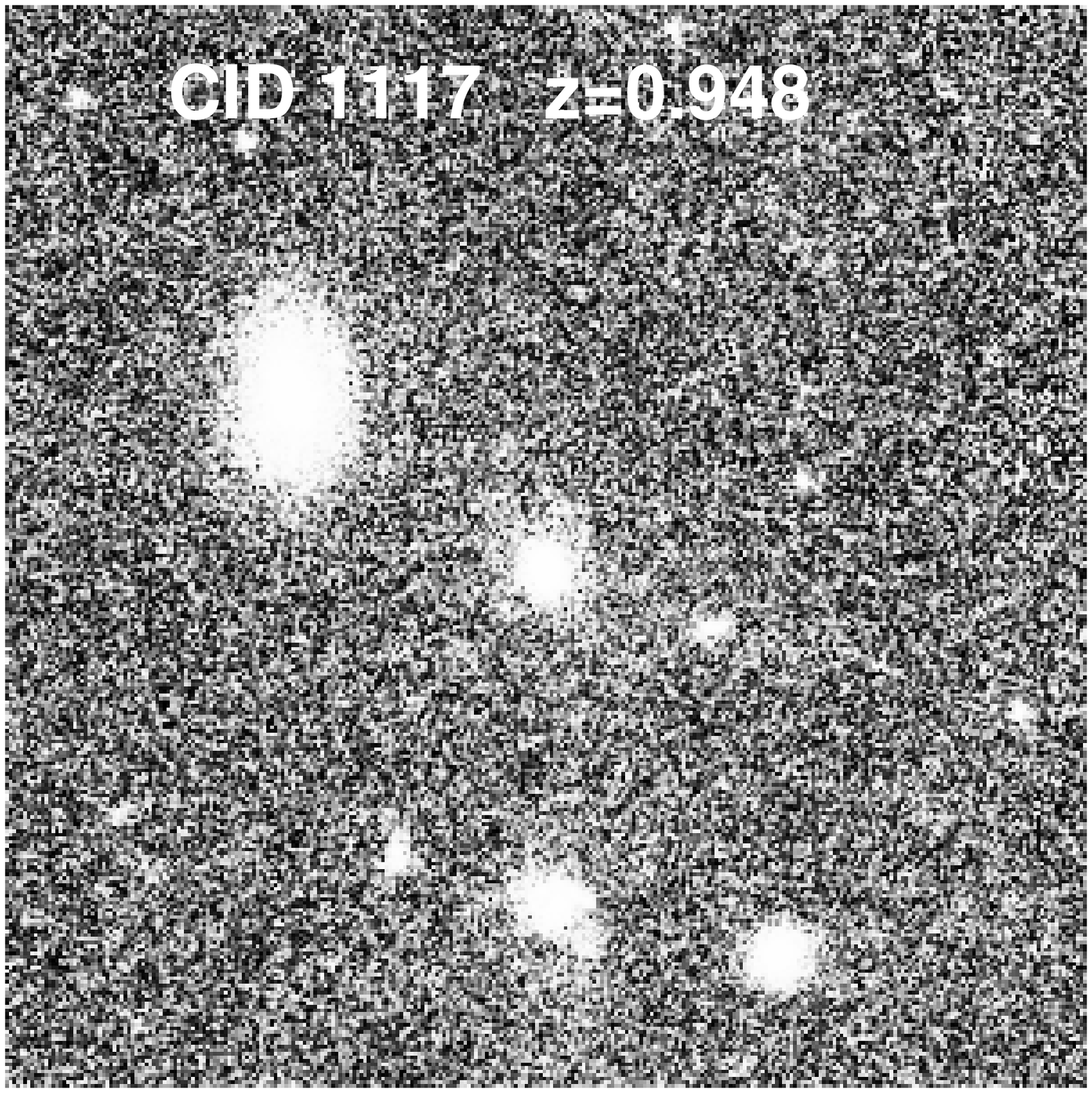}
\includegraphics[width=0.25\textwidth]{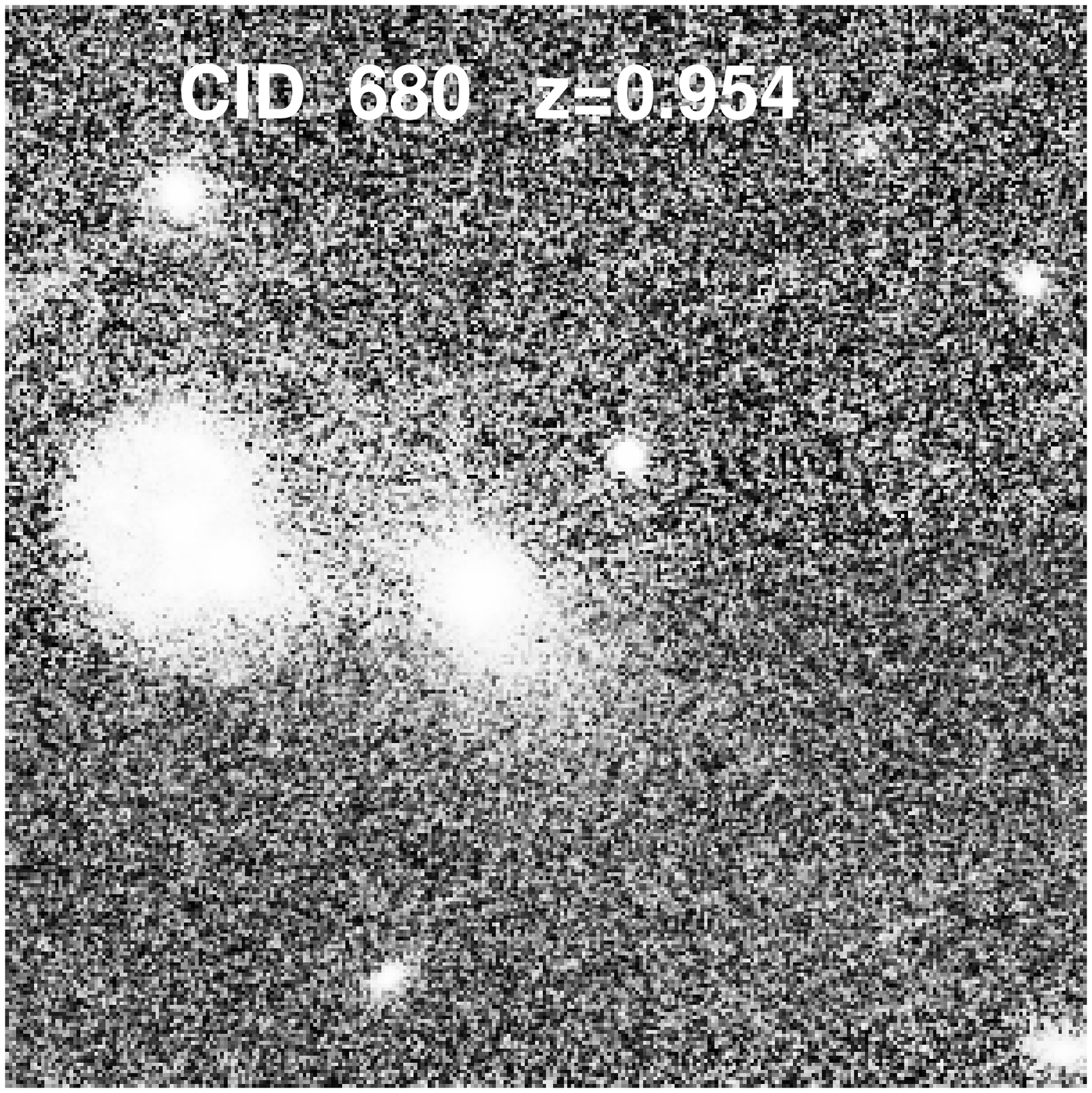}
\includegraphics[width=0.25\textwidth]{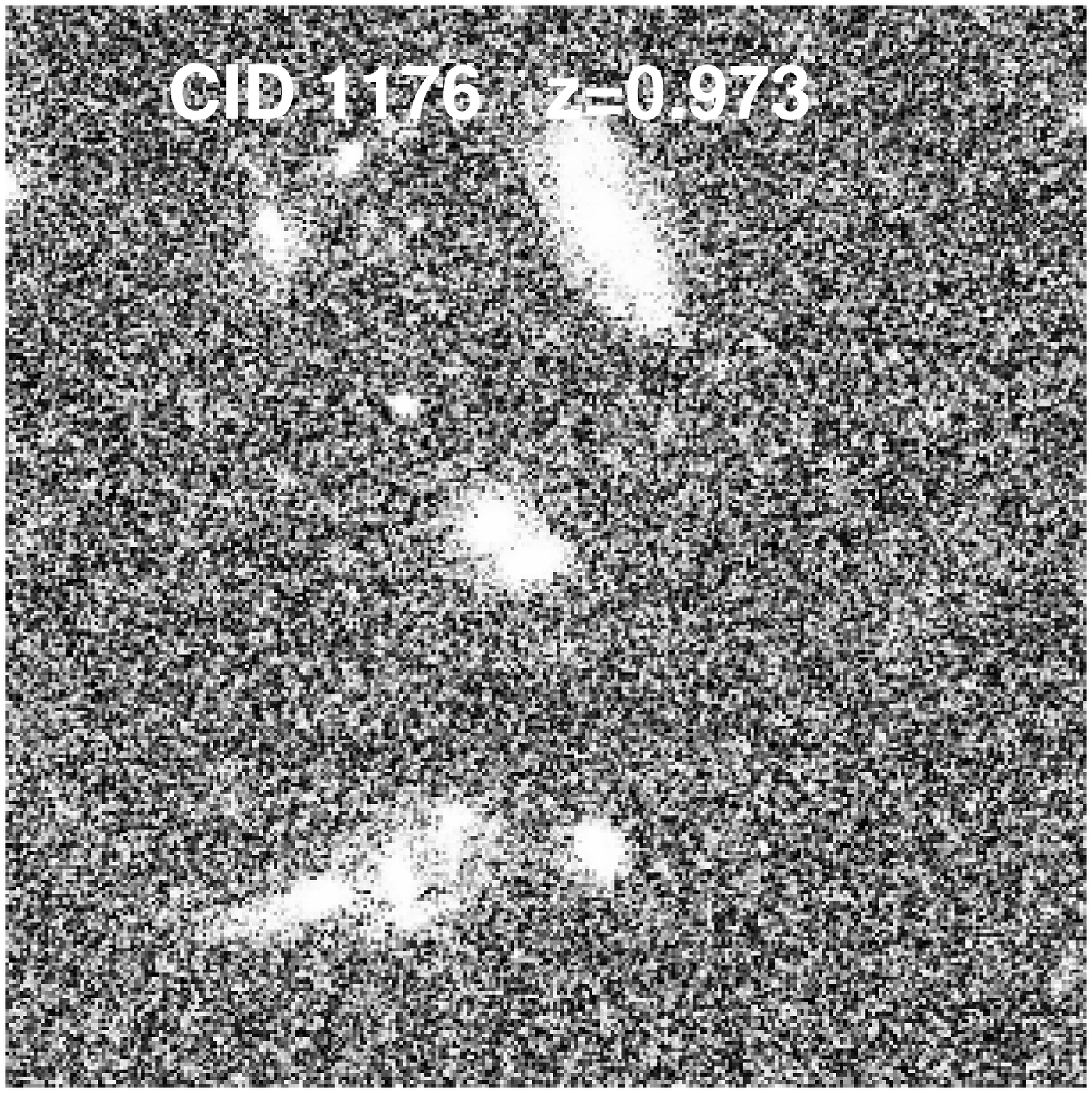}
\includegraphics[width=0.25\textwidth]{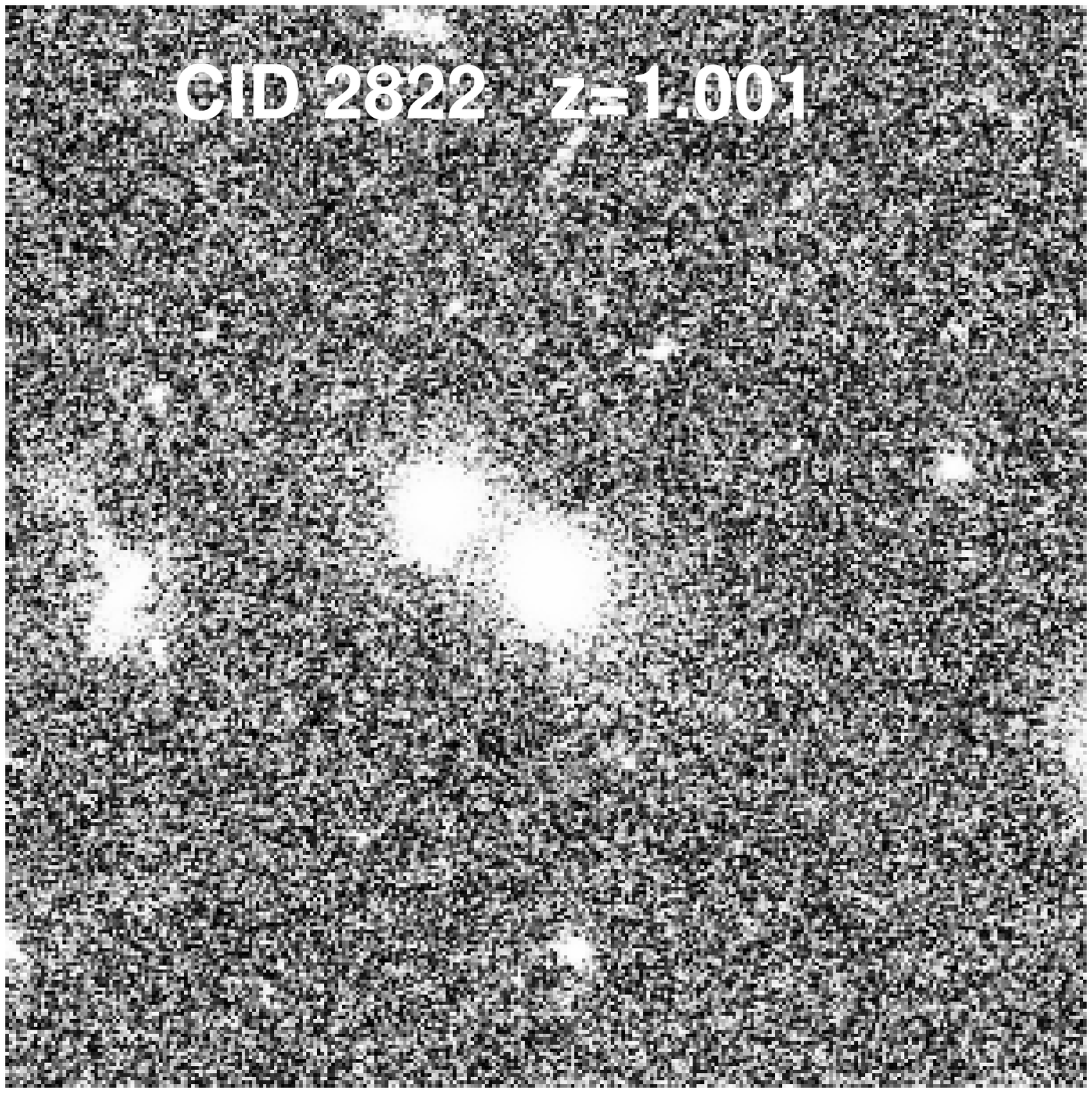}
\includegraphics[width=0.25\textwidth]{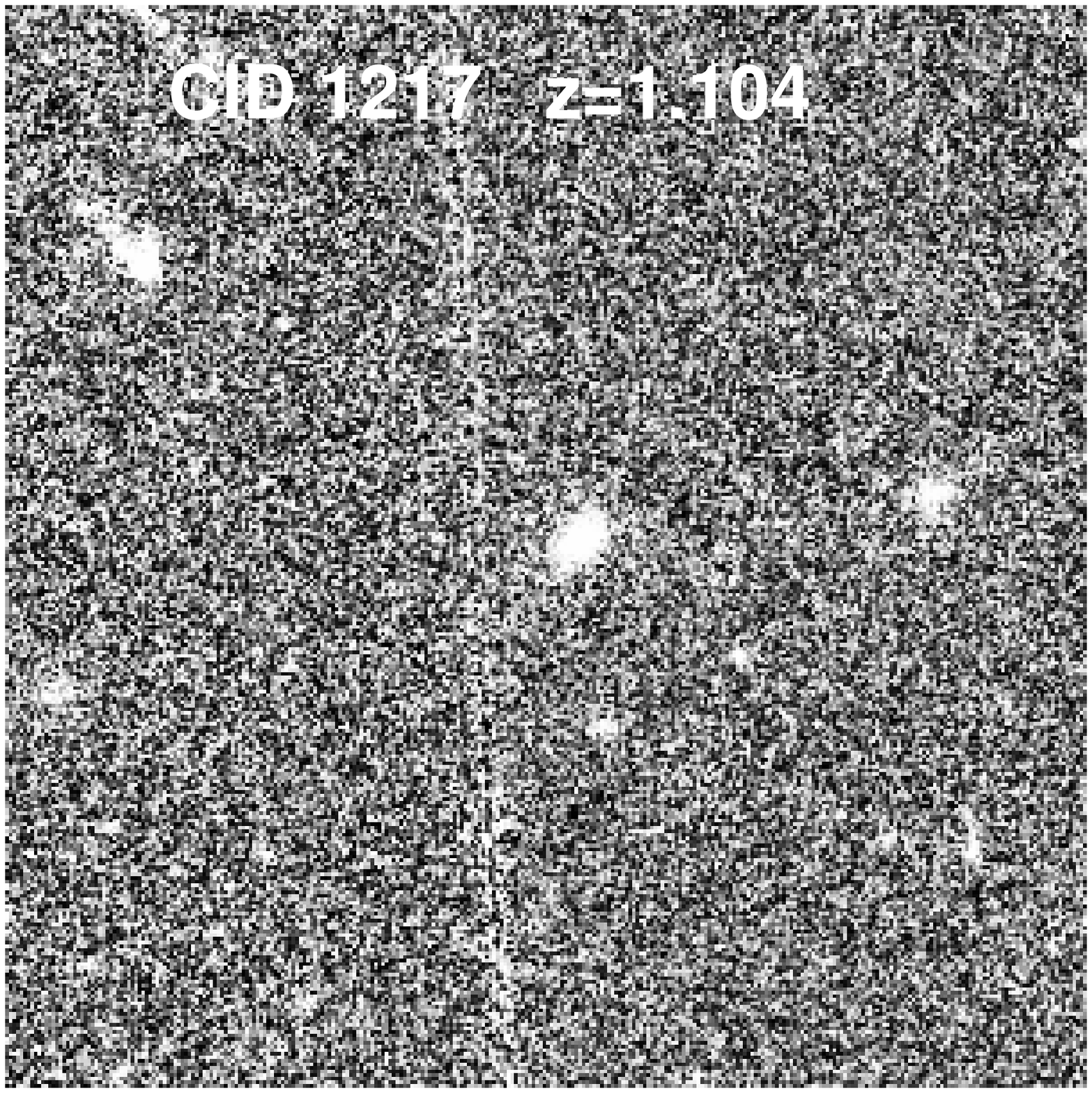}
\includegraphics[width=0.25\textwidth]{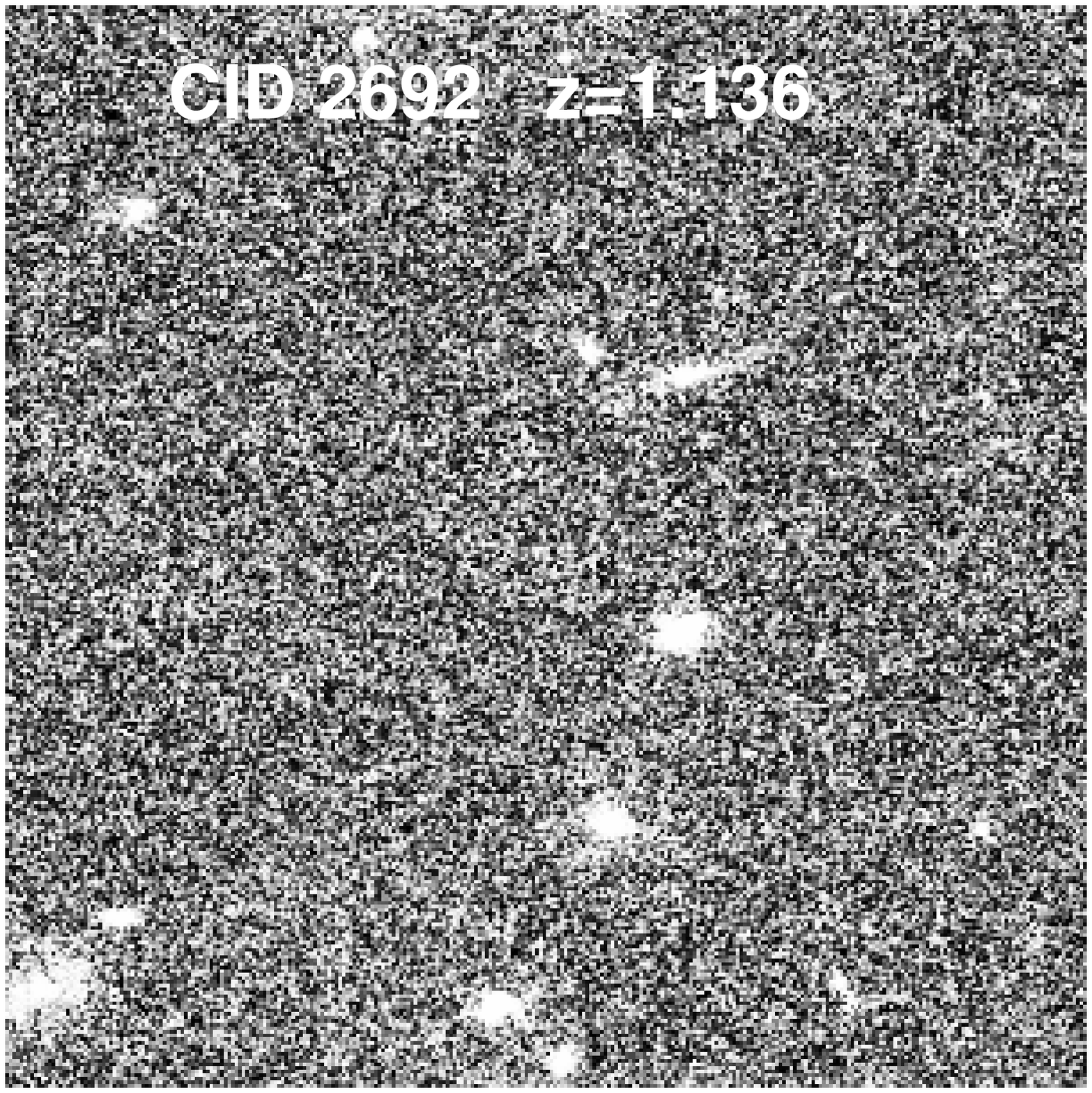}
\includegraphics[width=0.25\textwidth]{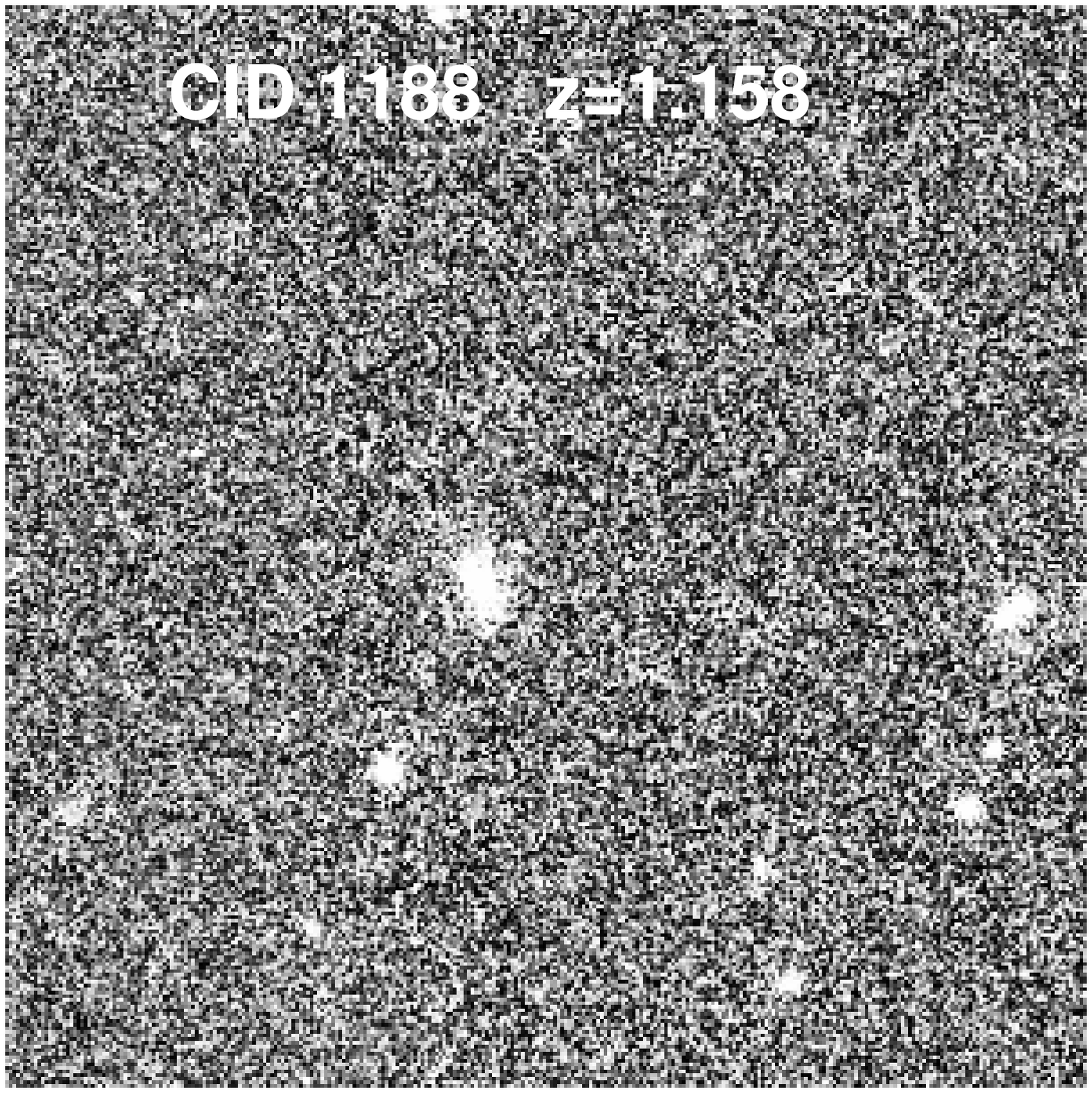}
\includegraphics[width=0.25\textwidth]{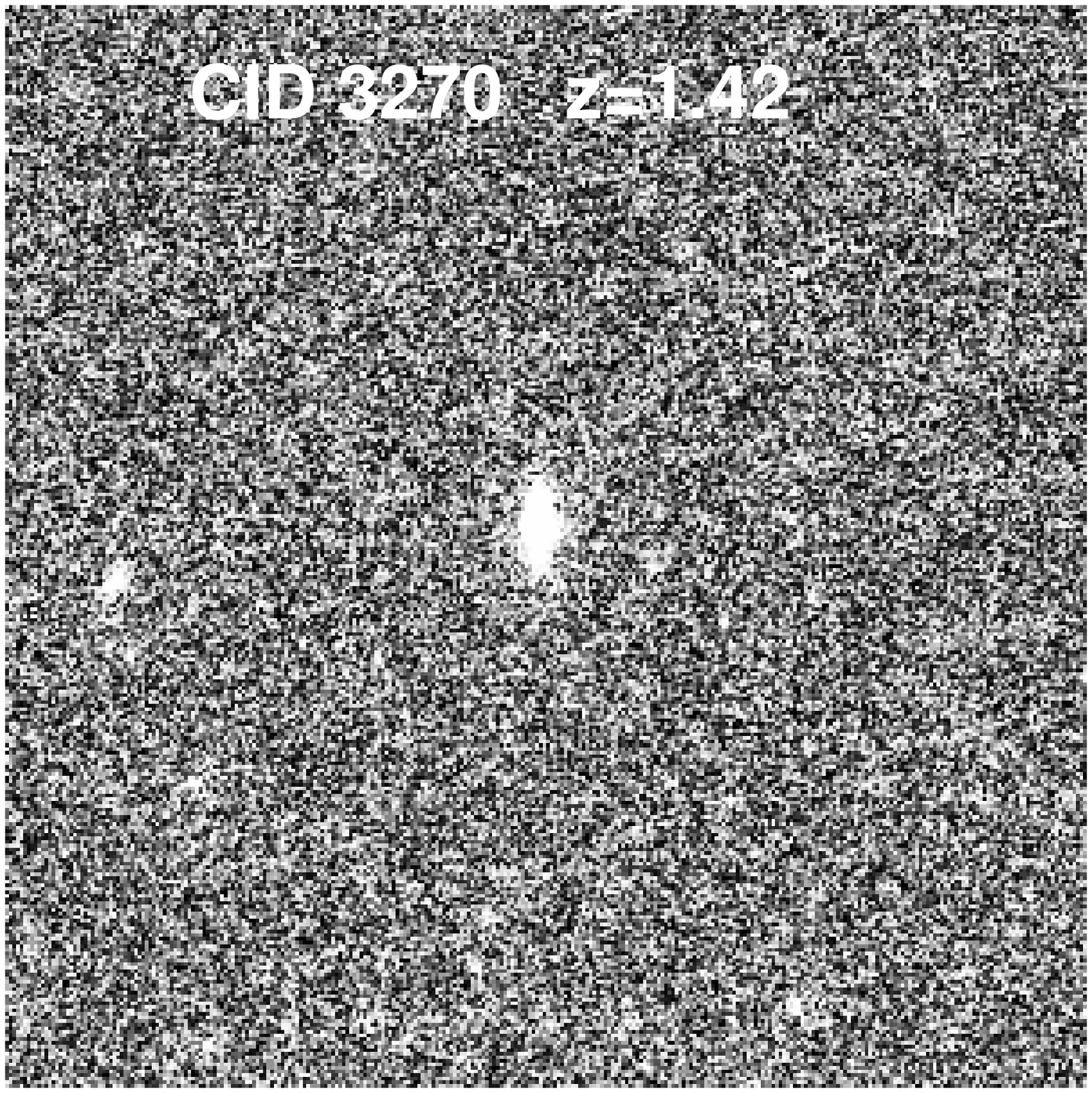}
\caption{\small HST ACS images (15$^{\prime\prime} \times$15$^{\prime\prime}$) of the X-ray ETGs in this paper sorted by increasing redshift.}
\label{fc1}
\end{figure}

\begin{acknowledgements}
The authors thank L. Pozzetti for providing the K-band luminosities as a function of time for the
Bruzual \& Charlot (2003) stellar population models.
FC acknowledges financial support by the NASA contract 11-ADAP11-0218, SP from MIUR grant PRIN 2010-2011, prot. 2010LY5N2T, 
AP by the NASA grant GO1-12125A.  This work was partially supported by NASA contract NAS8-03060 (CXC).

\end{acknowledgements}

\end{document}